\documentclass{fsdthesis}
\fsdthesisSetup{%
  language=en,  
  type=MSc,  
  title={Controller Design of an Airship},  
  subtitle={Reglerentwurf für ein Luftschiff},  
  authorLabelForm=m,  
  author={Manuel Schimmer},
  authorID={03763775},
  supervisorsLabelForm=D,  
  supervisors={Dr.-Ing. Agnes Steinert\\Jakob Bachler, M.Sc.},
  date={2025-06-02},  
  anonymize=false,  
}

\usepackage{blindtext}
\usepackage{currfile}
\usepackage{mathtools}
\usepackage{derivative}
\usepackage{lscape}
\usepackage{pgfplots}
\pgfplotsset{compat=newest}
\usepgfplotslibrary{groupplots}

\usetikzlibrary{shapes.geometric, shapes.arrows}

\DeclareMathOperator*{\argmin}{arg\,min}

\DeclareSIUnit{\rad}{rad}

\newcommand{\Vector}[1]{\bm{#1}}
\newcommand{\Matrix}[1]{\bm{#1}}
\newcommand{\VKR}{\left( \Vector{V}_K^R \right)^E_B}
\newcommand{\VKRdot}{\left( \Vector{\dot{V}}_K^R \right)^{EB}_B}
\newcommand{\om}{\left( \Vector{\omega}_K^{OB} \right)_B}
\newcommand{\omdot}{\left( \Vector{\dot{\omega}}_K^{OB} \right)^B_B}

\newcommand{\pseudo}{\Vector{\nu}}
\newcommand{\pseudodot}{\dot{\Vector{\nu}}}
\newcommand{\udot}{\dot{\Vector{u}}}

\newcommand{\Bsurf}{\Matrix{B}_{surf}}
\newcommand{\usurf}{\Vector{u}_{surf}}

\newcommand*{\imagewidth}{0.7\linewidth}

\newcommand*{\sgamma}{\sin{\gamma_i}}
\newcommand*{\cgamma}{\cos{\gamma_i}}

\newcommand*{\massmatrix}{\overline{\Matrix{M}}_a}

\newacronym{ERP}{ERP}{Fast Exact Redistributed Pseudoinverse Method}
\newacronym{CGI}{CGI}{Cascaded generalized inverse}
\newacronym{AMS}{AMS}{Attainable moment set}
\newacronym{RPI}{RPI}{Redistributed pseudoinverse}
\newacronym{RSPI}{RSPI}{Redistributed scaling pseudoinverse}
\newacronym{NDI}{NDI}{Nonlinear Dynamic Inversion}
\newacronym{INDI}{INDI}{Incremental Nonlinear Dynamic Inversion}
\newacronym{AMI-PGS}{AMI-PGS}{Air Mobility Initiative - Positioning \& Guidance Systems}
\newacronym{CFD}{CFD}{Computational Fluid Dynamics}
\newacronym{ESC}{ESC}{Electronic Speed Controller}
\newacronym{PCH}{PCH}{Pseudo Control Hedging}

\begin{document}


\maketitle


\begin{frontMatter}

  \makeStatutoryDeclaration{}

  \cleardoubleemptypage{}

  \begin{abstract}{en}
    Airships offer unique operational advantages due to their ability to generate lift via buoyancy, enabling low-speed flight and stationary hovering. These capabilities make them ideal for missions requiring endurance and precision positioning. However, they also present significant control challenges: their large, lightweight structures are highly sensitive to environmental disturbances, and conventional aerodynamic control surfaces lose effectiveness during low-speed or hover flight.

    The objective of this thesis is to develop a robust control strategy tailored to a vectored-thrust airship equipped with tiltable propellers. The proposed approach is based on an Extended Incremental Nonlinear Dynamic Inversion inner loop in combination with a high level outer loop, controlling the attitude and velocity of the airship. The proposed method is able to effectively control the airship over the whole envelope, including hover and high speed flight. For this, effective use of available actuators is key. This includes especially the tilt rotors, for which a control allocation method is presented.

    The controller's performance is validated through a series of simulation-based test scenarios, including aggressive maneuvering, gust rejection, atmospheric turbulence, and significant parameter mismatches. The controller is compared against an alternative controller developed at the institute, offering insight into the trade-offs between direct inversion and incremental control approaches. Results demonstrate that the proposed E-INDI controller achieves very good tracking performance and high high robustness against parameter uncertainties.
  \end{abstract}
  \vspace{2em}
  \begin{abstract}{de}
    Luftschiffe bieten aufgrund ihres Auftriebs durch statischen Auftrieb einzigartige operationelle Vorteile. Sie ermöglichen langsame Fluggeschwindigkeiten sowie stationäres Schweben und sind daher ideal für Einsätze geeignet, die Ausdauer und präzise Positionshaltung erfordern. Diese Fähigkeiten gehen jedoch mit besonderen Regelungsherausforderungen einher: Die große und leichte Struktur macht Luftschiffe besonders anfällig gegenüber Umwelteinflüssen, und bei niedrigen Geschwindigkeiten oder im Schwebeflug verlieren konventionelle aerodynamische Steuerflächen aufgrund des geringen Staudrucks ihre Wirkung.

    Ziel dieser Arbeit ist die Entwicklung einer robusten Regelungsstrategie für ein Luftschiff mit schubvektorisierter Antriebskonfiguration und schwenkbaren Propellern. Der vorgeschlagene Ansatz basiert auf einer erweiterten inkrementellen Nichtlinearen Dynamischen Inversion im inneren Regelkreis in Kombination mit einem überlagerten äußeren Regelkreis zur Regelung der Geschwindigkeit und Lage des Luftschiffs. Die entwickelte Methode ermöglicht eine effektive Regelung über den gesamten Flugbereich hinweg, einschließlich Schwebe- und Hochgeschwindigkeitsflug. Dabei ist die effiziente Nutzung der verfügbaren Aktuatoren entscheidend, insbesondere der Schwenkrotoren, für die ein Kontrollallokationsverfahren vorgestellt wird.

    Die Leistungsfähigkeit des Reglers wird anhand mehrerer simulationsbasierter Testszenarien validiert, darunter aggressive Flugmanöver, Windböen, atmosphärische Turbulenz sowie erhebliche Modellunsicherheiten. Der Regler wird mit einem alternativen, am Institut entwickelten Regelungsansatz verglichen. Die Ergebnisse zeigen, dass der vorgeschlagene E-INDI-Regler ein sehr gutes Führungsverhalten sowie eine hohe Robustheit gegenüber Modellunsicherheiten aufweist.
  \end{abstract}

  \makeTableOfContents{}
  \makeListOfFigures{}
  \makeListOfTables{}
  \makeListOfAlgorithms{}
  \makeListOfListings{}
  \makeListOfAcronyms{}
  \makeatletter
\chapter*{\fsdthesis@labels@listOfSymbolsTitle}
\addcontentsline{toc}{chapter}{\fsdthesis@labels@listOfSymbolsTitle}
\markboth{\fsdthesis@labels@listOfSymbolsTitle}{\fsdthesis@labels@listOfSymbolsTitle}%

\begin{longtable}[l]{@{}ll}

  \multicolumn{2}{@{}l}{\textbf{Latin Letters}}                   \\* \tabuphantomline
  \textbf{Symbol}       & \textbf{Description}                    \\*
  $\bm{B}$              & Allocation matrix                       \\
  $c$                   & ERP stage scaling factor                \\
  $d$                   & ERP total scaling factor                \\
  $\bm{f}$              & dynamics                                \\
  $\bm{F}$              & Force                                   \\
  $g$                   & gravitational constant                  \\
  $\bm{J}$              & Inertia matrix of airship               \\
  $\bm{J}_a$            & Apparent inertia matrix of airship      \\
  $\bm{J}_{Ba}$         & Apparent inertia matrix of buoyancy air \\
  $\bm{J}_v$            & Virtual inertia matrix                  \\
  $k$                   & Step counter                            \\
  $L$                   & Roll Moment                             \\
  $m$                   & Mass                                    \\
  $\bm{M}_a$            & Apparent mass matrix of airship         \\
  $\overline{\bm{M}}_a$ & Generalized mass matrix                 \\
  $\bm{M}_{Ba}$         & Apparent mass matrix of buoyancy air    \\
  $\bm{M}_v$            & Virtual mass matrix                     \\
  $\bm{M}$              & Moment                                  \\
  $M$                   & Pitch Moment                            \\
  $N$                   & Yaw Moment                              \\
  $n$                   & Load Factor                             \\
  $p$                   & roll rate                               \\
  $q$                   & pitch rate                              \\
  $r$                   & yaw rate                                \\
  $\bm{r}$              & Position vector                         \\
  $\bm{u}$              & Input vector                            \\
  $u$                   & velocity component in x-direction       \\
  $v$                   & velocity component in y-direction       \\
  $V_B$                 & airship volume                          \\
  $\bm{V}$              & Velocity vector                         \\
  $w$                   & velocity component in z-direction       \\
  $\bm{x}$              & state vector                            \\

  \\  
  \multicolumn{2}{@{}l}{\textbf{Greek Letters}}                   \\* \tabuphantomline
  \textbf{Symbol}       & \textbf{Description}                    \\*
  $\rho$                & density                                 \\
  $\omega$              & Bandwith                                \\
  $\bm{\omega}$         & angular velocity vector                 \\
  $\Omega$              & Rotor RPM                               \\
  $\gamma$              & Rotor tilt angle                        \\
  $\gamma_K$            & Kinematic Climb Angle                   \\
  $\eta$                & Control Surface Deflection              \\
  $\bm{\nu}$            & pseudocontrol vector                    \\
  $\bm{\varPhi}$        & Roll angle                              \\
  $\bm{\varTheta}$      & Pitch angle                             \\
  $\bm{\varPsi}$        & Yaw angle                               \\

  \\  
  \multicolumn{2}{@{}l}{\textbf{Super- and subscripts}}           \\* \tabuphantomline
  \textbf{Symbol}       & \textbf{Description}                    \\*
  $B$                   & body-fixed frame / center of buoyancy   \\
  $C$                   & C-frame                                 \\
  $\epsilon$            & reduced                                 \\
  $F$                   & Control Surfaces                        \\
  $G$                   & center of gravity                       \\
  $K$                   & kinematic                               \\
  $O$                   & NED-frame                               \\
  $P$                   & Propulsion                              \\
  $R$                   & reference point                         \\
  $T$                   & Total                                   \\
  $W$                   & wind                                    \\
\end{longtable}

\end{frontMatter}


\begin{mainContent}

  \chapter{Introduction}
This thesis was conducted at the Institute of Flight System Dynamics at the Technical University of Munich as part of the AMI-PGS research project. Unlike traditional fixed-wing aircraft, airships generate lift primarily through static buoyancy rather than dynamic aerodynamic forces. This fundamental difference allows them to operate efficiently at low speeds and even hover in place, making them particularly well-suited for missions requiring stationkeeping or persistent aerial observation.

However, these advantages are accompanied by distinct challenges. The large volume and light construction necessary for buoyant lift make airships inherently sensitive to wind disturbances. In addition, during low-speed flight or hover, the control surfaces lose effectiveness due to low dynamic pressure, placing a greater burden on the propulsion system for control. To address these issues, the investigated airship relies heavily on a vectored thrust systems, consisting of tiltable propellers, which must be carefully coordinated to maintain stability and maneuverability under various flight conditions.

The primary objective of this thesis is to design and evaluate a control strategy capable of controlling the airship over the full envelope, including hover and high speed flight, even under demanding scenarios and in the presence of disturbances or model uncertainties.

The remainder of this thesis is structured as follows:
Chapter 2 consists of the system description and derives the relevant equations of motion, with a focus on the differences with respect to fixed wing aircraft, as well as decribing the external forces acting on the airship.

Chapter 3 presents the proposed control strategy, describing the working principle of the inner and outer loop as well as introducing the control allocation strategy.

Chapter 4 consists of a comparison of the controller developed in this thesis with another airship controller that was developed at the institute. The champter introduces the working principle of the other controller and compares both controllers in four test scenarios, evaluating their performance in demanding flight conditions.
  \chapter{System Description}\label{chap:system_description}
The airship studied in this thesis was developed as part of the \gls{AMI-PGS} project. It has an overall length of approximately \qty{16}{\meter} and a maximum diameter of about \qty{3.2}{\meter}.
To enable maneuvering, particularly at low speeds and during hovering, the airship is equipped with \num{4} tiltable propellers mounted along its lower half.
Additionally, it features horizontal and vertical stabilizers arranged in an inverted Y-tail configuration at the rear of the airship. Specifically, a single vertical stabilizer extends upward, while two horizontal stabilizers are inclined downward at an angle of \ang{30} relative to the horizontal plane.
The stabilizers are outfitted with control surfaces to enhance maneuverability, especially at higher flight speeds.
An image of the discussed airship is shown in \cref{fig:airship}.

\begin{figure}
    \centering
    \includegraphics[width=\imagewidth]{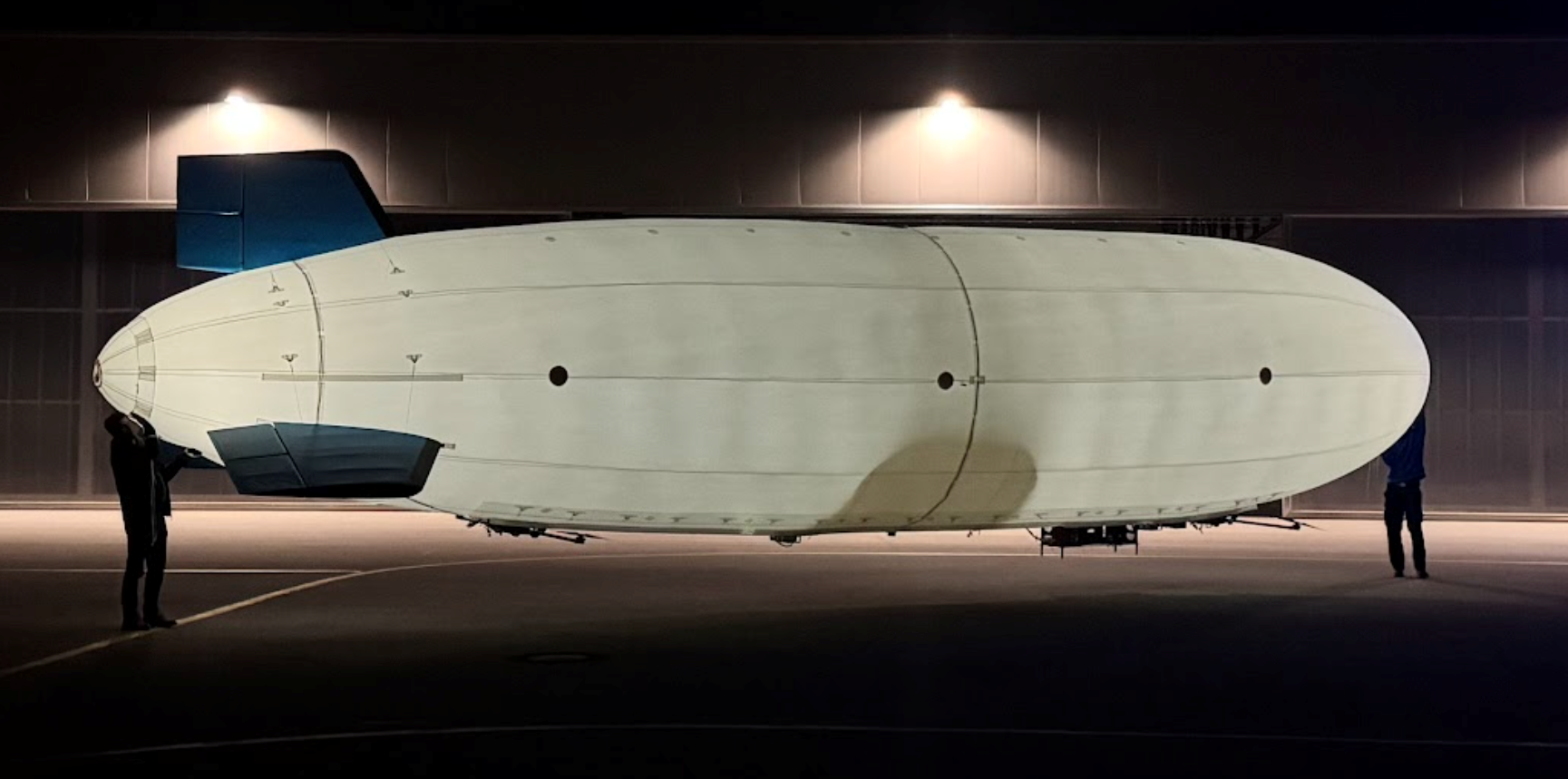}
    \caption{Image of the airship discussed in this thesis}
    \label{fig:airship}
\end{figure}

For the remainder of this chapter, the differences in modeling and flight dynamics between airships and traditional platforms are discussed, and the mathematical description of the system is presented.

\section{Special Characteristics of Airships}
Unlike conventional aircraft, which rely on wings to generate lift through aerodynamic forces, airships achieve lift primarily through buoyancy.
Instead of depending on dynamic lift created by airflow over wings, airships leverage static lift, also known as aerostatic force, to remain airborne.
This force arises from the principle of buoyancy, which describes the upward force exerted by a fluid on an object submerged within it, counteracting the force of gravity.
Unlike aerodynamic lift, which depends on dynamic pressure, static lift remains independent of the airship's velocity relative to the surrounding air.

According to Archimedes' principle, a body immersed in a fluid experiences an upward force equivalent to the weight of the displaced fluid.
In the case of an airship, this fluid is the surrounding atmosphere, and the airship itself behaves as the immersed object, moving through the air.
The magnitude of this buoyant force is given by \cite{kaempf}:

\begin{equation}\label{eq:archimedes} F_{Archimedes} = \rho_{air} V_B g \end{equation}

where $ \rho_{air} $ represents the density of air, $ V_B $ is the volume of displaced air (i.e., the airship's envelope), and $ g $ is the acceleration due to gravity.
Under normal conditions, $ V_B $ and $ g $ remain constant, meaning that variations in air density, which change with altitude, primarily influence the aerostatic force.
To achieve neutral buoyancy, the total mass of the airship must equal the mass of the air it displaces.
Consequently, an airship cannot be filled with ambient air, as the combined weight of its structure and contained gases would exceed that of the displaced volume.
Instead, it is filled with a lifting gas that has a lower density than air, ensuring that the overall weight remains close to the displaced air mass.
To maintain controlled flight, airships are typically designed to be slightly heavier than the displaced air, allowing them to return to the ground naturally.
Helium is the preferred lifting gas due to its inert nature, making it a safer alternative to hydrogen, which poses a fire hazard due to its flammability \cite{helium}.
The airship examined in this study also utilizes helium.

As the airship ascends, the decrease in ambient pressure causes the lifting gas to expand, increasing the internal pressure within the hull.
To regulate this pressure differential, airships are equipped with internal air chambers known as \emph{ballonets}.
These ballonets contain ambient air and serve to manage the expansion of the lifting gas.
At ground level, the ballonets are fully inflated; as the airship gains altitude, they gradually deflate, allowing more space for the lifting gas to expand while keeping the internal pressure stable \cite{buoyancy-issues,Zeppelin}.

The total mass of an airship includes both solid components (e.g. the hull, propulsion system, and structural elements), as well as gaseous components, namely the lifting gas and the air contained within the ballonets.
The net buoyant force, $F_{B,net}$, can be determined by subtracting the combined weight of the lifting gas and ballonet air from the total upward force described by Archimedes' principle, as follows:

\begin{align}\label{eq:f_b_net}
    F_{B,net} & = \rho_{air} V_B g - \overbrace{\rho_{air} V_{ballonet}}^{m_{air, ballonet}} g - \overbrace{\rho_{helium} V_{helium}}^{m_{helium}} g \nonumber \\
    F_{B,net} & = \rho_{air} g \left( V_B - V_{ballonet} \right) - \rho_{helium} V_{helium} g                                                        \nonumber \\
    F_{B,net} & = \rho_{air} g V_{helium} - \rho_{helium} V_{helium} g                                                                               \nonumber \\
    F_{B,net} & = g V_{helium} \left( \rho_{air} - \rho_{helium} \right)
\end{align}

For the remainder of this thesis, this force will be referred to as the \emph{buoyancy force} or \emph{static lift} $\bm{F}_B$.
This force acts through the airship's center of volume and is directed upward, following a north-east-down coordinate frame \cite{kaempf}:

\begin{equation}\label{eq:f_b_vec}
    \left( \bm{F}_B^R \right)_O = \begin{bmatrix}
        0 \\
        0 \\
        -F_{B,net}
    \end{bmatrix}_O
\end{equation}

\section{Mathematial Modeling of the Airship Dynamics}
To enable the simulation of the airship and the development of an appropriate control strategy, its dynamics must first be accurately modeled using mathematical equations.
The motion of the airship is inherently complex and can generally be described by a set of coupled, non-linear differential equations:
\begin{align}
    \dot{\Vector{x}}\left(t\right) & = \Vector{f}\left(\Vector{x}\left(t\right), \Vector{u}\left(t\right)\right) \\
    \Vector{y}\left(t\right)       & = \Vector{h}\left(\Vector{x}\left(t\right)\right)
\end{align}
Here, $\Vector{x}$ is the state vector, $\Vector{u}$ is the input vector and $\Vector{y}$ is the output vector. $\Vector{f}$ is a vector function and captures the dynamics of the system to relate the state and input vector to the rate of change of the state vector $\dot{\Vector{x}}$. $\Vector{h}$ is the output function.
For simplicity, the explicit notation of time dependence is omitted throughout the remainder of this thesis.

The state vector consists of the following \num{9} states:
\begin{equation}
    \Vector{x} = \left[ \VKR \quad \om \quad \Vector{\Phi} \right]^T
\end{equation}
$\VKR$ is the vector of the kinematic velocities in the reference point with respect to the earth expressed in the body-fixed frame, whereas $\om$ is the kinematic angular velocity of the body-fixed frame with respect to the vehicle fixed NED-frame expressed in the body-fixed frame. $\Vector{\Phi} = \left[ \varPhi \quad \varTheta \quad \varPsi \right]^T$ is the attitude vector consisting of the roll, pitch and azimuth angle.
To fully describe the motion of the airship, a position vector would also be required. However, since the controller developed in this thesis does not control the position, it will be omitted here. This is justified because none of the \num{9} states depend on the position.

The input vector consists of the four tilt angles of the propellers, the RPMs of each respective propeller and the deflection angles of the control surfaces:
\begin{equation}
    \Vector{u} = \left[ \gamma_1 \quad \gamma_2 \quad \gamma_3 \quad \gamma_4 \quad \Omega_1 \quad \Omega_2 \quad \Omega_3 \quad \Omega_4 \quad \eta_1 \quad \eta_2 \quad \eta_3 \right]^T
\end{equation}
Later in this thesis a different definition of the input vector is presented for control allocation purposes, as this different definition allows for a linear relationship between the input vector and the produced forces and moments.

The output vector consists of the same components as the state vector:
\begin{equation}
    \Vector{y} = \left[ \VKR \quad \om \quad \Vector{\Phi} \right]^T
\end{equation}

A distinguishing feature of airships compared to conventional aircraft is that the mass of the displaced air is of the same order of magnitude as the airship's own mass.
As the airship moves through the atmosphere, the surrounding air must also be set in motion.
This results in a coupling between the acceleration of the airship and that of the displaced air.
The associated change in linear momentum generates a force that opposes the airship's motion, effectively increasing its perceived inertia.
This phenomenon is referred to as \emph{added mass} or \emph{added inertia}.

The magnitude of the added mass effect depends on the direction of motion, introducing coupling effects between different states of movement \cite{kaempf}.
As a result, the equations of motion for an airship differ from those of a fixed-wing aircraft, as they must account for this additional inertial influence.
Mathematically, added mass and inertia can be represented as diagonal matrices in $ \mathbb{R}^{3 \times 3} $, which are added to the airship's physical properties.
Consequently, rather than being a simple scalar quantity, the airship's mass becomes a matrix, behaving as a direction-dependent property.

To derive the equations of motion for an airship, it is necessary to consider the angular and linear momentum of not only the airship itself but also the displaced air and the surrounding atmosphere.
A comprehensive derivation is provided in \cite{influence_of_wind_speed}, with a correction to the angular momentum equation published in \cite{erratum}. This thesis incorporates the findings from both sources.
The assumptions used for deriving these equations of motion are that the earth can be treated as flat and non-rotating, that the airship behaves like a rigid body with quasi steady mass and quasy steady mass distribution, which means that the rate of change of the mass and inertia tensor is so small that it can be neglected.
Additionally, the equations of motion are formulated relative to a fixed reference point which shall be the center of buoyancy.
The reason for this is that the center of gravity is not constant.
Finally, adjusted to the nomenclature used in this thesis, the equations of motion for the airship are:
{
\small
\begin{flalign}\label{eq:eom}
                    & \begin{bmatrix}
                          \bm{M}_a             & -m \bm{r}^{RG \times} \\
                          m \bm{r}^{RG \times} & \bm{J}_a
                      \end{bmatrix}
    \begin{bmatrix}
        \VKRdot \\
        \omdot
    \end{bmatrix} = & \notag                                                                                                                                                        \\
                    & \begin{bmatrix}
                          \left(\bm{F}^R_T\right)_B -m \om \times \left( \om \times \bm{r}^{RG} \right) - \om \times \bm{M}_a \VKR \\
                          \left(\bm{M}^R_T\right)_B -\om \times \bm{J}_a \om - \om \times \left( \bm{r}^{RG} \times m \VKR \right) + m \VKR \times \left( \bm{r}^{RG} \times \om \right)
                      \end{bmatrix} + \nonumber \\
                    & \begin{bmatrix}
                          \bm{M}_{Ba} \left(\dot{\bm{V}}_W^R\right)_B^{EB} + \om \times \bm{M}_{Ba} \left(\bm{V}_W^R\right)_B^E \\
                          \bm{J}_{Ba} \left(\dot{\bm{\omega}}_W^{OB}\right)_B + \om \times \bm{J}_{Ba} \left(\bm{\omega}_W^{OB}\right)_B
                      \end{bmatrix}
\end{flalign}
}

In the equation above, $\Matrix{M}_a$ and $\Matrix{J}_a$ denote the airship's apparent mass and inertia matrices, and $\Vector{r}^{RG}$ represents the vector from the reference point to the center of gravity. The terms $\Matrix{M}_{Ba}$ and $\Matrix{J}_{Ba}$ correspond to the apparent mass and inertia matrices of the buoyancy air, associated with the displaced air.
$\VKR$ is the kinematic velocity of the reference point relative to a flat, non-rotating Earth, while $\om$ describes the angular velocity of the body frame with respect to the vehicle-fixed NED frame.
The time derivatives $\VKRdot$ and $\omdot$ of $\VKR$ and $\om$ are expressed relative to the body-fixed frame.
Quantities indexed with a $W$ are related to the wind, where $\left(\bm{V}_W^R\right)_B^{EB}$ denotes the wind velocity and $\left(\bm{\omega}_W^{OB}\right)_B$ represents the wind's angular velocity.

The total external forces and moments acting on the airship consist of aerodynamic, aerostatic, gravitational, propulsive and control surfaces contributions:
\begin{align}
    \left(\bm{F}^R_T\right)_B & = \left(\bm{F}^R_A\right)_B + \left(\bm{F}^R_B\right)_B + \left(\bm{F}^R_G\right)_B + \left(\bm{F}^R_P\right)_B + \left(\bm{F}^R_F\right)_B \label{eq:total_forces}  \\
    \left(\bm{M}^R_T\right)_B & = \left(\bm{M}^R_A\right)_B + \left(\bm{M}^R_B\right)_B + \left(\bm{M}^R_G\right)_B + \left(\bm{M}^R_P\right)_B + \left(\bm{M}^R_F\right)_B \label{eq:total_moments}
\end{align}
where the subscripts $A$, $B$, $G$, $P$ and $F$ denote aerodynamic, aerostatic, gravitational, propulsive and control surface forces and moments, respectively.

In the leftmost matrix of \cref{eq:eom}, the term $\bm{r}^{RG \times}$ denotes the skew-symmetric matrix that represents the cross product operation of $\bm{r}^{RG}$ with another vector.
\begin{equation}
    \bm{r}^{RG \times} = \begin{bmatrix}
        0       & -z^{RG} & y^{RG}  \\
        z^{RG}  & 0       & -x^{RG} \\
        -y^{RG} & x^{RG}  & 0
    \end{bmatrix}.
\end{equation}

The apparent mass matrix $\bm{M}_a$ of the airship accounts for both its physical mass and the added mass effects from the surrounding air.
It is given by \begin{equation} \bm{M}_a = m \bm{E} + \bm{M}_v \end{equation} where $\bm{E}$ is the $3 \times 3$ identity matrix, and $\bm{M}_v$ represents the virtual (added) mass matrix.
Similarly, the airship's apparent inertia tensor $\bm{J}_a$ is the sum of its physical inertia tensor and the virtual inertia tensor $\bm{J}_v$: \begin{equation} \bm{J}_a = \bm{J} + \bm{J}v. \end{equation} In an analogous manner, the apparent mass and inertia matrices of the displaced air are expressed as
\begin{align}
    \bm{M}_{Ba} & = m_B \bm{E} + \bm{M}_v \\
    \bm{J}_{Ba} & = \bm{J}_B + \bm{J}_v
\end{align}
where $m_B$ denotes the mass and $\bm{J}_B$ the inertia tensor of the air displaced by the airship.

In the simplified special case where the hull of the airship is approximated as an ellipsoid, the virtual mass and inertia matrices can be expressed as:
\begin{equation}
    \Matrix{M}_v = V_B \begin{bmatrix}
        k_1 & 0   & 0   \\
        0   & k_2 & 0   \\
        0   & 0   & k_2
    \end{bmatrix} \qquad \Matrix{J}_v = \begin{bmatrix}
        0 & 0   & 0   \\
        0 & k_3 & 0   \\
        0 & 0   & k_3
    \end{bmatrix}
\end{equation}
where $V_B$ is the volume of the airship hull, and $k_1$, $k_2$, and $k_3$ are known as ellipsoid inertia factors, obtained from potential flow analysis around the ellipsoid.
For a more detailed discussion on the modeling of these factors, the interested reader is referred to \cite{lamb1918inertia,munk1924aerodynamics,tuckerman1924inertia}.

For the purpose of controller development, certain simplifications are applied to the equations of motion. First, the wind-related terms and the aerodynamic forces and moments are neglected from this point onward. Second, by applying the Jacobi identity to the angular dynamics in \cref{eq:eom}, a more compact and simplified form is obtained. The Jacobi identity is given by:
\begin{equation}
    \vec{a} \times \left(\vec{b} \times \vec{c}\right) + \vec{b} \times \left(\vec{c} \times \vec{a}\right) + \vec{c} \times \left(\vec{a} \times \vec{b}\right) = 0
\end{equation}
By setting $\vec{a} = \om$, $\vec{b} = \bm{r}^{RG}$ and $\vec{c} = m \VKR$, the Jacobi identity is:
{
\footnotesize
\begin{align}\label{eq:jacobi_identity}
     & 0                                                        = \om \times \left(\bm{r}^{RG} \times m \VKR\right) + \bm{r}^{RG} \times \left(m \VKR \times \om\right) + m \VKR \times \left(\om \times \bm{r}^{RG}\right) \nonumber \\
     & \bm{r}^{RG} \times \left(m \VKR \times \om\right)  = - \om \times \left(\bm{r}^{RG} \times m \VKR\right) - m \VKR \times \left(\om \times \bm{r}^{RG}\right)                 \nonumber                                         \\
     & \bm{r}^{RG} \times \left(m \VKR \times \om\right)  = - \om \times \left(\bm{r}^{RG} \times m \VKR\right) + m \VKR \times \left(\bm{r}^{RG} \times \om\right)
\end{align}
}
Applying the above-mentioned simplifications, the equations of motion reduce to:
\begin{equation}
    \begin{split}
         & \overline{\Matrix{M}}_a
        \begin{bmatrix}
            \VKRdot \\
            \omdot
        \end{bmatrix} = \begin{bmatrix}
                            \left(\bm{F}^R_B\right)_B + \left(\bm{F}^R_G\right)_B + \left(\bm{F}^R_P\right)_B+ \left(\bm{F}^R_F\right)_B \\
                            \left(\bm{M}^R_B\right)_B + \left(\bm{M}^R_G\right)_B + \left(\bm{M}^R_P\right)_B + \left(\bm{M}^R_F\right)_B
                        \end{bmatrix} + \\
         & \begin{bmatrix}
               -m \om \times \left( \om \times \bm{r}^{RG} \right) - \om \times \bm{M}_a \VKR \\
               -\om \times \bm{J}_a \om + \bm{r}^{RG} \times \left(m \VKR \times \om\right)
           \end{bmatrix}
    \end{split}
\end{equation}
where $\overline{\Matrix{M}}_a$ is the generalized mass matrix of the airship:
\begin{equation}
    \overline{\Matrix{M}}_a = \begin{bmatrix}
        \bm{M}_a             & -m \bm{r}^{RG \times} \\
        m \bm{r}^{RG \times} & \bm{J}_a
    \end{bmatrix}
\end{equation}

In addition to the equations of motion, the attitude differential equations must be considered to fully define the airship's state vector.
Unlike the equations of motion, the attitude differential equations describe a purely kinematic relationship.
As a result, they are identical to those used in the fixed-wing aircraft case.

The attitude differential equations are given by \cite{FSD2}:
\begin{equation}\label{eq:attitude_dgl}
    \dot{\Vector{\Phi}} = \begin{bmatrix}
        \dot{\varPhi}   \\
        \dot{\varTheta} \\
        \dot{\varPsi}
    \end{bmatrix} = \begin{bmatrix}
        1 & \sin{\varPhi} \tan{\varTheta}         & \cos{\varPhi} \tan{\varTheta}         \\
        0 & \cos{\varPhi}                         & -\sin{\varPhi}                        \\
        0 & \frac{\sin{\varPhi}}{\cos{\varTheta}} & \frac{\cos{\varPhi}}{\cos{\varTheta}}
    \end{bmatrix} \begin{bmatrix}
        p \\
        q \\
        r
    \end{bmatrix}
\end{equation}
It should be noted that the presence of the $\cos{\varTheta}$ term in the denominator of the last row introduces a singularity when $\varTheta = \ang{\pm 90}$.
This issue could be resolved by using a quaternion-based representation of attitude.
However, such an approach would increase the overall system complexity without offering practical advantages, as the airship typically operates at pitch angles well below this limit, generally not exceeding \ang{30}.

Regarding the actuators of the system, there are three types used on the airship: the tilt mechanisms, the electric motors driving the propellers, and the servos actuating the control surfaces on the fins.
Both the tilt actuators and the control surface actuators are modeled as first-order systems. In contrast, the electric motors for the propellers are modeled in greater detail, including the simulation of the \gls{ESC} with its underlying electrical circuits and the solution of the mechanical equations of motion of the motor.
However, because this level of detail is too complex for controller development, the motor dynamics are also approximated by a first-order system for control design purposes.
The natural frequency of the tilt actuators is \qty{5}{\radian\per\second}, while the control surface actuators operate significantly faster, with a natural frequency of \qty{25}{\radian\per\second}. The natural frequency of the propeller motors is approximated as \qty{20}{\radian\per\second}.
Each actuator has absolute and rate limits, i.e. limits on how far it can move and how fast it can move.
The respective limits are listed in \cref{tab:actuator_limits}.

\begin{table}
    \centering
    \begin{tabular}{|c|c|}
        \hline
        Quantity                                                              & Value                            \\
        \hline
        Maximum RPM $\Omega_{\text{max}}$                                     & \qty{340}{\radian\per\second}    \\
        Minimum RPM $\Omega_{\text{min}}$                                     & \qty{0}{\radian\per\second}      \\
        Maximum rate of change of RPM  $\dot{\Omega}_{\text{max}}$            & \qty{156}{\radian\per\second^2}  \\
        Minimum rate of change of RPM  $\dot{\Omega}_{\text{min}}$            & \qty{-135}{\radian\per\second^2} \\
        Maximum tilt angle $\gamma_{\text{max}}$                              & \ang{255}                        \\
        Minimum tilt angle $\gamma_{\text{min}}$                              & \ang{-75}                        \\
        Maximum rate of change of tilt angle  $\dot{\gamma}_{\text{max}}$     & \qty{45}{\degree\per\second}     \\
        Minimum rate of change of tilt angle  $\dot{\gamma}_{\text{min}}$     & \qty{-45}{\degree\per\second}    \\
        Maximum deflection angle $\eta_{\text{max}}$                          & \ang{40}                         \\
        Minimum deflection angle $\eta_{\text{min}}$                          & \ang{-40}                        \\
        Maximum rate of change of deflection angle  $\dot{\eta}_{\text{max}}$ & \qty{300}{\degree\per\second}    \\
        Minimum rate of change of deflection angle  $\dot{\eta}_{\text{min}}$ & \qty{-300}{\degree\per\second}   \\
        \hline
    \end{tabular}
    \caption{Physical limits of the actuators}
    \label{tab:actuator_limits}
\end{table}

To streamline the notation used throughout the remainder of this thesis, the following conventions are introduced:
\begin{enumerate}
    \item If not specified, the frame in which a quantity is expressed is the body-fixed frame
    \item The linear velocity is, if not specified otherwise, meant as the kinematic velocity in the reference point with respect to the earth, therefore $\bm{V} = \VKR$
    \item The angular velocity is, if not specified otherwise, meant as the kinematic angular velocity of the body-fixed frame $B$ with respect to the NED-frame $O$, therefore $\bm{\omega} = \om$
    \item Forces and moments are expressed in the reference point, if not mentioned otherwise
    \item If not specified, time derivatives are taken with respect to the body-fixed frame, i.e. $\dot{\Vector{V}} = \VKRdot$ and $\dot{\Vector{\omega}} = \omdot$
\end{enumerate}

Using these conventions, the simplified equations of motion are given by:
\begin{equation}\label{eq:eom_simple}
    \begin{bmatrix}
        \dot{\bm{V}} \\
        \dot{\bm{\omega}}
    \end{bmatrix} = \massmatrix^{-1} \begin{bmatrix}
        \bm{F}_B + \bm{F}_G + \bm{F}_P + \bm{F}_F -m \bm{\omega} \times \left( \bm{\omega} \times \bm{r}^{RG} \right) - \bm{\omega} \times \bm{M}_a \bm{V} \\
        \bm{M}_B + \bm{M}_G + \bm{M}_P + \bm{M}_F -\bm{\omega} \times \bm{J}_a \bm{\omega} + \bm{r}^{RG} \times \left(m \bm{V} \times \bm{\omega}\right)
    \end{bmatrix}
\end{equation}

\section{External Forces and Moments}
In this section the external forces and moments shall be briefly described.
As already outlined in \cref{eq:total_forces,eq:total_moments} the external forces and moments are made up of the aerodynamic, aerostatic, gravitational, propulsive and control surfaces contributions.

\subsection{Aerostatic Force}
The aerostatic force was already introduced in \cref{eq:f_b_vec} and only needs to be transformed into the body-fixed frame with the rotation matrix $\Matrix{M}_{BO}$ which transforms a vector from the NED-frame to the body-fixed frame.
The matrix can be contructed from 3 individual rotation matrices, in the sequence of the euler angles $\varPhi$, $\varTheta$, $\varPsi$ as follows \cite{FSD2}:
\begin{equation}
    \Matrix{M}_{BO} = \begin{bmatrix}
        1 & 0              & 0             \\
        0 & \cos{\varPhi}  & \sin{\varPhi} \\
        0 & -\sin{\varPhi} & \cos{\varPhi}
    \end{bmatrix} \begin{bmatrix}
        \cos{\varTheta} & 0 & -\sin{\varTheta} \\
        0               & 1 & 0                \\
        \sin{\varTheta} & 0 & \cos{\varTheta}
    \end{bmatrix} \begin{bmatrix}
        \cos{\varPsi}  & \sin{\varPsi} & 0 \\
        -\sin{\varPsi} & \cos{\varPsi} & 0 \\
        0              & 0             & 1
    \end{bmatrix}
\end{equation}
The aerostatic force expressed in the body-fixed frame therefore is
\begin{equation}
    \left(\bm{F}^R_B\right)_B = \Matrix{M}_{BO} \begin{bmatrix}
        0 \\
        0 \\
        -F_{B,net}
    \end{bmatrix}_O
\end{equation}
The aerostatic moment in the reference point is zero, since the force acts exactly in this point and therefore the lever arm vanishes.

\subsection{Gravity}
The gravitational acceleration is assumed to be constant with a value of \qty[per-mode=fraction]{9.81}{\meter\per\second^2}.
Since gravity force is conveniently expressed in the NED-frame, similar to the aerostatic force the transformation matrix $\Matrix{M}_{BO}$ needs to be applied.
The resulting moment at the reference point is obtained by the cross product of the vector pointing from the reference point to the center of gravity, where the gravitational force is acting.
Therefore, the gravitational force and moment are:
\begin{align}
    \left(\bm{F}^R_G\right)_B & = \Matrix{M}_{BO} \begin{bmatrix}
                                                      0 \\
                                                      0 \\
                                                      m g
                                                  \end{bmatrix}                        \\
    \left(\bm{M}^R_G\right)_B & = \Vector{r}^{RG} \times \Matrix{M}_{BO} \begin{bmatrix}
                                                                             0 \\
                                                                             0 \\
                                                                             m g
                                                                         \end{bmatrix}
\end{align}

\subsection{Aerodynamics}
The aerodynamic forces and moments are inherently complex and non-analytical, which means that no simple equations exist to model the aerodynamics accurately. In the Simulink plant model, the aerodynamics are represented using lookup tables containing aerodynamic force coefficients, which are calculated offline using \gls{CFD}.
For purposes of controller design, the aerodynamic forces and moments will be neglected, except for the control surface contributions.

\subsection{Propulsion}\label{subsec:propulsion}
The propulsion forces and moments are generated by the four previously mentioned tiltable propellers. In the simulation model, these forces and moments are modeled in high detail, taking into account the inflow conditions at each propeller and again using lookup tables to compute the corresponding forces and moments. Even the gyroscopic moments resulting from the rotation of the propellers are included in the model.

However, for controller development, a simplified relationship between the produced forces and moments and the corresponding tilt angles and RPMs of the propellers must be established, since these variables act as inputs to the system and must be determined in order to control the airship.

Each rotor produces a force along its axis and a moment about that same axis.
The magnitude of the thrust force is a function of the air density $\rho$, the rotational speed of the propeller $\Omega_i$, and the thrust coefficient $k_T$.

Similarly, the moment generated by each rotor also depends on the air density and the rotational speed.
However, instead of the thrust coefficient, it depends on the torque coefficient $k_N$ and the direction of rotation of the propeller, denoted by $\sigma_{dir,i}$.

Since the engine nacelles are tiltable, a propulsion-specific reference frame is introduced for each rotor. This frame originates at the rotor's position, with its $y$-axis aligned with the $y$-axis of the body-fixed frame.
Within this propulsion frame, the propulsive force and moment are modeled to act along and around the $z$-axis, respectively, as described below:
\begin{equation}\label{eq:propeller_model}
    \left(\Vector{F}_{P,i}^{Pi}\right)_{P_i} = \begin{bmatrix}
        0 \\
        0 \\
        \rho k_T \Omega_i^2
    \end{bmatrix},
    \quad
    \left(\Vector{M}_{P,i}^{Pi}\right)_{P_i} = \begin{bmatrix}
        0 \\
        0 \\
        -\rho k_N \sigma_{dir,i} \Omega_i^2
    \end{bmatrix}
\end{equation}

In this context, \(\left(\bm{F}_{P,i}^{Pi}\right)_{P_i}\) denotes the thrust force generated by the $i$-th rotor, evaluated at its own location and expressed in its associated propulsion frame (P-frame). Similarly, \(\left(\bm{M}_{P,i}^{Pi}\right)_{P_i}\) represents the corresponding moment, also expressed in the propulsion frame and acting at the same point.

When formulating the equations of motion for the airship, it is necessary to express all forces and moments in the body-fixed frame. This requires transforming quantities from the local propulsion frames to the body frame using appropriate rotation matrices. These transformations depend on the orientation of each nacelle relative to the body, which varies with the tilt angle of the rotor.

By convention, a tilt angle of \ang{0} corresponds to the rotor axis pointing vertically upward, aligned with the negative direction of the body-fixed $z$-axis. Increasing the tilt angle rotates the nacelle forward toward the body-fixed $x$-axis, while negative tilt angles tilt it rearward.

Each propulsion frame is defined locally at the pivot point of its nacelle. It is oriented such that its $y$-axis is parallel to the body frame's $y$-axis, the $z$-axis aligns with the thrust direction (changing with the tilt angle), and the $x$-axis completes a right-handed coordinate system.

Because the $y$-axes of the propulsion and body-fixed frames are aligned, the rotation between them can be described entirely as a rotation around this common $y$-axis. In the propulsion frame, the thrust force vector points along the positive $z$-axis. To align this vector with the correct direction in the body frame when the tilt angle is zero, a rotation of \ang{180} (i.e., $\pi$ radians) about the $y$-axis is required.

For arbitrary tilt angles $\gamma_i$, this \ang{180} rotation must be adjusted by the tilt, resulting in a total rotation of
\begin{equation}
    \theta_{P_i} = \pi - \gamma_i
\end{equation}

This ensures that the direction of the thrust is correctly represented in the body-fixed frame, taking into account the nacelle's orientation.

A visual representation of this transformation process is provided in \cref{fig:rotation}, which illustrates the rotation of the thrust vector from the propulsion frame into the body-fixed frame for a general tilt angle.

\begin{figure}
    \centering
    \includegraphics[width=\imagewidth]{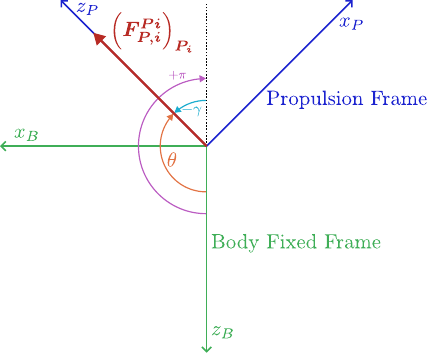}
    \caption{Illustration of the process of transforming a force from the P-frame to the B-frame}
    \label{fig:rotation}
\end{figure}

Thus, the transformation from the P-frame to the B-frame is a pure rotation around the shared y-axis of both frames, resulting in following rotation matrix
\begin{equation}
    \bm{M}_{BP_i} = \begin{bmatrix}
        \cos{\theta_{P_i}}  & 0 & \sin{\theta_{P_i}} \\
        0                   & 1 & 0                  \\
        -\sin{\theta_{P_i}} & 0 & \cos{\theta_{P_i}} \\
    \end{bmatrix},
    \theta_{P_i} = \pi - \gamma_i
\end{equation}
where $\gamma_i$ is the tilt angle of the i-th nacelle.
The transformation matrix can be simplified by applying the trigonometric relationship $\sin{\alpha} = -\sin{\left(\alpha+\pi\right)}$ and $\cos{\alpha} = -\cos{\left(\alpha+\pi\right)}$, $\cos{-\alpha} = \cos{\alpha}$, $\sin{-\alpha} = -\sin{\alpha}$.
The constant offset $\pi$ vanishes and the final result is:
\begin{equation}
    \bm{M}_{BP_i} = \begin{bmatrix}
        -\cos{\gamma_i} & 0 & \sin{\gamma_i}  \\
        0               & 1 & 0               \\
        -\sin{\gamma_i} & 0 & -\cos{\gamma_i} \\
    \end{bmatrix},
\end{equation}

Up to this point, the forces and moments have been expressed either in the propulsion frame or in the body-fixed frame, but they are still evaluated at the respective installation points of the rotors.
For further analysis, it is necessary to evaluate their contribution at the system's reference point, where the total forces and moments acting on the airship are summed.

While the force acting at the reference point remains identical to the force applied at the installation point, an additional moment must be considered. This moment results from the lever arm between the installation point and the reference point.

Therefore, the total force and moment exerted by a single propeller, evaluated at the reference point, are given by:
\begin{align}\label{eq:b_for_and_mom}
    \left(\bm{F}_{P_i}^R\right)_B & = \bm{M}_{BP_i} \begin{bmatrix}
                                                        0 \\
                                                        0 \\
                                                        \rho k_T \Omega_i^2
                                                    \end{bmatrix}\nonumber                                                \\
    \left(\bm{M}_{P_i}^R\right)_B & = \bm{M}_{BP_i} \begin{bmatrix}
                                                        0 \\
                                                        0 \\
                                                        -\rho k_N \sigma_{dir,i} \Omega_i^2
                                                    \end{bmatrix} + \bm{r}^{RP_i} \times \bm{M}_{BP_i} \begin{bmatrix}
                                                                                                           0 \\
                                                                                                           0 \\
                                                                                                           \rho k_T \Omega_i^2
                                                                                                       \end{bmatrix}
\end{align}
where $\bm{r}^{RP_i}$ is the vector from the reference point to the installation point of the i-th propeller.
After applying the rotation matrix and accounting for the cross product due to the lever arm, the forces and moments by a propeller acting at the reference point can be written as:
\begin{align}
    \left(\bm{F}_{P_i}^R\right)_B & = \begin{bmatrix}
                                          \sin{\gamma_i} \rho k_T \Omega_i^2 \\
                                          0                                  \\
                                          -\cos{\gamma_i} \rho k_T \Omega_i^2
                                      \end{bmatrix}\label{eq:Force_at_cg}                                                              \\
    \left(\bm{M}_{P_i}^R\right)_B & = \begin{bmatrix}
                                          -\sin{\gamma_i} \rho k_N \sigma_{dir,i} \Omega_i^2 - y^{RP_i} \cos{\gamma_i} \rho k_T \Omega_i^2 \\
                                          \left(z^{RP_i} \sin{\gamma_i} + x^{RP_i} \cos{\gamma_i} \right) \rho k_T \Omega_i^2              \\
                                          \cos{\gamma_i} \rho k_N \sigma_{dir,i} \Omega_i^2 - y^{RP_i} \sin{\gamma_i} \rho k_T \Omega_i^2
                                      \end{bmatrix}
\end{align}
As there are four propellers in total, their individual contribution can simply be added together to form the total propulsive forces and moments at the reference point:
{
\newcommand*{\Force}[1]{\left( \tilde{\bm{F}}_{P_#1}^R\left(\gamma_{#1}\right) \right)_B \Omega_#1^2}
\newcommand*{\Moment}[1]{\left( \tilde{\bm{M}}_{P_#1}^R\left(\gamma_{#1}\right) \right)_B \Omega_#1^2}
\begin{equation}\label{eq:total_forces_and_moments}
    \begin{split}
        \left(\bm{F}_{P}^R\right)_B   = \Force{1} + \Force{2} + \\ \Force{3} + \Force{4} \\[3pt]
        \left(\bm{M}_{P}^R\right)_B = \Moment{1} + \Moment{2} + \\ \Moment{3} + \Moment{4}
    \end{split}
\end{equation}
}
For improved readability, the terms $\left(\tilde{\bm{F}}_{P_i}^R\right)_B$ and $\left(\tilde{\bm{M}}_{P_i}^R\right)_B$ were introduced in the equation above. These terms group together all components that are independent of the rotor RPM:
\begin{align}
    \left(\tilde{\bm{F}}_{P_i}^R\right)_B & = \begin{bmatrix}
                                                  \sin{\gamma_i} \rho k_T \\
                                                  0                       \\
                                                  -\cos{\gamma_i} \rho k_T
                                              \end{bmatrix}                                                                \\
    \left(\tilde{\bm{M}}_{P_i}^R\right)_B & = \begin{bmatrix}
                                                  -\sin{\gamma_i} \rho k_N \sigma_{dir,i} - y^{RP_i} \cos{\gamma_i} \rho k_T \\
                                                  \left(z^{RP_i} \sin{\gamma_i} + x^{RP_i} \cos{\gamma_i} \right) \rho k_T   \\
                                                  \cos{\gamma_i} \rho k_N \sigma_{dir,i} - y^{RP_i} \sin{\gamma_i} \rho k_T\end{bmatrix}
\end{align}
\cref{eq:total_forces_and_moments} can be reformulated as a single matrix vector product, where the resulting vector represents the forces and moments exerted at the reference point by the propulsion system.
The side force component can be omitted, as the propellers are only capable of generating thrust within the x-z-plane of the body-fixed frame.
This is not only physically intuitive but is also evident from the second entry in \cref{eq:Force_at_cg}, which remains zero.
By retaining only the first and third rows of the force vector in \cref{eq:Force_at_cg}, the matrix vector formulation becomes:
\begin{equation}\label{eq:B_matrix}
    \begin{bmatrix}
        \left( L_{P}^R \right) \\
        \left( M_{P}^R \right) \\
        \left( N_{P}^R \right) \\
        \left( X_{P}^R \right) \\
        \left( Z_{P}^R \right)
    \end{bmatrix}_B = \begin{bmatrix}
        \left(\tilde{\bm{M}}_{P_1}^R\right)_B & \left(\tilde{\bm{M}}_{P_2}^R\right)_B & \left(\tilde{\bm{M}}_{P_3}^R\right)_B & \left(\tilde{\bm{M}}_{P_4}^R\right)_B \\
        \left(\bar{\bm{F}}_{P_1}^R\right)_B   & \left(\bar{\bm{F}}_{P_2}^R\right)_B   & \left(\bar{\bm{F}}_{P_3}^R\right)_B   & \left(\bar{\bm{F}}_{P_4}^R\right)_B
    \end{bmatrix} \begin{bmatrix}
        \Omega_1^2 \\
        \Omega_2^2 \\
        \Omega_3^2 \\
        \Omega_4^2
    \end{bmatrix}
\end{equation}
where $\left(\bar{\bm{F}}_{P_i}^R\right)_B$ is the vector that contains the parts of the first and third row of \cref{eq:Force_at_cg} that are independent of the RPM:
\begin{equation}
    \left(\bar{\bm{F}}_{P_i}^R\right)_B = \begin{bmatrix}
        1 & 0 & 0 \\
        0 & 0 & 1
    \end{bmatrix} \left(\tilde{\bm{F}}_{P_i}^R\right)_B = \begin{bmatrix}
        1 & 0 & 0 \\
        0 & 0 & 1
    \end{bmatrix} \begin{bmatrix}
        \sin{\gamma_i} \rho k_T \\
        0                       \\
        -\cos{\gamma_i} \rho k_T
    \end{bmatrix} = \begin{bmatrix}
        \sin{\gamma_i} \rho k_T \\
        -\cos{\gamma_i} \rho k_T
    \end{bmatrix}
\end{equation}
In \cref{tab:rotor_data}, the numerical values of the rotor positions, the respective rotation directions and the force coefficient $k_T$ and the moment coefficient $k_N$ are presented.

\begin{table}
    \centering
    \begin{tabular}{|c|c|}
        \hline
        Quantity [Unit]                  & Value                 \\
        \hline
        $\Vector{r}^{RP_1} \ [\text{m}]$ &
        $\left[\begin{array}{@{}S[table-format=2.0]@{\hskip 15pt}
                           S[table-format=1.3]@{\hskip 15pt}
                           S[table-format=1.2]@{}}
                           4 & 1.338 & 1.49
                       \end{array}\right]^T$ \\

        $\Vector{r}^{RP_2} \ [\text{m}]$ &
        $\left[\begin{array}{@{}S[table-format=2.0]@{\hskip 15pt}
                           S[table-format=1.3]@{\hskip 15pt}
                           S[table-format=1.2]@{}}
                           -4 & 1.338 & 1.38
                       \end{array}\right]^T$ \\

        $\Vector{r}^{RP_3} \ [\text{m}]$ &
        $\left[\begin{array}{@{}S[table-format=2.0]@{\hskip 15pt}
                           S[table-format=1.3]@{\hskip 15pt}
                           S[table-format=1.2]@{}}
                           -4 & -1.338 & 1.38
                       \end{array}\right]^T$ \\

        $\Vector{r}^{RP_4} \ [\text{m}]$ &
        $\left[\begin{array}{@{}S[table-format=2.0]@{\hskip 15pt}
                           S[table-format=1.3]@{\hskip 15pt}
                           S[table-format=1.2]@{}}
                           4 & -1.338 & 1.49
                       \end{array}\right]^T$ \\

        $\sigma_{dir,1} [\text{-}]$      & 1                     \\
        $\sigma_{dir,2} [\text{-}]$      & 1                     \\
        $\sigma_{dir,3} [\text{-}]$      & 1                     \\
        $\sigma_{dir,4} [\text{-}]$      & 1                     \\
        $k_T [\text{-}]$                 & \num{5d-4}            \\
        $k_N [\text{-}]$                 & \num{1.7d-5}          \\
        \hline
    \end{tabular}
    \caption{Rotor positions, rotation directions and force and moment coefficients}
    \label{tab:rotor_data}
\end{table}

\subsection{Control Surfaces}\label{subsec:control_surfaces}
The final source of external forces and moments considered in this model comes from the control surfaces located at the rear of the airship.
\Cref{fig:control_surfaces} illustrates the arrangement of the fins. The three fins are evenly spaced with an angular separation of \ang{120}, with the rudder mounted vertically on the top fin.
As a result, the remaining two fins are inclined slightly downward relative to the horizon, forming an angle of \ang{30} with the horizontal plane.
\begin{figure}
    \centering
    \includegraphics[width=\imagewidth]{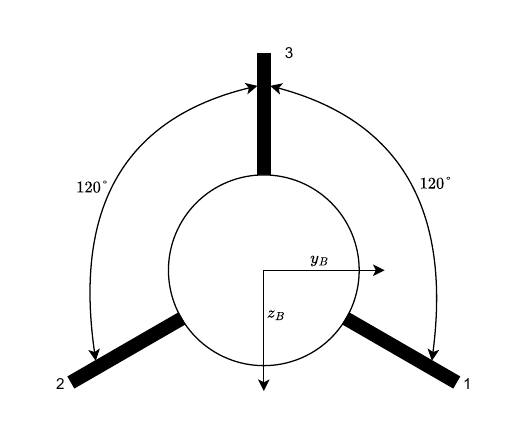}
    \caption{Arrangement of the fins viewed from the back of the airship}
    \label{fig:control_surfaces}
\end{figure}

For simulation purposes, the aerodynamic effects of the fins and their control surfaces are evaluated using lookup tables, which are precomputed using \gls{CFD}.
These tables contain force and moment coefficients that are dependent on the current flow conditions, accounting for parameters such as the angle of attack and sideslip angle.
The aerodynamic influence of the fixed fins is taken as a baseline, while the additional contributions from the deflected control surfaces, dependent on their individual deflection angles, are added to determine the total force and moment coefficients.

Since the control surfaces are intended to be used during high-speed flight, a simplified relationship between control surface deflection and the resulting forces and moments is required, similar to the case with the propellers.

To simplify the modeling for control purposes, only the additional effect of the control surface deflections will be considered, while the baseline contribution of the fixed fins is disregarded.

Although the control surfaces generate both forces and moments, the resulting moments are approximately an order of magnitude larger due to the lever arm between the control surfaces and the airship's center of gravity.
Therefore, only the moments induced by the control surface deflections are taken into account.

\cref{fig:control_surfaces_coefficients} shows the incremental moment coefficients for roll, pitch, and yaw as functions of the control surface deflection angle, for the case in which the local angle of attack at the fin is zero.
It is important to note that the moment coefficient increments are expressed in the local fin coordinate system, not in the body-fixed frame. As such, a coordinate transformation is required to incorporate them into the global model.

\begin{figure}
    \centering
    \includegraphics[width=\imagewidth]{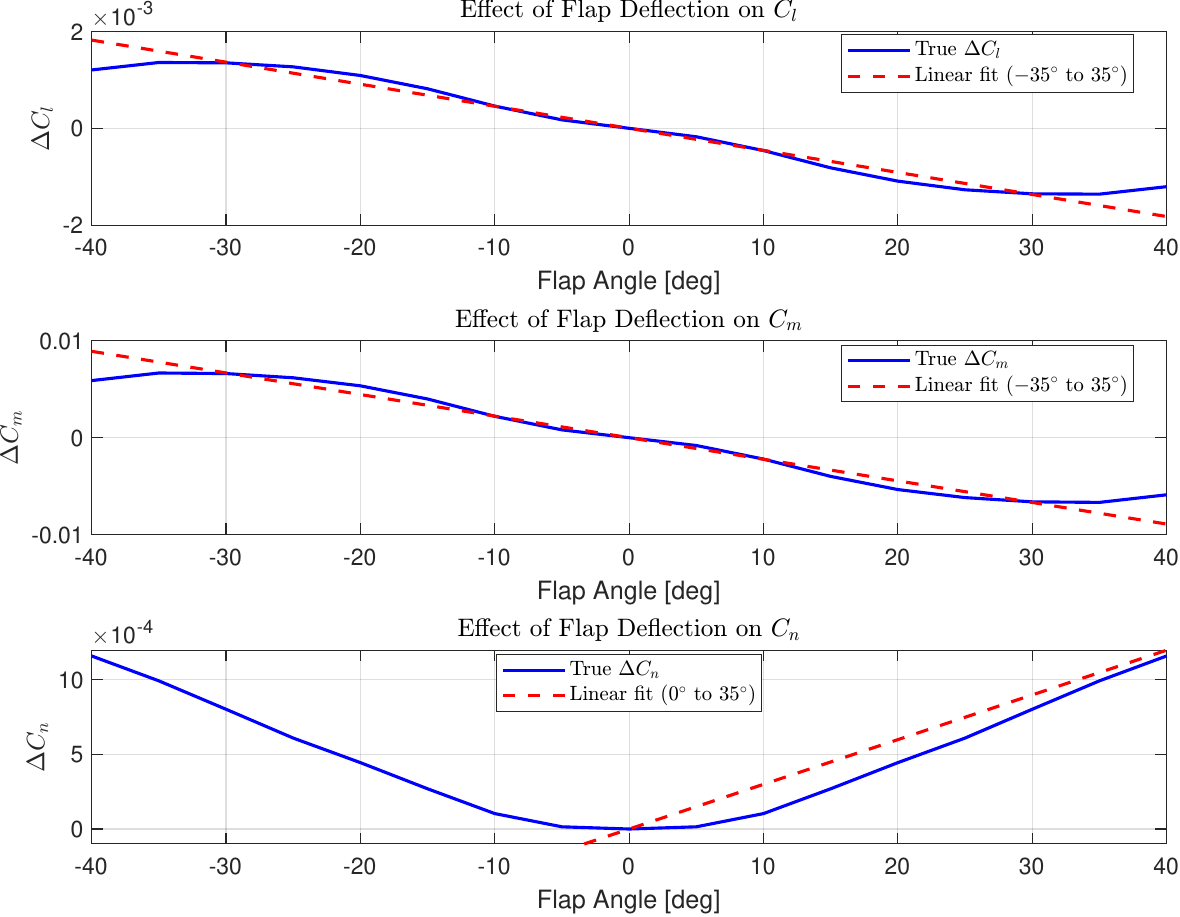}
    \caption{Delta moment coefficient of the control surfaces wrt. deflection angle}
    \label{fig:control_surfaces_coefficients}
\end{figure}

From the plot, it can be observed that the roll and pitch moment coefficient increments vary almost linearly across a wide range of deflection angles.
The maximum values are reached before the limits of the deflection range, at approximately \ang{35} and \ang{-35}, respectively.
In contrast, the yaw moment coefficient increment remains close to zero for small deflections,  specifically within the first \ang{5} from the neutral position. Beyond that point, it begins to increase in an approximately linear manner.
However, its overall magnitude remains at least an order of magnitude smaller than the corresponding roll and pitch moment coefficients.

This is because, in the local fin coordinate system, the yaw moment primarily results from the additional drag caused by the control surface deflection, multiplied by the lateral distance to the airship's centerline.
Since airfoils generally produce significantly more lift than drag, the contribution to the yaw moment is inherently small.

As a result, the yaw moment in the fin coordinate system is neglected in the model.
It is important to emphasize, however, that this does not imply the control surfaces don't generate yaw moments in the body-fixed frame. They certainly do and this will become evident once the transformation matrix is employed.

To get to moment increments from the moment coefficients, they have to be multiplied with dynamic pressure, the reference area and the reference length:
\begin{equation}\label{eq:cs_moment_increments}
    \begin{bmatrix}
        \Delta L_{F,i} \\
        \Delta M_{F,i} \\
        \Delta N_{F,i}
    \end{bmatrix}_{Fin} = \frac{\rho}{2} V^2 S_{ref} l_{ref} \begin{bmatrix}
        \Delta c_{l,i} \\
        \Delta c_{m,i} \\
        0
    \end{bmatrix}_{Fin}
\end{equation}
It is noted that the moment increments are still expressed in the local fin coordinate system.
The values for the reference area and the reference length are found in \cref{tab:fin_data}.

The fact that the moment coefficient increments behave very linear can be used to calculate the moment coefficient derivative with respect to the deflection angle, which is equal to the gradient of the linear function fitted to the curves.
As the yaw moment in the local fin frame is discarded, the two derivatives are $c_{l \eta}$, $c_{m \eta}$.
Their numerical values are found in \cref{tab:fin_data}.
\cref{eq:cs_moment_increments} can then be rewritten:
\begin{equation}
    \begin{bmatrix}
        \Delta L_{F,i} \\
        \Delta M_{F,i} \\
        \Delta N_{F,i}
    \end{bmatrix}_{Fin} = \frac{\rho}{2} V^2 S_{ref} l_{ref} \begin{bmatrix}
        c_{l \eta} \\
        c_{m \eta} \\
        0
    \end{bmatrix}_{Fin} \eta_i
\end{equation}

Next, the moment increments must be transformed from the local fin coordinate system into the body-fixed frame.
Since the fin and body frames share the same x-axis, this transformation involves a single rotation about that axis.

The resulting transformation matrix from the i-th fin frame to the body-fixed frame is given by:
\begin{equation}
    \Matrix{M}_{BFin_i} = \Matrix{M}_{Fin_i B}^T = \begin{bmatrix}
        1 & 0                & 0               \\
        0 & \cos{\varphi_i}  & \sin{\varphi_i} \\
        0 & -\sin{\varphi_i} & \sin{\varphi_i}
    \end{bmatrix}^T = \begin{bmatrix}
        1 & 0               & 0                \\
        0 & \cos{\varphi_i} & -\sin{\varphi_i} \\
        0 & \sin{\varphi_i} & \sin{\varphi_i}
    \end{bmatrix}
\end{equation}
Here, $\varphi_i$ denotes the angle by which the i-th fin is rotated relative to the body-fixed frame.
These angles are given in the \cref{tab:fin_data}, where the numbering is according to the drawing in \cref{fig:control_surfaces}.
\begin{table}
    \centering
    \begin{tabular}{|c|c|}
        \hline
        Quantity    & Value                 \\
        \hline
        $\varphi_1$ & \ang{30}              \\
        $\varphi_2$ & \ang{150}             \\
        $\varphi_3$ & \ang{270}             \\
        $S_{ref}$   & \qty{17.93}{\meter^2} \\
        $l_{ref}$   & \qty{14.53}{\meter}   \\
        $c_{l\eta}$ & \num{-2.61d-3}        \\
        $c_{m\eta}$ & \num{-1.28d-2}        \\
        \hline
    \end{tabular}
    \caption{Values required to calculate the moment contributions of the fins}
    \label{tab:fin_data}
\end{table}
Finally, the contributions of each of the fins can be transformed in the body-fixed frame and then be summed:
{
\small
\begin{equation}
    \begin{bmatrix}
        \Delta L_{F} \\
        \Delta M_{F} \\
        \Delta N_{F}
    \end{bmatrix}_{B} = \frac{\rho}{2} V^2 S_{ref} l_{ref} \left( \Matrix{M}_{BFin_1} \begin{bmatrix}
        c_{l \eta} \\
        c_{m \eta} \\
        0
    \end{bmatrix}_{Fin_1} \eta_1 + \Matrix{M}_{BFin_2} \begin{bmatrix}
        c_{l \eta} \\
        c_{m \eta} \\
        0
    \end{bmatrix}_{Fin_2} \eta_2 + \Matrix{M}_{BFin_3} \begin{bmatrix}
        c_{l \eta} \\
        c_{m \eta} \\
        0
    \end{bmatrix}_{Fin_3} \eta_3 \right)
\end{equation}
}
After applying the transformation matrix to each term:
\begin{equation}
    \begin{bmatrix}
        \Delta L_{F} \\
        \Delta M_{F} \\
        \Delta N_{F}
    \end{bmatrix}_{B} = \frac{\rho}{2} V^2 S_{ref} l_{ref} \left( \begin{bmatrix}
        c_{l \eta}                 \\
        \cos{\varphi_1} c_{m \eta} \\
        \sin{\varphi_1} c_{m \eta}
    \end{bmatrix}_{B} \eta_1 + \begin{bmatrix}
        c_{l \eta}                 \\
        \cos{\varphi_2} c_{m \eta} \\
        \sin{\varphi_2} c_{m \eta}
    \end{bmatrix}_{B} \eta_2 + \begin{bmatrix}
        c_{l \eta}                 \\
        \cos{\varphi_3} c_{m \eta} \\
        \sin{\varphi_3} c_{m \eta}
    \end{bmatrix}_{B} \eta_3 \right)
\end{equation}
This result can be conveniently rewritten as a matrix vector expression:
\begin{equation}
    \begin{bmatrix}
        \Delta L_{F} \\
        \Delta M_{F} \\
        \Delta N_{F}
    \end{bmatrix}_{B} = \frac{\rho}{2} V^2 S_{ref} l_{ref} \begin{bmatrix}
        c_{l \eta}                 & c_{l \eta}                 & c_{l \eta}                 \\
        \cos{\varphi_1} c_{m \eta} & \cos{\varphi_2} c_{m \eta} & \cos{\varphi_3} c_{m \eta} \\
        \sin{\varphi_1} c_{m \eta} & \sin{\varphi_2} c_{m \eta} & \sin{\varphi_3} c_{m \eta}
    \end{bmatrix}_{B} \begin{bmatrix}
        \eta_1 \\
        \eta_2 \\
        \eta_3
    \end{bmatrix}
\end{equation}
Since $\varphi_3=\ang{270}$ and $\cos{\ang{270}}=0$, there is one last simplification:
\begin{equation}\label{eq:b_matrix_surf}
    \begin{bmatrix}
        \Delta L_{F} \\
        \Delta M_{F} \\
        \Delta N_{F}
    \end{bmatrix}_{B} = \frac{\rho}{2} V^2 S_{ref} l_{ref} \begin{bmatrix}
        c_{l \eta}                 & c_{l \eta}                 & c_{l \eta}                 \\
        \cos{\varphi_1} c_{m \eta} & \cos{\varphi_2} c_{m \eta} & 0                          \\
        \sin{\varphi_1} c_{m \eta} & \sin{\varphi_2} c_{m \eta} & \sin{\varphi_3} c_{m \eta}
    \end{bmatrix}_{B} \begin{bmatrix}
        \eta_1 \\
        \eta_2 \\
        \eta_3
    \end{bmatrix}
\end{equation}
This observation is physically intuitive, since the rudder is mounted on top of the airship, it cannot generate pitching moments in the body-fixed frame.

This concludes the description of all external forces acting on the system, as well as the overall system dynamics.
The following section presents the control strategy developed for the airship.
  \chapter{Controller}
The controller developed in this thesis consists of an outer loop and an inner loop controller.
A high-level schematic of the controller is shown in \cref{fig:controller_high_level}, illustrating its basic functionality.
This section provides a brief overview of the working principle of both loops, without going into detail.
Each loop will be explained more thoroughly in its respective section.

As shown in \cref{fig:controller_high_level}, the pilot's stick inputs are first converted into outer loop commands. These commands are the forward and vertical velocities expressed in the control frame (C-frame), as well as the time derivative of the azimuth angle, $\dot{\varPsi}$.

The C-frame is a rotated version of the NED frame that moves with the vehicle. It is obtained by rotating the NED frame around its $z$-axis by the azimuth angle $\varPsi$.
As a result, the $x$-axis of the C-frame always points in the direction of the vehicle's heading.

The forward and vertical velocities in the C-frame are fed into the flight path controller, which computes the required pitch angle and transforms the velocities into the body-fixed frame (B-frame).
The calculated pitch angle, along with the desired azimuth rate $\dot{\varPsi}$, is passed to the attitude controller, which outputs the desired angular velocity $\Vector{\omega}_{cmd}$.

Together with the body-frame velocity commands $\left[ u \quad w \right]^T_{B,cmd}$, this angular velocity forms the input to the inner loop controller.
To ensure smooth tracking of these inputs, reference models and error controllers are used. These generate commands that are one dynamic level below the inputs of the inner loop controller, namely $\left[ \dot{u} \quad \dot{w} \right]^T_{B,cmd}$ and $\dot{\Vector{\omega}}_{cmd}$.
These quantities are processed by two consecutive blocks: a model-based block called \emph{Inversion}, and a second block that implements an $\pseudo$ reference model and error controller, in order to generate the desired time derivative $\dot{\Vector{\nu}}_{cmd}$ of the pseudo-control signal, which is then used by the control allocation module to compute the plant input commands.

The \emph{Kinematic Calculations / Filtering} block is responsible for deriving and estimating the necessary signals when they are not directly available from sensor measurements.
\begin{figure}
    \centering
    \includegraphics[width=\linewidth]{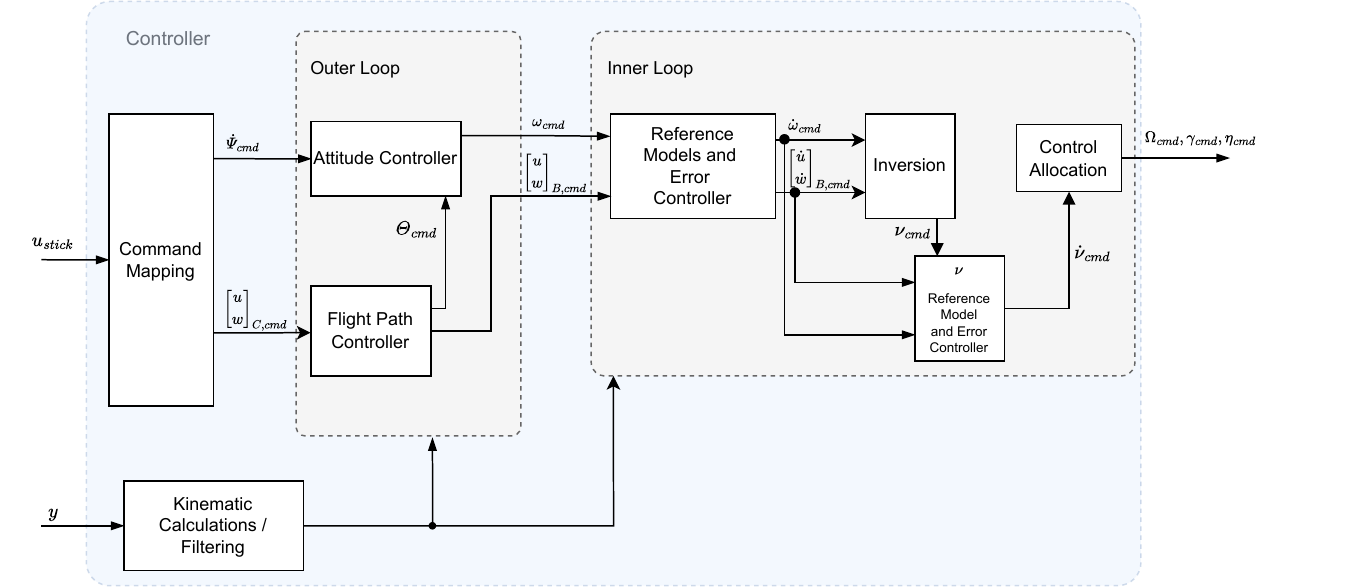}
    \caption{Overview of the control law structure}
    \label{fig:controller_high_level}
\end{figure}
\section{Stick Mapping}
A standard remote control with two sticks can be used to control the airship within the Simulink simulation environment.
The right stick controls the forward and backward velocity by moving the stick up and down.
The left stick controls the heave motion via up and down input, and also controls the airship's turn rate through left and right deflection.

Stick deflections are interpreted by the \emph{Command Mapping} subsystem, which converts them into outer loop commands using simple piecewise linear functions.
These mappings are not necessarily symmetric. For example, higher forward velocities than backward velocities are allowed, due to the inherently destabilizing behavior of the control surfaces during backward flight.
Similarly, greater climb rates than descent rates are permitted.
A visual depiction of the command mapping is shown in \cref{fig:stick_mapping} where the x-axis shows the stick deflection and the y-axis shows the respective commanded variable.
As illustrated, the vertical motion allows for a maximum climb rate of \qty{3}{\meter\per\second} and a maximum descent rate of \qty{1}{\meter\per\second}.
The forward velocity can reach up to \qty{10}{\meter\per\second}, while the maximum backward velocity is limited to \qty{3}{\meter\per\second}.
The maximum turn rate is symmetric in both directions and limited to \qty{10}{\degree\per\second}.
Commanded values are linearly interpolated between the neutral stick position and the respective maximum deflections.
\begin{figure}
    \centering
    \begin{tikzpicture}
        \begin{groupplot}[
                group style={
                        group size=3 by 1,
                        horizontal sep=1.5cm
                    },
                width=6cm,
                height=6cm,
                axis lines=middle,
                grid=both
            ]

            \nextgroupplot[
                ylabel={$w_{C,cmd}$},
                ymin=-4, ymax=2,
                xmin=-1.2, xmax=1.2,
                title={Heave Mapping}
            ]
            \addplot[thick, blue] coordinates {
                    (-1, -3)
                    (0, 0)
                    (1, 1)
                };

            \nextgroupplot[
                ylabel={$\dot{\varPsi}_{cmd}$ [°/s]},
                ymin=-12, ymax=12,
                xmin=-1.2, xmax=1.2,
                title={Yaw Rate Mapping}
            ]
            \addplot[thick, red] coordinates {
                    (-1, -10)
                    (0, 0)
                    (1, 10)
                };

            \nextgroupplot[
                ylabel={$u_{C,cmd}$},
                ymin=-4, ymax=11,
                xmin=-1.2, xmax=1.2,
                title={Forward Velocity Mapping}
            ]
            \addplot[thick, green!60!black] coordinates {
                    (-1, -3)
                    (0, 0)
                    (1, 10)
                };

        \end{groupplot}
    \end{tikzpicture}
    \caption{Mapping of stick deflections to outer-loop command values.}
    \label{fig:stick_mapping}
\end{figure}
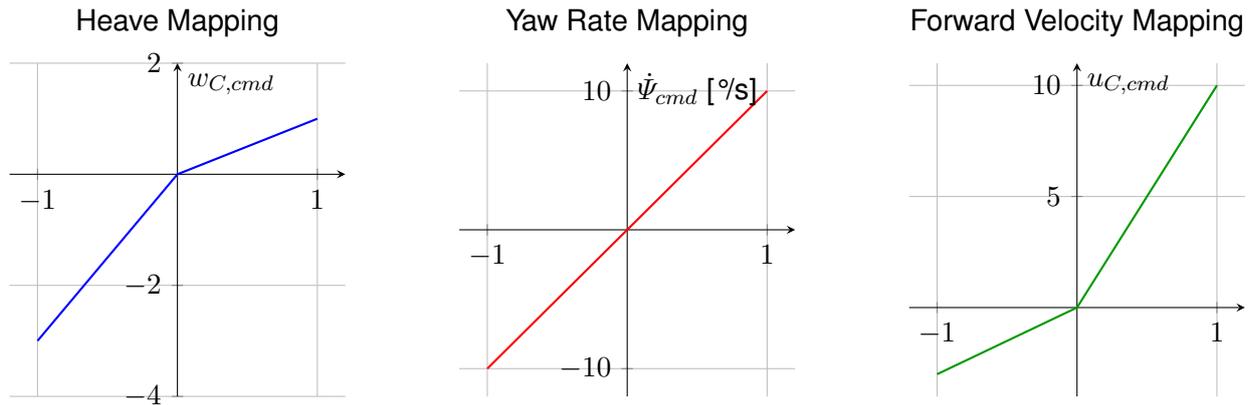

\section{Inner Loop}
This section presents the design and implementation of the inner loop controller, which is responsible for tracking commanded linear and angular velocities by generating appropriate actuator inputs.

\subsection{Background}
The inner loop presented in this thesis is an evolution of the work done in \cite{schimmer}.
In \cite{schimmer}, a nonlinear dynamic inversion based controller was presented that inverts the jerk level dynamics to obtain the rate of change of the propulsion forces and moments.
The latter are then supplied to the control allocation module which calculates the actuator commands.
The inversion at the jerk level is needed because the tilt actuators are rather slow, making it necessary to take their dynamics and specifically their rate limits into account.
In order to do this, the allocation effectively works on the $\dot{\Vector{u}}$ level, allocating $\dot{\gamma}$ and $\dot{\Omega}$.
Through actuator inversion, $\gamma_{cmd}$ and $\Omega_{cmd}$ values are calculated.
The tilt rotors play a central role in controlling the airship and are used extensively throughout this section. To simplify notation, the input vector $\Vector{u}$ will, unless otherwise specified, refer exclusively to the propulsion input vector. When necessary, the input vector associated with control surfaces will be explicitly denoted as $\Vector{u}_{surf}$. The following definition of $\Vector{u}$ will apply consistently throughout the remainder of this thesis:
\begin{equation}
    \Vector{u} = \left[ \Omega_1 \quad \Omega_2 \quad \Omega_3 \quad \Omega_4 \quad \gamma_1 \quad \gamma_2 \quad \gamma_3 \quad \gamma_4 \right]^T
\end{equation}

\subsection{Pseudo Control Concept}
In this context, the term \emph{pseudo control} is introduced.
In classical nonlinear dynamic inversion, pseudo control refers to the lowest-order time derivative of the output vector $\Vector{y}$ that is directly influenced by the input vector $\Vector{u}$.
For the airship system, the input vector generates forces and moments as described in \cref{subsec:propulsion,subsec:control_surfaces}.
Specifically, moments can be produced around all three body axes, while forces can only be generated in the body-fixed $x$–$z$ plane.

From the equations of motion, it becomes clear that these forces and moments directly influence the time derivative $\dot{\Vector{x}}$ of the state vector $\Vector{x}$.
If the system output is defined as the velocities in the $x$- and $z$-directions and the angular velocities, then the pseudo controls would correspond to the linear accelerations in $x$ and $z$, and the angular accelerations, as these are the quantities directly influenced by the control inputs through propulsion and aerodynamic effects.

However, in practical applications, pseudo controls do not necessarily need to be the actual time derivatives of the outputs.
Any quantities that lie on the same dynamic level and maintain a purely algebraic relationship with the input vector can also be used \cite{raab_consideration_2019}.
In this case, instead of using accelerations, the forces and moments themselves are selected as pseudo controls, since they are algebraically related to the inputs.

Going forward, the pseudo control vector is defined as the propulsive moments and the propulsive forces in the body-fixed x- and z-direction:
\begin{equation}
    \pseudo = \begin{bmatrix}
        L_P \\
        M_P \\
        N_P \\
        X_P \\
        Z_P
    \end{bmatrix}_B
\end{equation}
Its time derivative is the rate of change of the propulsive moments and forces with respect to the body-fixed frame:
\begin{equation}
    \dot{\pseudo} = \begin{bmatrix}
        \dot{L}_P^B \\
        \dot{M}_P^B \\
        \dot{N}_P^B \\
        \dot{X}_P^B \\
        \dot{Z}_P^B
    \end{bmatrix}_B
\end{equation}
Consider the equations of motion in \cref{eq:eom_simple}, which are repeated here for convenience. In this version, the contributions from aerostatic moments and control surface forces have been removed, as these findings were introduced after the original derivation of the equations of motion:
\begin{equation}
    \begin{bmatrix}
        \dot{\bm{V}} \\
        \dot{\bm{\omega}}
    \end{bmatrix} = \begin{bmatrix}
        \bm{M}_a             & -m \bm{r}^{RG \times} \\
        m \bm{r}^{RG \times} & \bm{J}_a
    \end{bmatrix}^{-1} \begin{bmatrix}
        \bm{F}_B + \bm{F}_G + \bm{F}_P -m \bm{\omega} \times \left( \bm{\omega} \times \bm{r}^{RG} \right) - \bm{\omega} \times \bm{M}_a \bm{V} \\
        \bm{M}_G + \bm{M}_P + \bm{M}_F -\bm{\omega} \times \bm{J}_a \bm{\omega} + \bm{r}^{RG} \times \left(m \bm{V} \times \bm{\omega}\right)
    \end{bmatrix}
\end{equation}
Through inversion of the equations of motion, the pseudo control vector $\pseudo$ can be calculated, given commanded values for the linear and angular acceleration:
\begin{equation}
    \pseudo_{cmd} = \underbrace{\begin{bmatrix}
            0 & 0 & 0 & 1 & 0 & 0 \\
            0 & 0 & 0 & 0 & 1 & 0 \\
            0 & 0 & 0 & 0 & 0 & 1 \\
            1 & 0 & 0 & 0 & 0 & 0 \\
            0 & 0 & 1 & 0 & 0 & 0
        \end{bmatrix}}_{\text{\scriptsize{\shortstack{\begin{math}
                    \Matrix{P}
                \end{math}: for aligning with structure\\of pseudo control vector}}}} \left(\overline{\Matrix{M}}_a \begin{bmatrix}
        \dot{\bm{V}}_{cmd} \\
        \dot{\bm{\omega}}_{cmd}
    \end{bmatrix} + \begin{bmatrix}
        -\bm{F}_B - \bm{F}_G  +m \bm{\omega} \times \left( \bm{\omega} \times \bm{r}^{RG} \right) + \bm{\omega} \times \bm{M}_a \bm{V} \\
        -\bm{M}_G - \bm{M}_F +\bm{\omega} \times \bm{J}_a \bm{\omega} - \bm{r}^{RG} \times \left(m \bm{V} \times \bm{\omega}\right)
    \end{bmatrix}\right)
\end{equation}
In the equation above, all terms are either known or measurable, with the exception of the moment generated by the control surfaces, $\Vector{M}_F$.
This moment depends on the control surface deflections, which are themselves determined by the desired moments, and is therefore not known a priori.

To handle this, the total force and moment vector is split into two parts: the desired forces and moments, and the portion attributed to the control surfaces.
The pseudo-control vector is then defined as the difference between these two components:
\begin{equation}\label{eq:pseudo_cmd}
    \pseudo_{cmd} = \Matrix{P} \Bigg(
    \underbrace{
        \overline{\Matrix{M}}_a \begin{bmatrix}
            \dot{\bm{V}}_{cmd} \\
            \dot{\bm{\omega}}_{cmd}
        \end{bmatrix} +
        \begin{bmatrix}
            -\bm{F}_B - \bm{F}_G  +m \bm{\omega} \times \left( \bm{\omega} \times \bm{r}^{RG} \right) + \bm{\omega} \times \bm{M}_a \bm{V} \\
            -\bm{M}_G +\bm{\omega} \times \bm{J}_a \bm{\omega} - \bm{r}^{RG} \times \left(m \bm{V} \times \bm{\omega}\right)
        \end{bmatrix}
    }_{\begin{bmatrix}
            \Vector{F}_{des} \\
            \Vector{M}_{des}
        \end{bmatrix}} -
    \begin{bmatrix}
        0 \\
        \bm{M}_F
    \end{bmatrix}
    \Bigg)
\end{equation}
For now, the control surface moment $\Vector{M}_F$ is considered to be known. Its computation will be detailed later in this chapter in \cref{subsec:surface_allocation}.

\subsection{Pseudo Control Reference Model}\label{subsec:pseudo_ref_model}
To determine the time derivative of the pseudo-control vector, a first-order linear reference model is employed.
An overview of this reference model is shown in \cref{fig:nu_reference_model}.

In equation form, the time derivative of the pseudo-control vector is given by:
\begin{equation}\label{eq:nu_ref_sys}
    \dot{\pseudo}_{ref} = \Vector{K}_{\pseudo} \left( \pseudo_{cmd} - \pseudo_{ref} \right)
\end{equation}
\begin{figure}
    \centering
    \includegraphics{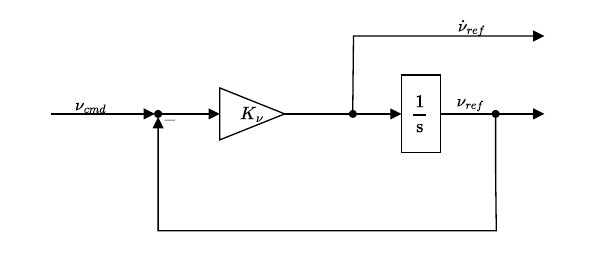}
    \caption{Reference Model to obtain time derivative of pseudo control vector}
    \label{fig:nu_reference_model}
\end{figure}

\begin{figure}
    \centering
    \includegraphics{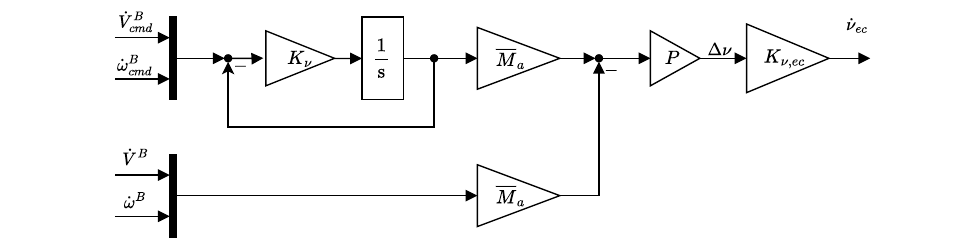}
    \caption{Error controller for $\pseudo$}
    \label{fig:indi_ec}
\end{figure}
Due to model uncertainties, unmodeled dynamics, or limited control authority, errors will arise between the reference and the actual pseudo control vector.
To compensate for these deviations, an error controller is employed.
In theory, the error controller for the pseudo control vector $\pseudo$ would require computing the difference $\pseudo_{ref} - \pseudo$, where $\pseudo$ is derived from the full equations of motion. However, calculating $\pseudo$ this way is complex, as it involves all external forces, moments, and cross-coupling terms.

To simplify this, an alternative structure is used as shown in \cref{fig:indi_ec}. Instead of explicitly computing $\pseudo$, the approach uses the commanded and measured accelerations $\dot{\Vector{V}}^B$ and $\dot{\bm{\omega}}^B$ directly. These are transformed into the pseudo control domain using the generalized mass-inertia matrix $\overline{\Matrix{M}}_a$.

To ensure that the error signal behaves consistently with the pseudo control reference model (which is a first-order system), the commanded accelerations are first passed through the same reference dynamics before comparison.

This structure is derived from difference between the desired and measured propulsion forces and moments:
\begin{flalign}
    \begin{bmatrix}
        \Delta \Vector{F}_P \\
        \Delta \Vector{M}_P
    \end{bmatrix} & = \begin{bmatrix}
                          \Vector{F}_{P,des} \\
                          \Vector{M}_{P,des}
                      \end{bmatrix} - \begin{bmatrix}
                                          \Vector{F}_{P} \\
                                          \Vector{M}_{P}
                                      \end{bmatrix}                                                                                                                                                          \nonumber \\ &=  \overline{\Matrix{M}}_a \begin{bmatrix}
        \dot{\bm{V}}_{des} \\
        \dot{\bm{\omega}}_{des}
    \end{bmatrix} + \begin{bmatrix}
        -\bm{F}_B - \bm{F}_G  +m \bm{\omega} \times \left( \bm{\omega} \times \bm{r}^{RG} \right) + \bm{\omega} \times \bm{M}_a \bm{V} \\
        -\bm{M}_G - \bm{M}_F +\bm{\omega} \times \bm{J}_a \bm{\omega} - \bm{r}^{RG} \times \left(m \bm{V} \times \bm{\omega}\right)
    \end{bmatrix} -\nonumber \\
                           & \hphantom{=\ \  } \overline{\Matrix{M}}_a \begin{bmatrix}
                                                                           \dot{\bm{V}} \\
                                                                           \dot{\bm{\omega}}
                                                                       \end{bmatrix} - \begin{bmatrix}
                                                                                           -\bm{F}_B - \bm{F}_G  +m \bm{\omega} \times \left( \bm{\omega} \times \bm{r}^{RG} \right) + \bm{\omega} \times \bm{M}_a \bm{V} \\
                                                                                           -\bm{M}_G - \bm{M}_F +\bm{\omega} \times \bm{J}_a \bm{\omega} - \bm{r}^{RG} \times \left(m \bm{V} \times \bm{\omega}\right)
                                                                                       \end{bmatrix} \nonumber  \\
                           & = \overline{\Matrix{M}}_a \begin{bmatrix}
                                                           \dot{\bm{V}}_{des} - \dot{\bm{V}} \\
                                                           \dot{\bm{\omega}}_{des} - \dot{\bm{\omega}}
                                                       \end{bmatrix}
\end{flalign}
As shown in the equations, many nonlinear and model-based terms cancel out when calculating the difference. What remains is a simple expression.

The corrective signal $\dot{\pseudo}_{ec}$ can then be calculated as:
\begin{equation}
    \dot{\pseudo}_{ec} = \Matrix{K}_{\pseudo,ec} \Delta \pseudo =    \Matrix{K}_{\pseudo,ec} \Matrix{P} \begin{bmatrix}
        \Delta \Vector{F}_P \\
        \Delta \Vector{M}_P
    \end{bmatrix} = \Matrix{K}_{\pseudo,ec} \Matrix{P} \overline{\Matrix{M}}_a \begin{bmatrix}
        \dot{\bm{V}}_{des} - \dot{\bm{V}} \\
        \dot{\bm{\omega}}_{des} - \dot{\bm{\omega}}
    \end{bmatrix}
\end{equation}
yielding the command signal for the pseudo control rate as the sum of the reference model and the error controller:
\begin{equation}
    \dot{\pseudo}_{cmd} = \dot{\pseudo}_{ref} + \dot{\pseudo}_{ec}
\end{equation}

\subsection{Extended INDI Inversion Law}
Next, the inversion law is introduced, following the approach presented in \cite{raab_consideration_2019}.
In that work, an extended \gls{INDI} framework is proposed, which accounts for actuator dynamics and saturation limits by deriving the inversion law from a continuous-time perspective, contrary to the original formulation introduced in \cite{sieberling_robust_2010}, which is based on a Taylor series expansion.
First, consider the system rewritten in terms of the pseudo control vector $\pseudo$ (compare \cref{eq:pseudo_cmd}):
\begin{equation}
    \pseudo = \Vector{F}(\Vector{x}, \Vector{u})
\end{equation}
Now, instead of the usual Taylor series approximation, the true time derivative is calculated:
\begin{equation}\label{eq:eindi_derivative}
    \dot{\pseudo} = \frac{\partial \Vector{F}(\Vector{x}, \Vector{u})}{\partial \Vector{x}} \dot{\Vector{x}} + \frac{\partial \Vector{F}(\Vector{x}, \Vector{u})}{\partial \Vector{u}} \dot{\Vector{u}}
\end{equation}
An approximation is made here by omitting the first term, which is a usual simplification made in \gls{INDI} \cite{sieberling_robust_2010}.

In the next step, the actuator dynamics are taken into account.
In this thesis, the actuators are approximated as first order systems with rate and absolute limits.
In the frequency domain, the transfer function for a general linear first order system is
\begin{equation}
    G(s) = \frac{\omega_0}{s + \omega_0}
\end{equation}
where $\omega_0$ is the natural frequency or the bandwith of the system.
The transfer function relates a commanded input to the current or achieved input of the system:
\begin{equation}
    u(s) = \frac{\omega_0}{s + \omega_0} u_{cmd}(s)
\end{equation}
The nature of \gls{INDI} is to calculate incremental input commands instead of absolute input commands.
Consider the input increment as
\begin{equation}
    \Delta u(s) = u_{cmd}(s) - u(s)
\end{equation}
and rewrite the achieved input in terms of the input increment:
\begin{align}
    u(s)                                                                                 & = \frac{\omega_0}{\omega_0 + s} u_{cmd}(s) \nonumber           \\
    u(s)                                                                                 & = \frac{\omega_0}{\omega_0 + s} (\Delta u(s) + u(s)) \nonumber \\
    \left( 1 - \frac{\omega_0}{\omega_0 + s} \right)u(s)                                 & = \frac{\omega_0}{\omega_0 + s} \Delta u(s)\nonumber           \\
    \left( \frac{\omega_0 + s}{\omega_0 + s} - \frac{\omega_0}{\omega_0 + s} \right)u(s) & = \frac{\omega_0}{\omega_0 + s} \Delta u(s)\nonumber           \\
    \frac{s}{\omega_0 + s}u(s)                                                           & = \frac{\omega_0}{\omega_0 + s} \Delta u(s)\nonumber           \\
    u(s)                                                                                 & = \frac{\omega_0}{s} \Delta u(s)
\end{align}
Next, the time derivative of the last line is calculated, which is just a multiplication by $s$ in the frequency domain.
Additionally, the equation can be transformed back into the time domain (the explicit notation of the dependence on time is dropped for simplicity):
\begin{equation}
    \dot{u} = \omega_0 \Delta u
\end{equation}
This expression can be inverted and solved for $\Delta u$ which gives the incremental command:
\begin{equation}
    \Delta u = \frac{\dot{u}}{\omega_0}
\end{equation}
Adding the incremental command to the current actuator position finally results in the absolute actuator command:
\begin{equation}
    u_{cmd} = u + \Delta u = u + \frac{\dot{u}}{\omega_0}
\end{equation}
This can be generalized for the MIMO case:
\begin{equation}
    \Vector{u}_{cmd} = \Vector{u} + \Delta \Vector{u} = \Vector{u} + \Matrix{K}_{act}^{-1} \Vector{\dot{u}}
\end{equation}
where $\Matrix{K}_{act}$ is the matrix containing the actuator bandwiths on its main diagonal.
This result is combined with \cref{eq:eindi_derivative} to derive the inversion law:
\begin{equation}\label{eq:inversion_law}
    \Vector{u}_{cmd} = \Vector{u} + \Matrix{K}_{act}^{-1} inv\left( \frac{\partial \Vector{F}(\Vector{x}, \Vector{u})}{\partial \Vector{u}} \right) \dot{\pseudo}_{cmd}
\end{equation}
Here, $inv(.)$ indicates that the matrix $\frac{\partial \Vector{F}(\Vector{x}, \Vector{u})}{\partial \Vector{u}}$, which is the control effectiveness matrix, needs to be inverted.
In practice, this could be achieved by a simple pseudo inverse, however in this thesis a more sophisticated algorithm is used, which is presented in \cref{sec:control_allocation}.

\subsection{Reference Models for Acceleration Commands}
The command values for the linear acceleration $\dot{\Vector{V}}_{cmd}$ and the angular acceleration $\dot{\Vector{\omega}}_{cmd}$ are generated via second order reference systems and linear error controllers.
The reference models for both the linear and angular acceleration follow the structure proposed in \cite{rupprecht_indi_2024} and depicted in \cref{fig:vdot_ref} and \cref{fig:omdot_ref} respectively.
Since the structure is the same in both cases, the functionality is explained for the linear velocity case.
\begin{figure}
    \centering
    \includegraphics[width=\linewidth]{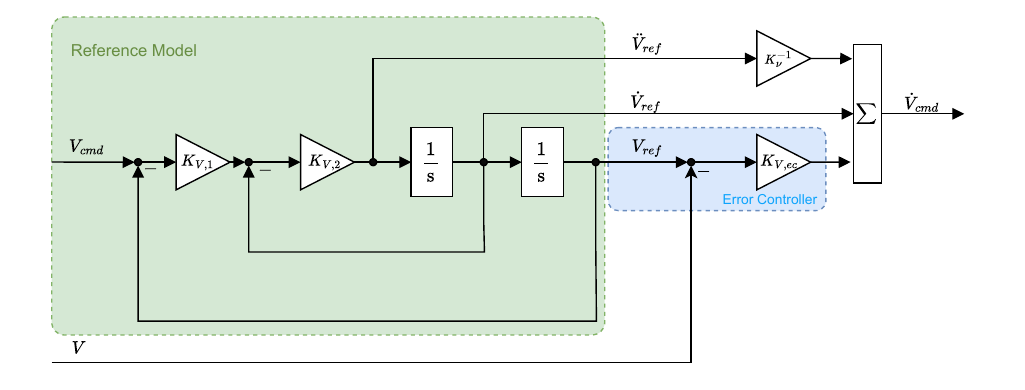}
    \caption{Reference model and error controller used for generating $\dot{\Vector{V}}_{cmd}$ commands}
    \label{fig:vdot_ref}
\end{figure}

\begin{figure}
    \centering
    \includegraphics[width=\linewidth]{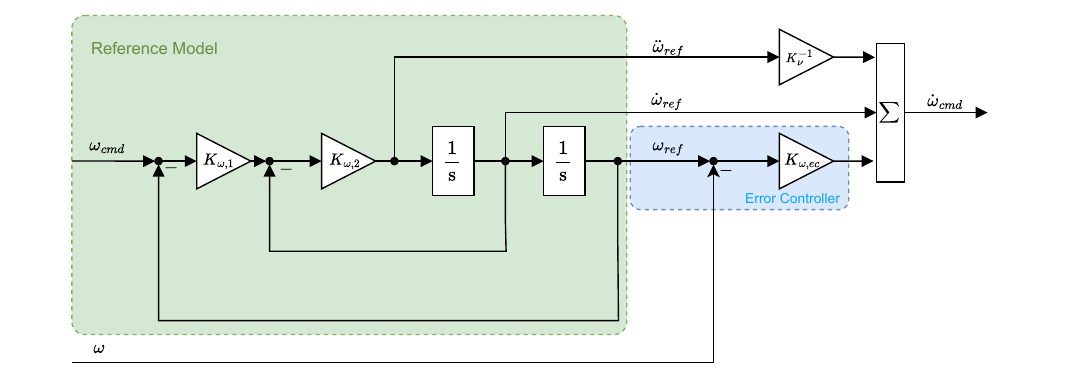}
    \caption{Reference model and error controller used for generating $\dot{\Vector{\omega}}_{cmd}$ commands}
    \label{fig:omdot_ref}
\end{figure}

The dynamics of the reference system can be described by an ordinary differential equation.
The differential equation for $\ddot{\Vector{V}}_{ref}$ is
\begin{align}
    \ddot{\Vector{V}}_{ref} & = \Matrix{K}_{\Vector{V},2} \left( \Matrix{K}_{\Vector{V},1} \left( \Vector{V}_{cmd} - \Vector{V}_{ref} \right) - \dot{\Vector{V}}_{ref} \right)                                                  \nonumber \\
    \ddot{\Vector{V}}_{ref} & = -\Matrix{K}_{\Vector{V},1} \Matrix{K}_{\Vector{V},2} \Vector{V}_{ref} - \Matrix{K}_{\Vector{V},2} \dot{\Vector{V}}_{ref} + \Matrix{K}_{\Vector{V},1} \Matrix{K}_{\Vector{V},2} \Vector{V}_{cmd}
\end{align}
When compared to a classical linear second order system, the equation can also be expressed in terms of natural frequency $\omega_0$ and relative damping $\zeta$
\begin{equation}
    \ddot{\Vector{V}}_{ref} = - \Matrix{\omega}^2_0 \Vector{V}_{ref} - 2 \Matrix{\omega}_0 \Matrix{\zeta} \dot{\Vector{V}}_{ref} + \Matrix{\omega}^2_0 \Vector{V}_{cmd}
\end{equation}
and the gains $\Matrix{K}_{\Vector{V},1}$ and $\Matrix{K}_{\Vector{V},2}$ can be used to tune the dynamics of the reference system in an intuitive way via
\begin{align}
    \Matrix{\omega}^2_0                & = \Matrix{K}_{\Vector{V},1} \Matrix{K}_{\Vector{V},2} \\
    2 \Matrix{\zeta} \Matrix{\omega}_0 & = \Matrix{K}_{\Vector{V},2}
\end{align}
The relative damping of the reference system is chosen as $\Matrix{\zeta} = \Matrix{E}$, with $\Matrix{E}$ being the identity matrix, such that the system is critically damped.
Given desired values for the natural frequency, the gains can then be chosen as:
\begin{align}
    \Matrix{K}_{\Vector{V},2} & = 2 \Matrix{\omega}_0                                                                                                                         \\
    \Matrix{K}_{\Vector{V},1} & = \Matrix{K}_{\Vector{V},2}^{-1} \Matrix{\omega}_0^2 = \frac{1}{2} \Matrix{\omega}_0^{-1} \Matrix{\omega}_0^2 = \frac{1}{2} \Matrix{\omega}_0
\end{align}
The command signal $\dot{\Vector{V}}_{cmd}$ is composed of three terms: the reference signal $\dot{\Vector{V}}_{ref}$, the correction from the error controller $\dot{\Vector{V}}_{ec}$, and an additional feedforward term $\Matrix{K}_\nu^{-1} \ddot{\Vector{V}}_{ref}$.
This feedforward term is required to compensate for the dynamics introduced by the first-order $\pseudo$ reference model and ensures that the plant tracks the desired signal $\dot{\Vector{V}}_{ref}$ accurately.
It therefore acts as a direct command feedthrough, enabling tracking of $\Vector{V}$ and $\Vector{\omega}$ with no additional delays.

The reasoning behind this can be explained as follows.
Consider the differential equation for the first-order reference system of the pseudo-control vector $\Vector{\nu}$, as defined in \cref{eq:nu_ref_sys}.
The input to this system is $\Vector{\nu}_{cmd}$, which includes the commanded propulsion moments and the commanded forces in the $x$-$z$ plane.
These quantities are algebraically related to the angular and linear accelerations as they lie on the same dynamic level. This means that moments and forces can be seen as transformed versions of angular and linear accelerations.

Because of this equivalence, the dynamics of the command signals for forces and moments are shaped by the reference models for angular and linear accelerations, shown in \cref{fig:omdot_ref,fig:vdot_ref}.
To illustrate why the feedforward term $\Matrix{K}_\nu^{-1} \ddot{\Vector{V}}_{ref}$ is needed, consider a simplified case where the pseudo-controls are defined directly as angular accelerations.

As previously discussed, this is valid since angular accelerations and propulsion moments are equivalent in dynamic order.
Under this assumption, the pseudo-control reference system becomes:

\begin{equation}
    \ddot{\Vector{\omega}}_{ref} = \Matrix{K}_{\Vector{\nu}} \left( \dot{\Vector{\omega}}_{cmd} - \dot{\Vector{\omega}}_{ref} \right)
\end{equation}

Now, substitute $\dot{\Vector{\omega}}_{cmd}$ with the output of the angular acceleration reference model from \cref{fig:omdot_ref}, and assume that the angular velocity perfectly tracks its reference, such that the error controller output can be neglected.

This substitution yields:

\begin{align}
    \ddot{\Vector{\omega}}_{ref} & = \Matrix{K}_{\Vector{\nu}} \left( \dot{\Vector{\omega}}_{ref} + \Matrix{K}_{\Vector{\nu}}^{-1} \ddot{\Vector{\omega}}_{ref} -  \dot{\Vector{\omega}}_{ref} \right) \\
    \ddot{\Vector{\omega}}_{ref} & = \ddot{\Vector{\omega}}_{ref}
\end{align}

This result confirms that, by including the feedforward term $\Matrix{K}_\nu^{-1} \ddot{\Vector{\omega}}_{ref}$ in the command signal, the dynamic behavior introduced by the pseudo control reference system is fully cancelled.
Consequently, the output of the system can exactly follow the desired reference trajectory.
The same reasoning is valid for the linear acceleration.

In \cref{fig:ref_mdl_comp}, the effect of incorporating the feedforward term $\Matrix{K}_\nu^{-1} \ddot{\Vector{\omega}}_{ref}$ into the $\pseudo$ reference model is illustrated. The left plot shows the system response when this feedforward term is not used. In this case, the input to the $\pseudo$ reference model consists solely of $\dot{\Vector{\omega}}_{ref}$, resulting in noticeable lag introduced by the $\pseudo$ reference model dynamics.

In contrast, the right plot includes the feedforward term. Here, the input to the $\pseudo$ reference model becomes $\dot{\Vector{\omega}}_{ref} + \Matrix{K}_\nu^{-1} \ddot{\Vector{\omega}}_{ref}$. As shown, this compensation effectively cancels the internal dynamics of the $\pseudo$ reference model, allowing the $\pseudo$ trajectory to follow the desired dynamics specified by the $\Vector{\omega}$ reference model exactly with no additional delay.

\begin{figure}
    \centering
    \includegraphics{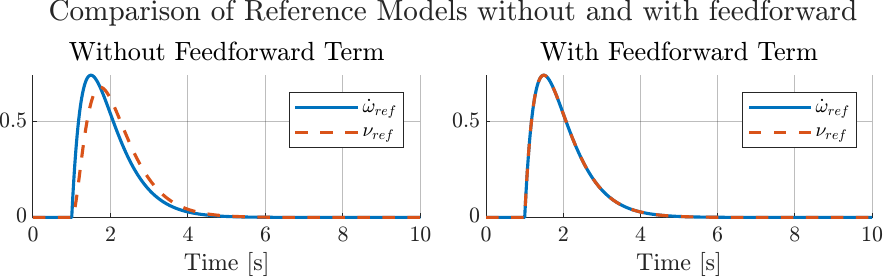}
    \caption{Comparison of $\pseudo$ dynamics with and without feedforward term in $\omega$ reference model}
    \label{fig:ref_mdl_comp}
\end{figure}

\subsection{Pseudo Control Hedging}
Because the actuators are subject to both rate and absolute limits, there are situations in which the control demand exceeds the physical capabilities of the plant.
In such cases, the commanded pseudo control rate $\dot{\pseudo}_{cmd}$ cannot be fully realized by the system, resulting in an error between the commanded and achieved pseudo control rates, i.e., $\dot{\pseudo}_{cmd} > \dot{\pseudo}_{ach}$.

When this happens, the reference models continue to integrate the desired input as if the plant were able to follow it perfectly.
This leads to integrator windup, as the models are unaware of the actuator limitations and thus accumulate unrealistic internal states.

To mitigate this issue, \gls{PCH} is employed.
Originally introduced in \cite{johnson_pseudo-control_2001} for adaptive flight control applications, PCH was designed to prevent the adaptive element from responding to certain system input characteristics, whether originating from the plant or the controller.

In the context of airship control, \gls{PCH} is adapted to limit the effect of unattainable control actions on the reference model.
Specifically, it slows down (hedges) the reference model response by the amount of pseudo control that the plant failed to produce.
This is done by subtracting the difference between the commanded and achieved pseudo control, ensuring that the internal states of the reference model change in line with the achievable system response, preventing the wind up of unachievable command trajectories.

\begin{figure}
    \centering
    \includegraphics{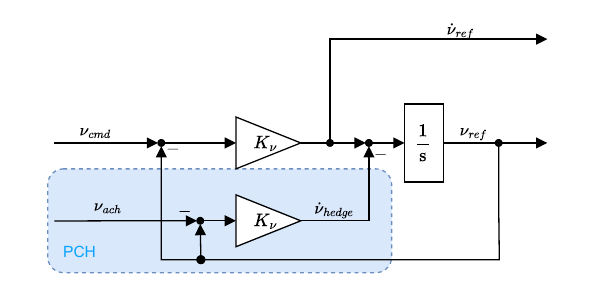}
    \caption{Pseudo control reference model with hedging}
    \label{fig:nu_ref_model_hedge}
\end{figure}

Concretely, the pseudo control reference model is adjusted as shown in \cref{fig:nu_ref_model_hedge}.
The hedging signal $\pseudo_{hedge}$ is computed as the difference between the achieved and the reference pseudo control vectors, resulting in a quantity denoted as $\Delta \pseudo_{hedge}$.
To bring this signal to the same dynamic level as $\dot{\pseudo}_{ref}$, it is multiplied by the gain matrix $\Matrix{K}_{\pseudo}$ in a manner similar to an error controller:
\begin{equation}
    \dot{\pseudo}_{hedge} = \Matrix{K}_{\pseudo} (\pseudo_{ref} - \pseudo_{ach})
\end{equation}

The differential equation for $\dot{\pseudo}_{ref}$ remains unchanged.
However, the presence of the hedging signal effectively slows down, or hedges, $\pseudo_{ref}$ in situations where the plant is unable to meet the control demand.

Next, \gls{PCH} is introduced for the $\Vector{\omega}$ and $\Vector{V}$ reference models.
To implement \gls{PCH} in both the angular and linear velocity reference models, the unachievable portion of the pseudo control command is transformed back into the corresponding acceleration domain.
This transformation enables the reference models to be slowed down by an amount proportional to the part of the command that the plant is physically unable to achieve.

The hedging signal $\dot{\pseudo}_{hedge}$ operates on the rates of change of the propulsion forces and moments.
Since forces and moments are on the same dynamic level as accelerations, their time derivatives, i.e. $\dot{\pseudo}_{hedge}$, are on the same dynamic level as linear and angular jerks.

To convert $\dot{\pseudo}_{hedge}$ into linear and angular jerks, the signal is multiplied by the inverse of the generalized mass inertia matrix:
\begin{align}
    \begin{bmatrix}
        \ddot{u}_{hedge} \\
        \ddot{w}_{hedge} \\
        \ddot{p}_{hedge} \\
        \ddot{q}_{hedge} \\
        \ddot{r}_{hedge}
    \end{bmatrix} & = \left( \tilde{\Matrix{P}} \overline{\Matrix{M}}_a \tilde{\Matrix{P}}^T \right)^{-1}  \begin{bmatrix}
                                                                                                               0 & 0 & 0 & 1 & 0 \\
                                                                                                               0 & 0 & 0 & 0 & 1 \\
                                                                                                               1 & 0 & 0 & 0 & 0 \\
                                                                                                               0 & 1 & 0 & 0 & 0 \\
                                                                                                               0 & 0 & 1 & 0 & 0
                                                                                                           \end{bmatrix} \dot{\pseudo}_{hedge} \\
    \tilde{P}           & = \begin{bmatrix}
                                1 & 0 & 0 & 0 & 0 & 0 \\
                                0 & 0 & 1 & 0 & 0 & 0 \\
                                0 & 0 & 0 & 1 & 0 & 0 \\
                                0 & 0 & 0 & 0 & 1 & 0 \\
                                0 & 0 & 0 & 0 & 0 & 1
                            \end{bmatrix} \nonumber
\end{align}
Here, the matrix $\tilde{\Matrix{P}}$ is used to exclude the second row and column of the full mass-inertia matrix $\overline{\Matrix{M}}_a$, as these correspond to lateral force and lateral acceleration, which are not actively controlled in this configuration.

The reference models are adjusted accordingly, to incorporate the hedging signals to the innermost loops.
The adjusted reference models can be seen in \cref{fig:vdot_ref_hedge,fig:omdot_ref_hedge}.

\begin{figure}
    \centering
    \includegraphics[width=\textwidth]{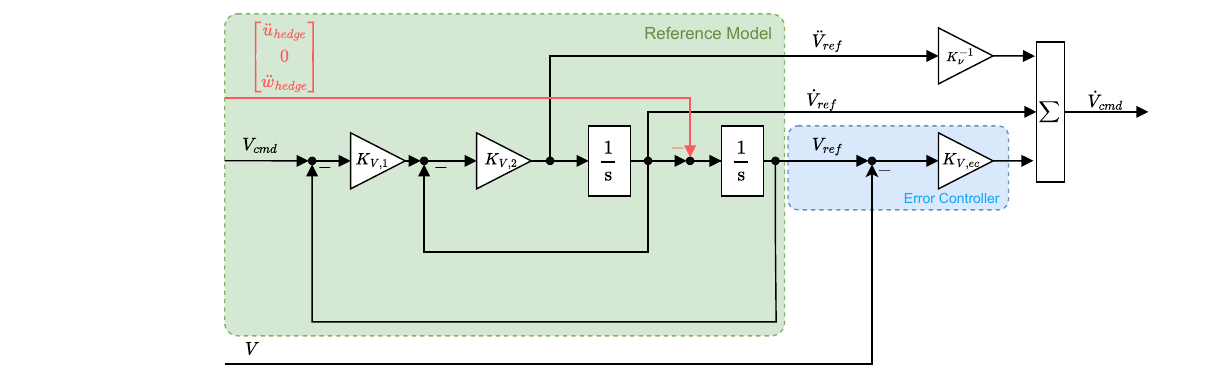}
    \caption{$\Vector{V}$ reference model adjusted with \gls{PCH}}
    \label{fig:vdot_ref_hedge}
\end{figure}

\begin{figure}
    \centering
    \includegraphics[width=\textwidth]{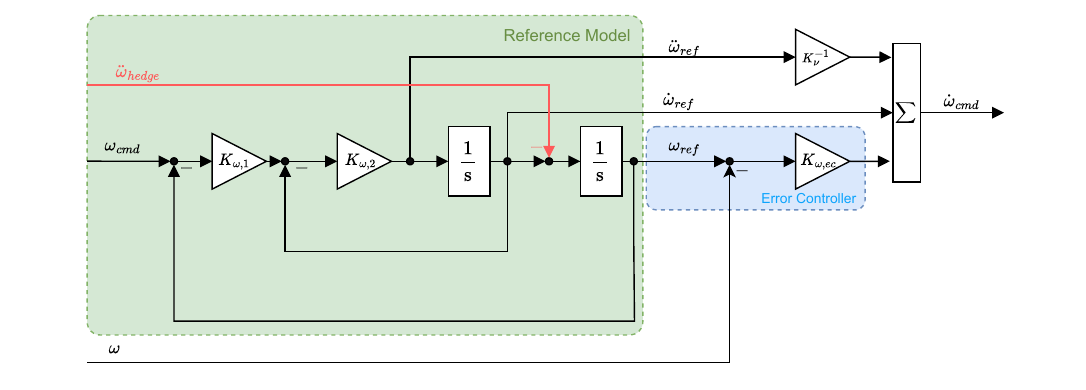}
    \caption{$\Vector{\omega}$ reference model adjusted with \gls{PCH}}
    \label{fig:omdot_ref_hedge}
\end{figure}

The effect of the \gls{PCH} shall be demonstrated with a simple example.
In \cref{fig:example_pch_inactive} and \cref{fig:example_pch_active} the response of the inner loop can be seen for a forward velocity step command of \qty{9}{\meter\per\second}.
In \cref{fig:example_pch_inactive}, the \gls{PCH} is disabled.
The plant can not keep up with the reference model which leads to larger errors between the two. The large errors lead to large corrective actions of the error controller, resulting in subtle, damped oscillations around the setpoint.
In contrast, in \cref{fig:example_pch_active} the \gls{PCH} is active.
The effect is easily observed, as the \gls{PCH} slows down the reference model, making the reference achievable by the plant, leading to overall more accurate and stable tracking of the reference.

\begin{figure}
    \centering
    \includegraphics{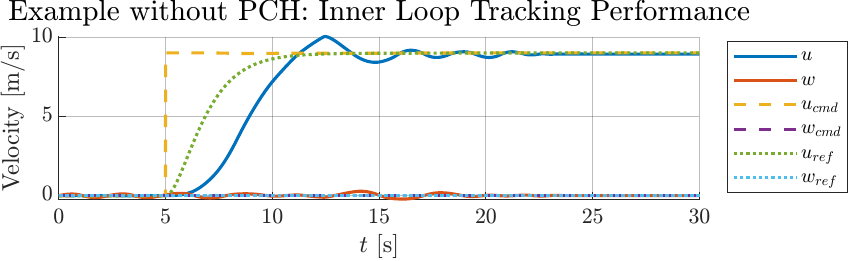}
    \caption{Velocity tracking without PCH}
    \label{fig:example_pch_inactive}
\end{figure}

\begin{figure}
    \centering
    \includegraphics{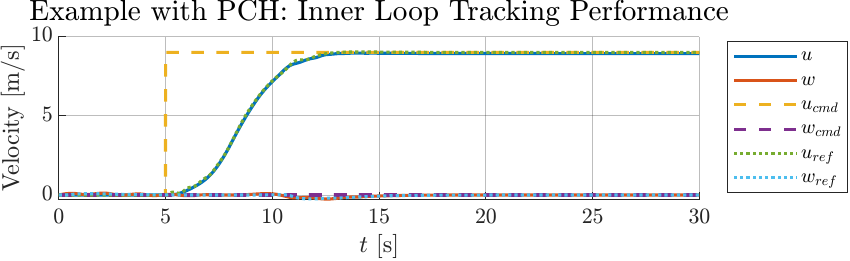}
    \caption{Velocity tracking with PCH enabled}
    \label{fig:example_pch_active}
\end{figure}

To summarize this section, the inner loop is responsible for tracking the desired body-fixed velocities and angular rates by generating actuator commands. To do this, the system uses a model-based approach that includes nonlinear dynamic inversion and an \gls{INDI}-based strategy.

A key concept in this context is the use of pseudo controls. These are forces and moments generated by the actuators, which are directly influenced by the control inputs. Instead of working with accelerations, the controller uses forces and moments because they are on the same dynamic level and easier to handle for control allocation purposes.

To ensure smooth tracking, reference models are used that generate the required accelerations. These accelerations are then converted into pseudo control commands. Because model errors and actuator limits can prevent the plant from exactly following these commands, an error controller corrects the pseudo control signal based on the actual system response.

An extended version of \gls{INDI} is used to compute the required actuator rate commands. This method includes the effect of actuator dynamics and rate limits. It works by calculating the time derivative of the pseudo control and translating it into actuator commands using the known control effectiveness matrix.

Finally, to avoid integrator windup when the actuator limits are reached, pseudo control hedging is used. This method slows down the internal reference models by subtracting the part of the command the plant could not achieve. This ensures realistic internal states, even when full control authority is not available.

\section{Outer Loop}
In this section, the outer loop control structure of the airship controller is presented.
The outer loop takes the azimuth angle rate $\dot{\varPsi}_{cmd}$ and the forward and vertical velocity in the C-frame $\left[u \ w  \right]^T_{cmd}$ as inputs and outputs an angular velocity command and the x- and z-components of the body-fixed velocity to the inner loop.
As a short recap, the C-frame results from the NED-frame by a rotation around the common z-axis by the azimuth angle $\varPsi$.
This is graphically illustrated in \cref{fig:C_Frame}

\begin{figure}
    \centering
    \includegraphics[width=\imagewidth]{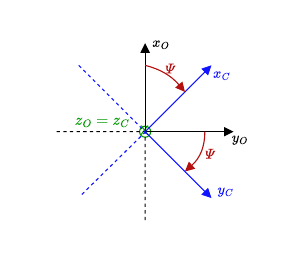}
    \caption{Illustration of how the C-Frame results from the NED-frame}
    \label{fig:C_Frame}
\end{figure}

Internally, the outer loop consists of three submodules:
\begin{enumerate}
    \item \emph{Flight Path Controller}: Has the velocity command in the C-frame as input and calculates the necessary pitch angle $\varTheta$ to achieve the desired trajectory. This pitch angle is forwarded to the attitude controller. Also transforms the velocity command from the C- to the B-frame as input for the inner loop
    \item \emph{Attitude Controller}: Is responsible for tracking the desired pitch angle and bank angle. The latter is internally calculated as a function of turn rate. Additionally transforms desired euler angle rates into body-fixed angular velocities.
    \item \emph{Wind Compensation}: Rejecting wind disturbances by using lateral loadfactor feedback to calculate a compensating $\dot{\varPsi}$ command that is added to the pilot commanded turn rate command
\end{enumerate}

\subsection{Flight Path Controller}
In this section the flight path controller is presented, which is responsible for calculating the desired pitch angle for the attitude controller and the velocity components in the body-fixed frame for the inner loop.
A block diagram of the flight path controller is depicted in \cref{fig:flight_path_controller}.

\begin{figure}
    \centering
    \includegraphics[width=\linewidth]{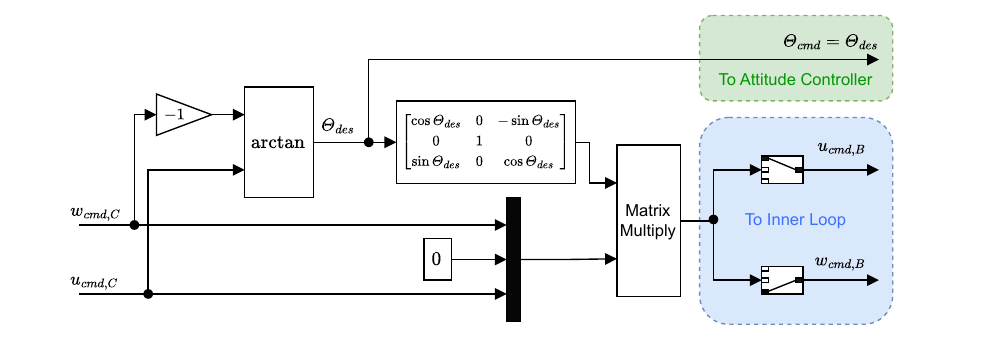}
    \caption{Block Diagram of the Flight Path Controller}
    \label{fig:flight_path_controller}
\end{figure}

In order to obtain the desired pitch angle, the required climb angle must be determined.
The kinematic velocity components in x- and z-direction of the C-frame define the kinematic climb angle of the airship approximately via the following trigonometric relationship:
\begin{equation}
    \tan{\gamma_K} \approx \frac{-w_C}{u_c}
\end{equation}
where $\gamma_K$ is the kinematic climb angle.
In \cref{fig:gamma}, the relationship is graphically shown.
Assuming only forward and vertical velocity components in the C-frame, the velocity vector $\Vector{V}$ is inclined relative to the C-frame with the kinematic climb angle $\gamma_K$.

\begin{figure}
    \centering
    \includegraphics{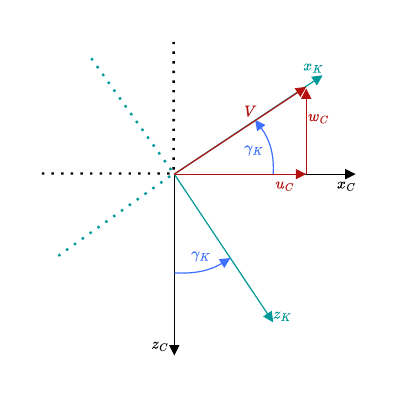}
    \caption{Definition of the kinematic climb angle $\gamma_K$ via the velocity components in the C-frame}
    \label{fig:gamma}
\end{figure}

The relationship can be derived via the following: the kinematic velocity components in the NED-frame can be expressed as \cite{FSD2}:
\begin{equation}
    \left( \Vector{V}_K^R \right)^{E}_O = V_K \begin{bmatrix}
        \cos{\chi_K} \cos{\gamma_K} \\
        \sin{\chi_K} \cos{\gamma_K} \\
        -\sin{\gamma_K}             \\
    \end{bmatrix}
\end{equation}
where $V_K$ is the absolute value of the kinematic velocity and $\chi_K$ is the flight path azimuth angle.
The velociy components can be transformed from the NED-frame to the C-frame via a rotation around the common z-axis with the azimuth angle $\varPsi$:
\begin{align}
    \left( \Vector{V}_K^R \right)^{E}_C & = \Matrix{M}_{CO} \left( \Vector{V}_K^R \right)^{E}_O = V_K \begin{bmatrix}
                                                                                                          \cos{\varPsi}  & \sin{\varPsi} & 0 \\
                                                                                                          -\sin{\varPsi} & \cos{\varPsi} & 0 \\
                                                                                                          0              & 0             & 1
                                                                                                      \end{bmatrix} \begin{bmatrix}
                                                                                                                        \cos{\chi_K} \cos{\gamma_K} \\
                                                                                                                        \sin{\chi_K} \cos{\gamma_K} \\
                                                                                                                        -\sin{\gamma_K}             \\
                                                                                                                    \end{bmatrix}\nonumber \\
                                        & = V_K \begin{bmatrix}
                                                    \cos{\varPsi} \cos{\chi_K} \cos{\gamma_K} + \sin{\varPsi} \sin{\chi_K} \cos{\gamma_K}  \\
                                                    -\sin{\varPsi} \cos{\chi_K} \cos{\gamma_K} + \cos{\varPsi} \sin{\chi_K} \cos{\gamma_K} \\
                                                    -\sin{\gamma_K}
                                                \end{bmatrix}
\end{align}
In the following it is assumed that the flight path azimuth angle $\chi_K$ is approximately equal to the azimuth angle $\varPsi$, which is true if the kinematic sideslip angle is very small.
The kinematic sideslip angle is defined as \cite{FSD2}:
\begin{equation}
    \beta^R_K = \frac{(v_K^R)^{E}_B}{\sqrt{\left[(u_K^R)^{E}_B\right]^2 + \left[(w_K^R)^{E}_B\right]^2}}
\end{equation}
Therefore, assumung $(v_K^R)^{E}_B << \sqrt{\left[(u_K^R)^{E}_B\right]^2 + \left[(w_K^R)^{E}_B\right]^2}$, the sideslip angle can be neglected.
By setting $\chi_K \approx \varPsi$ the velocity components in the C-frame can be written in a simplified manner:
\begin{align}
    \left( \Vector{V}_K^R \right)^{E}_C & = V_K \begin{bmatrix}
                                                    \cos{\varPsi} \cos{\chi_K} \cos{\gamma_K} + \sin{\varPsi} \sin{\chi_K} \cos{\gamma_K}  \\
                                                    -\sin{\varPsi} \cos{\chi_K} \cos{\gamma_K} + \cos{\varPsi} \sin{\chi_K} \cos{\gamma_K} \\
                                                    -\sin{\gamma_K}
                                                \end{bmatrix}         \\
    \left( \Vector{V}_K^R \right)^{E}_C & \approx V_K \begin{bmatrix}
                                                          \cos{\varPsi} \cos{\varPsi} \cos{\gamma_K} + \sin{\varPsi} \sin{\varPsi} \cos{\gamma_K}  \\
                                                          -\sin{\varPsi} \cos{\varPsi} \cos{\gamma_K} + \cos{\varPsi} \sin{\varPsi} \cos{\gamma_K} \\
                                                          -\sin{\gamma_K}
                                                      \end{bmatrix} \\
    \left( \Vector{V}_K^R \right)^{E}_C & \approx V_K \begin{bmatrix}
                                                          \cos{\gamma_K} (\cos^2{\varPsi} + \sin^2{\varPsi})                                       \\
                                                          -\sin{\varPsi} \cos{\varPsi} \cos{\gamma_K} + \cos{\varPsi} \sin{\varPsi} \cos{\gamma_K} \\
                                                          -\sin{\gamma_K}
                                                      \end{bmatrix} \\
    \left( \Vector{V}_K^R \right)^{E}_C & \approx V_K \begin{bmatrix}
                                                          \cos{\gamma_K}                                                                           \\
                                                          -\sin{\varPsi} \cos{\varPsi} \cos{\gamma_K} + \cos{\varPsi} \sin{\varPsi} \cos{\gamma_K} \\
                                                          -\sin{\gamma_K}
                                                      \end{bmatrix}
\end{align}
Finally, the kinematic climb angle can be calculated by considering the ratio of the third and first element:
\begin{align}
    \frac{(w_K^R)^{E}_C}{(u_K^R)^{E}_C} = \frac{-\sin{\gamma_K}}{\cos{\gamma_K}} \\
    \frac{-(w_K^R)^{E}_C}{(u_K^R)^{E}_C} = \frac{\sin{\gamma_K}}{\cos{\gamma_K}} \\
    \frac{-(w_K^R)^{E}_C}{(u_K^R)^{E}_C} = \tan{\gamma_K}
\end{align}

Assuming symmetric flight (no kinematic sideslip angle, no bank angle), the pitch angle can be expressed as the sum of the kinematic climb angle and the kinematic angle of attack \cite{FSD2}:
\begin{equation}
    \varTheta = \gamma_K + \alpha_K
\end{equation}
Since the airship does not need an angle of attack to stay airborne and a non-zero angle of attack increases drag, the goal is to fly with zero angle of attack.
When the angle of attack is zero, the pitch angle becomes the kinematic climb angle:
\begin{equation}
    \varTheta = \gamma_K + \underbrace{\alpha_K}_{=0} = \gamma_K
\end{equation}
Therefore, the desired pitch angle $\varTheta_{des}$ can be directly calculated from the velocity command in the C-frame:
\begin{equation}\label{eq:theta_des}
    \varTheta_{des} = \arctan{\frac{-(w_{cmd})_C}{(u_{cmd})_C}}
\end{equation}
The velocities can be transformed from the C-frame to the B-frame by a simple rotation around the y-axis with the desired pitch angle (assuming $\varPhi \approx 0$):
\begin{equation}\label{eq:vb_cmd_theta}
    V_{B,cmd} = \begin{bmatrix}
        \cos{\varTheta_{des}} & 0 & -\sin{\varTheta_{des}} \\
        0                     & 1 & 0                      \\
        \sin{\varTheta_{des}} & 0 & \cos{\varTheta_{des}}
    \end{bmatrix} \begin{bmatrix}
        u_{cmd} \\
        0       \\
        w_{cmd}
    \end{bmatrix}_C
\end{equation}
Using the desired pitch angle instead of the current pitch angle offers a key advantage, which can be illustrated with a simple example:

Assume the airship is flying straight and level at a constant forward velocity with zero vertical velocity. Now, a climb is commanded by specifying an upward velocity in the C-frame. If the transformation from the C-frame to the body-fixed frame were performed using the current pitch angle instead of the desired one, the vertical velocity command would initially result in a significant velocity component in the body-fixed z-direction. This causes a buildup of a negative angle of attack, which is undesirable as it degrades aerodynamic performance. The z-component of velocity in the body frame would only gradually diminish as the airship pitches up over time to align with the commanded flight path. Only once the desired pitch angle is fully reached would the velocity vector in the body frame align with the x-axis.

Using the desired pitch angle for the frame transformation ensures that the commanded velocity is immediately aligned with the body-fixed x-axis. This prevents the transient angle-of-attack excursion and ensures smoother, more aerodynamically efficient transitions during maneuvers.

A graphical illustration of the benefit of using the desired pitch angle for transforming velocity commands is provided in \cref{fig:climb1} and \cref{fig:climb2}. These figures depict the same climb maneuver, but differ in the frame transformation approach: \cref{fig:climb1} uses the actual pitch angle for the transformation, while \cref{fig:climb2} employs the desired pitch angle.

In \cref{fig:climb1}, the scenario begins with the airship in steady, level flight. This is shown in the first frame, where the airship has a forward velocity and zero pitch angle. In the second frame, a climb command is issued, resulting in a new commanded velocity vector that includes both a forward and an upward component in the C-frame.
The commanded velocity vector in the C-frame is now pointed upwards, revealing the desired pitch angle $\varTheta_{des}$.

This commanded velocity is then transformed into the body-fixed frame using the current pitch angle, which in this time instance is still zero. As a result, the body-fixed frame and the C-frame are equal and the commanded velocity in the body frame includes a significant vertical component. This vertical velocity leads to a negative angle of attack: the airship appears to "push" into the airflow from below. A corresponding pitch rate command is also generated to initiate nose-up rotation.

This angle of attack vanishes when the airship reaches the desired pitch angle, as shown in the last box.
When the desired pitch angle is reached, the velocity command in the body-fixed frame only consists of a forward component.

\begin{figure}
    \centering
    \includegraphics{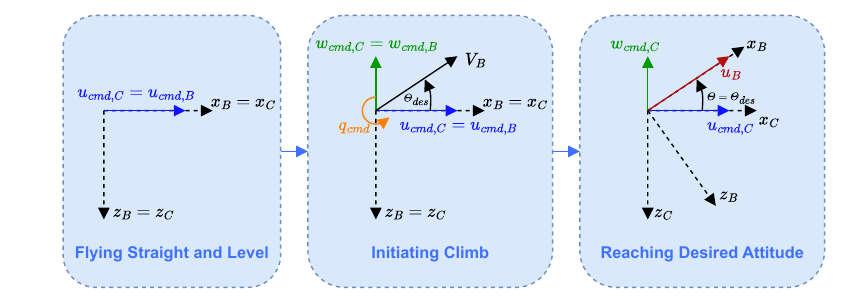}
    \caption{Illustration of the climb maneuver using the actual pitch angle for velocity transformation}
    \label{fig:climb1}
\end{figure}

In order to avoid the transient angle of attack during climb or descent maneuvers, the desired pitch angle is used for transforming the velocity components into the body-fixed frame.
\cref{fig:climb2} shows how the scenario differs from the previous case.
The first box shows the airship again in steady-state straight and level flight, with only a forward velocity.
Again, a climb command is issued.
For the sake of illustration, consider the forward velocity command in the C-frame to be $u_{cmd,C} = \qty{10}{\meter\per\second}$ and the vertical velocity command to be $w_{cmd,C} = \qty{-3}{\meter\per\second}$.
According to \cref{eq:theta_des}, this leads to a desired climb angle of $\varTheta_{des} = \arctan{-\frac{-3}{10}} \approx \ang{16.7}$.
Using this desired pitch angle, the velocity components are transformed from the C-frame to the body-fixed frame according to \cref{eq:vb_cmd_theta}:
\begin{equation}
    V_{B,cmd} = \begin{bmatrix}
        \cos{\ang{16.7}} & 0 & -\sin{\ang{16.7}} \\
        0                & 1 & 0                 \\
        \sin{\ang{16.7}} & 0 & \cos{\ang{16.7}}
    \end{bmatrix} \begin{bmatrix}
        \qty{10}{\meter\per\second} \\
        0                           \\
        \qty{-3}{\meter\per\second}
    \end{bmatrix}_C =
    \begin{bmatrix}
        \qty{10.44}{\meter\per\second} \\
        0                              \\
        0
    \end{bmatrix}_B
\end{equation}
As expected, the vertical velocity command in the body-fixed frame is zero and stays constant during the pitch up maneuver, not creating an angle of attack.
This is shown in the second box. The climb command does only lead to a forward velocity command in the body-fixed frame along with a commanded pitch rate to achieve the desired pitch attitude.
The vertical velocity command in the C-frame is therefore solely achieved by the pitch up maneuver.
In the last box shows the airship in the desired pitch attitude, which is identical to the previous case.

\begin{figure}
    \centering
    \includegraphics{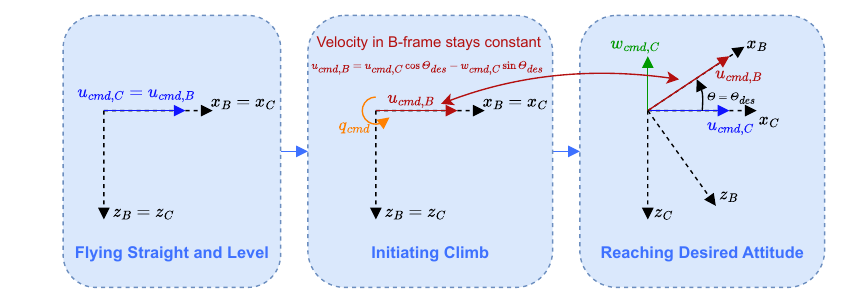}
    \caption{Illustration of the climb maneuver using the desired pitch angle for velocity transformation}
    \label{fig:climb2}
\end{figure}

\subsection{Attitude Controller}
The attitude controller has the task of tracking the desired attitude of the airship.
The desired attitude comes from two sources: the desired climb angle is calculated by the flight path controller, as just outlined. The desired roll angle is calculated from the desired turn rate and the forward velocity.
The azimuth angle is not actively controlled, rather the azimuth angle rate is controlled.

The attitude controller generates desired euler angle rates, which are in turn transformed into body-fixed angular velocities that serve as inputs for the inner loop.

A high level block diagram of the attitude controller is shown in \cref{fig:attitude_controller}.
The individual components will be further explained in this subsection.

\begin{figure}
    \centering
    \includegraphics[width=\linewidth]{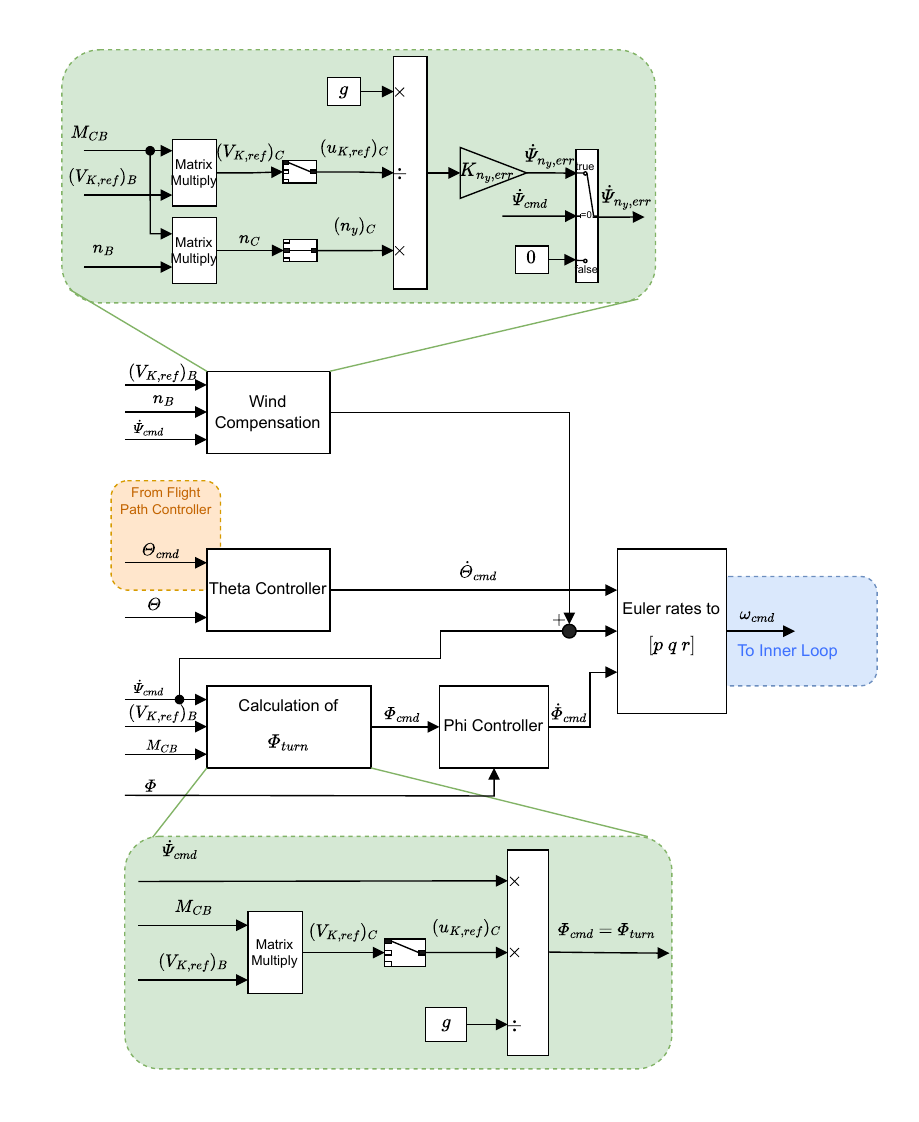}
    \caption{Block Diagram of the Attitude Controller}
    \label{fig:attitude_controller}
\end{figure}

\subsubsection{Calculation of the desired roll angle}
Unlike conventional fixed-wing aircraft, airships do not rely on banking to generate lateral lift during turning. Instead, they achieve turning flight primarily through sideslip, where the hull generates aerodynamic side force. This lateral force replaces the role of a tilted lift vector and enables curved flight without significant banking.

In fixed-wing aircraft, coordinated turns are achieved by rolling the aircraft so the lift vector tilts inward, providing the centripetal force needed for the turn. As a result, the apparent acceleration aligns with the body-fixed z-axis, creating a balanced and comfortable turn.

For airships, coordinated flight follows a different principle. Because they remain mostly level and generate side force via sideslip, the required lateral force leads to a small natural roll angle. This roll angle does not stem from banking but arises due to the tilt of the apparent acceleration vector caused by the lateral load factor. The controller must account for this effect to ensure the airship avoids unnecessary corrective actions that would counteract this natural behavior \cite{kaempf}.

\cref{fig:force_equi_turn} illustrates the force equilibrium during turn flight of the airship.
\begin{figure}
    \centering
    \includegraphics[width=\imagewidth]{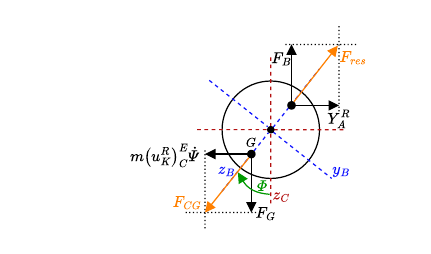}
    \caption{Force equilibrium during coordinated turn flight (adapted from \cite{kaempf})}
    \label{fig:force_equi_turn}
\end{figure}
The airship is assumed to fly at constant altitude while performing a turn with the turn rate $\dot{\varPsi}$.
In the vertical plane there is the gravitational force $F_G$ and the buoyancy force $F_B$. For simplicity they are assumed to be equal but in opposite directions.
In the horizontal plane there's the aerodynamic side force $Y_A^R$ generated by the angle of sideslip, which is acting inward toward the turn center.
The aerodynamic force is balanced by the inertial force $\Vector{F}_{centrifugal}$, which arises due to the curved flightpath, acting outward from the center of gravity.
The centrifugal force relates to the turn rate via $F_{centrifugal} = m\left(u_K^R\right)^E_C \dot{\varPsi}$.

Due to the side forces and the position of the center of gravity being below the center of buoyancy, a rolling moment is induced which generates a rolling motion until the resultant force vector is aligned with the body-fixed z-axis.
This leads to a natural roll angle that arises in turning flight.

An approximation for the natural roll angle in steady, coordinated turn flight can be derived from the ratio of centrifugal and gravitational forces \cite{kaempf}:
\begin{equation}\label{eq:load_factor_turn}
    \varPhi_{turn} \approx \arctan{\frac{F_{centrifugal}}{F_G}} = \arctan{\frac{m\left(u_K^R\right)^E_C \dot{\varPsi}}{mg}} \approx \frac{\left(u_K^R\right)^E_C \dot{\varPsi}}{g}
\end{equation}

where $\left(u_K^R\right)^E_C$ is the forward kinematic velocity, $\dot{\Psi}$ is the turn rate, and $g$ is the gravitational acceleration.

This natural roll angle is calculated by the controller and used as a feedforward term in the attitude command. This prevents the controller from fighting the natural roll angle, which in turn preserves control authority for other tasks.

\subsection{Phi and Theta Controller}
Simple proportional controllers are used to generate the $\dot{\varPhi}_{cmd}$ and $\dot{\varTheta}_{cmd}$ command values. Both controllers are shown in \cref{fig:phi_ref} and \cref{fig:theta_ref}, respectively.

The pitch command $\varTheta_{cmd}$ is provided by the flight path controller, while the roll command $\varPhi_{cmd}$ corresponds to $\varPhi_{turn} = \frac{\left(u_K^R\right)^E_C \dot{\varPsi}}{g}$.

\begin{figure}
    \centering
    \includegraphics[width=\imagewidth]{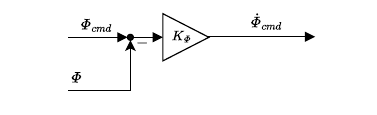}
    \caption{$\varPhi$ controller}
    \label{fig:phi_ref}
\end{figure}

\begin{figure}
    \centering
    \includegraphics[width=\imagewidth]{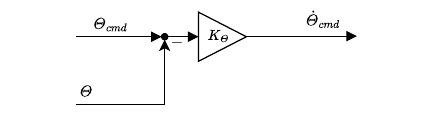}
    \caption{$\varTheta$ controller}
    \label{fig:theta_ref}
\end{figure}
\subsection{Transforming from Euler Rates to Angular Velocities}
The attitude controller employs nonlinear dynamic inversion to compute the body-fixed angular velocity command that is passed to the inner loop. Specifically, the controller inverts the attitude differential equations given in \cref{eq:attitude_dgl}. Based on the desired rates of change for the Euler angles, the corresponding body-fixed angular velocity command can be calculated as \cite{FSD2}:
\begin{equation}
    \Vector{\omega}_{cmd} = \begin{bmatrix}
        \dot{\varPhi}_{cmd} - \dot{\varPsi}_{cmd} \sin{\varTheta}                               \\
        \dot{\varTheta}_{cmd} \cos{\varPhi} + \dot{\varPsi}_{cmd} \sin{\varPhi} \cos{\varTheta} \\
        -\dot{\varTheta}_{cmd} \sin{\varPhi} + \dot{\varPsi}_{cmd} \cos{\varPhi} \cos{\varTheta}
    \end{bmatrix}
\end{equation}

\subsection{Wind Compensation}
Airships are particularly prone to atmospheric wind disturbances due to their relatively low flight speed, low mass, and large surface area. The low flight speed means that even moderate wind velocities can significantly alter the flow angles, specifically the angle of attack and angle of sideslip. These changes in flow angles affect the pressure distribution along the hull. Given the large surface area of the hull, even small variations in pressure can generate considerable aerodynamic forces and moments. Due to the airship's low mass, these forces have the potential to yield relatively large accelerations, making precise control more challenging in windy conditions.

To ensure stable and efficient flight, the airship should always be aligned with the relative wind direction, meaning the angle of sideslip should be kept close to zero. This condition minimizes unwanted aerodynamic forces on the hull and ensures controllability. Wind disturbances are detected through feedback of the lateral load factor. When sideslip occurs, it generates a lateral aerodynamic force, which in turn results in a measurable lateral load factor. The controller uses this information to compute a turn rate command. This command is derived based on the current velocity reference and the measured lateral loadfactor, effectively creating a restoring command that turns the airship back into the wind. This mechanism helps maintain stable flight and suppresses drift caused by crosswinds.
A graphical representation of this is presented in \cref{fig:airship_wind}.

\begin{figure}
    \centering
    \includegraphics[width=\imagewidth]{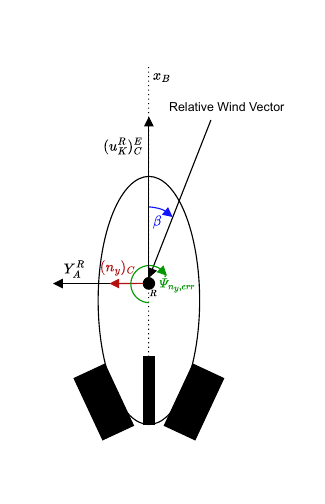}
    \caption{Sketch of the sideslip correction mechanism}
    \label{fig:airship_wind}
\end{figure}

The lateral load factor in turning flight is related to the measured forwards kinematic velocity and turn rate via (compare \cref{eq:load_factor_turn}):
\begin{equation}
    (n_y)_C = \frac{(u^R_K)_C \dot{\varPsi}}{g}
\end{equation}
Solving this equation for $\dot{\varPsi}$ and using a tunable gain, a restoring command based on measured acceleration and measured velocity can be computed:
\begin{equation}
    \dot{\varPsi}_{n_y,err} = K_{n_y,err} \frac{g}{(u^R_K)_C} (n_y)_C
\end{equation}
During pilot commanded turn flight, lateral forces naturally develop as a result of the curved trajectory. These forces generate a non-zero lateral load factor, even in the absence of wind disturbances. This presents a challenge for the wind rejection controller, which also uses lateral load factor as an indicator of sideslip caused by crosswind. If not accounted for, the controller may incorrectly interpret the lateral force from intentional turning as a wind disturbance, resulting in a control action that opposes the commanded turn.

To avoid such interference, the wind rejection controller is selectively enabled. Specifically, it is only active when the pilot's commanded turn rate is zero. This ensures that the controller does not conflict with intentional maneuvering. When no turn is commanded, any observed lateral load factor can be attributed to wind-induced sideslip, and the controller will generate a corrective yaw rate command to align the airship with the apparent wind. In contrast, when a turn is actively commanded by the pilot, the lateral load factor is expected and permitted, and the wind controller is temporarily disabled to prevent undesired counteraction.

\section{Kinematic Calculations / Filtering}
The inner loop requires feedback of the angular acceleration $\dot{\omega}$. While sensors for linear acceleration are commonly available and used on the airship, sensors capable of directly measuring angular acceleration are rare and not typically employed in aerospace applications \cite{sieberling_robust_2010}. Consequently, angular acceleration must be estimated from other available measurements.

A common approach for signal estimation involves the use of state observers, such as Kalman filters or Luenberger observers. These observers combine system model knowledge with sensor data to provide robust and accurate estimates, effectively leveraging the strengths of both.

However, for the airship application, a complementary filter was chosen instead. This decision was driven by several practical advantages: the complementary filter is computationally efficient, straightforward to implement, and easy to tune, as it avoids the need to define or optimize covariance matrices. Despite its simplicity, it provides sufficiently accurate estimates of angular acceleration for control purposes.

The complementary filter method combines the strengths of two different estimation strategies: a model-based prediction and a signal-based differentiation.

The model-based prediction method uses the known equations of motion of the airship to estimate angular accelerations from measured states and applied forces and moments. This approach is advantageous due to its low noise and fast response, as it avoids numerical differentiation. However, it suffers from poor robustness against model uncertainties and external disturbances, as it lacks any form of feedback correction.

In contrast, the differentiation-based method estimates angular acceleration by numerically differentiating angular velocity measurements and applying a low pass filter filter to reduce high frequency noise introduced by numerical differentiation. While this method theoretically yields the "real" angular acceleration due to not relying on a model, it also amplifies high-frequency measurement noise and introduces phase lag due to the low pass filter, which can then reduce its responsiveness.

The complementary filter addresses these limitations by fusing both estimation results. It applies a low-pass filter to the differentiated measurement, preserving low-frequency accuracy, while applying a high-pass filter to the model-based estimate, providing a fast and noise-free response. Since the two filters are complementary (i.e., their frequency responses sum to one), the resulting estimate maintains both lower noise and good responsiveness \cite{jiali_angular_2016}.

The estimate for the angular acceleration by the complementary filter can be written as:
\begin{equation}
    \hat{\dot{\Vector{\omega}}}(s) = \underbrace{\frac{1}{\tau_c s + 1}}_{\text{lowpass}} \dot{\Vector{\omega}}_{meas} + \underbrace{\frac{\tau_c s}{\tau_c s + 1}}_{\text{highpass}} \dot{\Vector{\omega}}_{mdl}
\end{equation}
Where $\tau_c$ is the time constant of the filter, $\dot{\Vector{\omega}}_{meas}$ is the angular acceleration resulting from differentiation of the measured angular velocity and $\dot{\Vector{\omega}}_{mdl}$ is the angular acceleration as calculated by the model dependent part.

The measured angular velocity is passed through a high-pass filter to approximate differentiation, since pure differentiation is not physically realizable due to its improper transfer function:
\begin{equation}
    \dot{\Vector{\omega}}_{meas} = \frac{s}{\tau_d s + 1} \Vector{\omega}_{meas}
\end{equation}
where $\Vector{\omega}_{meas}$ is the measured angular velocity and $\tau_d$ is the time constant of the filter.
For low frequencies ($s \rightarrow 0$) this filter approximates the derivative of the input signal while for high frequencies ($s \rightarrow \infty$) the gain is limited to approximately $\frac{1}{\tau_d}$.

The model dependend $\dot{\Vector{\omega}}_{mdl}$ is calculated according to \cref{eq:eom_simple}.

The time constants of the filters were chosen empirically as $\tau_d = \frac{1}{25} s$ and  $\tau_c = \frac{1}{15} s$.
For implementation purposes, the filters were implemented in matlab as transfer functions which were then converted into state space formulation and discretized via the built-in bilinear transformation, which is also known as Tustin's method.
The discreted state space models are then implemented in Simulink.

In addition to angular acceleration, linear acceleration is also used as feedback in the inner loop. Since linear accelerations can be directly measured using accelerometers, no estimation is required. However, kinematic corrections are necessary for two main reasons. First, the accelerometer provides acceleration data at the center of gravity, whereas the inner loop requires the acceleration at the reference point, i.e. the center of buoyancy. Second, accelerometer measurements include not only the accelerations caused by motion, but also the gravitational component. As a result, the raw measurement must be corrected to remove the influence of gravity.

The accelerometer measurement is denoted as $\Vector{f}^G$, representing the specific force measured at the center of gravity. When corrected for gravity, this measurement corresponds to the linear acceleration of the center of gravity relative to the inertial frame $I$.
\begin{equation}\label{eq:specific_forces}
    \Vector{f}^G + \Vector{g} = (\Vector{a}^G)^{II} = (\dot{\Vector{V}}_K^G)^{II}
\end{equation}
Here, $(\dot{\Vector{V}}_K^G)^{II}$ denotes the time derivative of the kinematic velocity of the center of gravity, taken with respect to the inertial frame. In other words, it represents the second time derivative—again with respect to the inertial frame—of the position vector $\Vector{r}^G$ of the center of gravity. This relationship can be written as:
\begin{equation}
    (\dot{\Vector{V}}_K^G)^{II} = \left( \odv{}{t} \right)^{I} \left[ \left( \odv{}{t} \right)^{I} \Vector{r}^G  \right]
\end{equation}
The position vector $\Vector{r}^G$ can be expressed as the sum of the vector to the reference point $\Vector{r}^R$ and the vector from the reference point to the center of gravity $\Vector{r}^{RG}$:
\begin{equation}
    \left( \odv{}{t} \right)^{I} \left[ \left( \odv{}{t} \right)^{I} \Vector{r}^G  \right] = \left( \odv{}{t} \right)^{I} \left[ \left( \odv{}{t} \right)^{I} \left( r^R + r^{RG} \right)  \right]
\end{equation}
When differentiating a vector quantity, the Euler differentiation rule can be applied to express the derivative in a different reference frame. This rule introduces an additional term: the cross product of the angular velocity between the two frames and the vector being differentiated. In this case, the derivative of $\Vector{r}^R$ is taken with respect to the Earth-fixed frame, because the velocity with respect to the earth surface is the desired quantity, while the derivative of $\Vector{r}^{RG}$ is computed in the body-fixed frame, which allows some simplification:
\begin{equation}
    \left( \odv{}{t} \right)^{I} \left[ \left( \odv{}{t} \right)^{I} \left( \Vector{r}^R + \Vector{r}^{RG} \right)  \right] = \left( \odv{}{t} \right)^{I} [ \underbrace{(\Vector{\Vector{V}}_K^R)^{E} + \overbrace{\Vector{\omega}^{IE}}^{=0} \times \Vector{r}^R}_{\left( \odv{}{t} \right)^{I} \Vector{r}^R} + \underbrace{\overbrace{(\dot{\Vector{r}}^{RG})^B}^{=0} + \Vector{\omega}^{IB} \times \Vector{r}^{RG}}_{\left( \odv{}{t} \right)^{I} \Vector{r}^{RG}}]
\end{equation}
Here, the term $\Vector{\omega}^{IE}$ vanishes because Earth's rotation is neglected in this model. Similarly, the term $(\dot{\Vector{r}}^{RG})^B$ becomes zero under the rigid body assumption, as the relative position between the reference point and the center of gravity does not change over time in the body-fixed frame.

The second time derivative is then applied individually to the two remaining terms.
\begin{align}
    \left( \odv{}{t} \right)^{I} (\Vector{\Vector{V}}_K^R)^{E}                 & = \left( \dot{\Vector{V}}_K^R \right)^{EB} + \Vector{\omega}^{IB} \times \left( \Vector{V}_K^R \right)^E                                                                                                     \\
    \left( \odv{}{t} \right)^{I} [\Vector{\omega}^{IB} \times \Vector{r}^{RG}] & = \left( \dot{\Vector{\omega}}^{IB} \right)^I \times \Vector{r}^{RG} + \Vector{\omega}^{IB} \times [ \overbrace{\left( \dot{\Vector{r}}^{RG} \right)^B}^{=0} + \Vector{\omega}^{IB} \times \Vector{r}^{RG} ]
\end{align}
And the results can be summed to yield the total derivative:
\begin{gather}
    \left( \odv{}{t} \right)^{I} [ (\Vector{\Vector{V}}_K^R)^{E} + \Vector{\omega}^{IB} \times \Vector{r}^{RG}] = \\
    = \left( \dot{\Vector{V}}_K^R \right)^{EB} + \Vector{\omega}^{IB} \times \left( \Vector{V}_K^R \right)^E + \left( \dot{\Vector{\omega}}^{IB} \right)^I \times \Vector{r}^{RG} + \Vector{\omega}^{IB} \times [ \Vector{\omega}^{IB} \times \Vector{r}^{RG} ]
\end{gather}
The term $\left( \dot{\Vector{\omega}}^{IB} \right)^I$ can be transformed using the Euler derivative:
\begin{equation}
    \left( \dot{\Vector{\omega}}^{IB} \right)^I = \left( \dot{\Vector{\omega}}^{IB} \right)^B + \underbrace{\Vector{\omega}^{IB} \times \Vector{\omega}^{IB}}_{=0} = \left( \dot{\Vector{\omega}}^{IB} \right)^B
\end{equation}
The cross product $\Vector{\omega}^{IB} \times \Vector{\omega}^{IB}$ equals zero, as the cross product of any vector with itself always results in a zero vector. Furthermore, under the assumption of a non-rotating and flat Earth, both the Earth's angular velocity $\Vector{\omega}^{IE}$ and the transport rate $\Vector{\omega}^{EO}$ are zero. As a result, the angular velocity $\Vector{\omega}^{IB}$ simplifies to:
\begin{equation}
    \Vector{\omega}^{IB} = \underbrace{\Vector{\omega}^{IE}}_{=0} + \underbrace{\Vector{\omega}^{EO}}_{=0} + \Vector{\omega}^{OB}
\end{equation}
The derived expression for $(\dot{\Vector{V}}_K^G)^{II}$ is now substituted into \cref{eq:specific_forces}, yielding:
\begin{equation}
    \Vector{f}^G + \Vector{g} = \left( \dot{\Vector{V}}_K^R \right)^{EB} + \Vector{\omega}^{OB} \times \left( \Vector{V}_K^R \right)^E + \left( \dot{\Vector{\omega}}^{OB} \right)^B \times \Vector{r}^{RG} + \Vector{\omega}^{OB} \times [ \Vector{\omega}^{OB} \times \Vector{r}^{RG} ]
\end{equation}
Solving this equation for the desired quantity $\left( \dot{\Vector{V}}_K^R \right)^{EB}$ gives:
\begin{align}
    \left( \dot{\Vector{V}}_K^R \right)^{EB} & = \Vector{f}^G + \Vector{g} - \Vector{\omega}^{OB} \times \left( \Vector{V}_K^R \right)^E - \left( \dot{\Vector{\omega}}^{OB} \right)^B \times \Vector{r}^{RG} - \Vector{\omega}^{OB} \times [ \Vector{\omega}^{OB} \times \Vector{r}^{RG} ]                              \\
    \left( \dot{\Vector{V}}_K^R \right)^{EB} & = \Vector{f}^G + \Vector{g} - \left( \dot{\Vector{\omega}}^{OB} \right)^B \times \Vector{r}^{RG} - \Vector{\omega}^{OB} \times \underbrace{\left[\left( \Vector{V}_K^R \right)^E + \Vector{\omega}^{OB} \times \Vector{r}^{RG} \right]}_{\left( \Vector{V}_K^G \right)^E} \\
    \left( \dot{\Vector{V}}_K^R \right)^{EB} & = \Vector{f}^G + \Vector{g} - \left( \dot{\Vector{\omega}}^{OB} \right)^B \times \Vector{r}^{RG} - \Vector{\omega}^{OB} \times \left( \Vector{V}_K^G \right)^E \label{eq:sforces}
\end{align}
With this, the final expression for $\left( \dot{\Vector{V}}_K^R \right)^{EB}$ is obtained. In this formulation, the angular acceleration is provided by the complementary filter estimate described earlier, while $\left( \Vector{V}_K^G \right)^E$,  $\Vector{\omega}^{OB}$ and $\Vector{f}^G$ are measurements and $\Vector{r}^{RG}$ is a constant.

This completes the explanation of the filtering and kinematic transformations applied to derive the linear and angular acceleration at the reference point.

\section{Control Allocation}\label{sec:control_allocation}
The airship's inner loop controller determines the required rates of change for propulsion forces and moments as shown in previous sections. These rates act as pseudo control inputs, which must then be translated into actual physical actuator commands to realize the desired control effect.
This process is referred to as control allocation.
In this section, the general control allocation problem statement and its applciation to the airship flight controller shall be explained.

\subsection{Control Allocation in General}
As already outlined, control allocation deals with allocating pseudo control commands to real actuator commands.
Introducing pseudo controls to decouple the physical inputs from the system dynamics offers a key advantage: the pseudo control vector typically has a lower dimension than the full input vector, simplifying the controller design process \cite{ERP}. As a result, the system dynamics can now be reformulated in terms of the pseudo control inputs instead of the physical actuators \cite{ERP}:
\begin{align}
    \dot{\Vector{x}} & = \Vector{f}(\Vector{x}, \pseudo) \\
    \Vector{y}       & = \Vector{h}(\Vector{x})
\end{align}
In the most general case, the relationship between the real physical inputs and the pseudo control can be described by a nonlinear vector function \cite{JOHANSEN20131087}:
\begin{equation}
    \pseudo = \Vector{g}(\Vector{u},\Vector{x},t)
\end{equation}
In many technical systems, the relationship between
pseudo controls and physical inputs can be reasonably approximated as linear. This assumption simplifies the control allocation process considerably. Define the linear mapping as
\begin{equation}
    \pseudo = \Matrix{B} \Vector{u}
\end{equation}
where $\Matrix{B}$ is the so called \emph{control effectiveness matrix}, which is a linear mapping from the input space to the pseudo control space.

\subsection{Derivation of the Control Effectiveness Matrix}
To derive the control effectiveness matrix for the airship, recall that the inner loop computes the time derivative of the pseudo control vector, $\pseudodot$, as the control command. As a result, control allocation must also be performed on this derivative level, therefore allocating the rate of change of pseudo controls to the rate of change of the physical input vector $\udot$. In this context, the input vector $\Vector{u}$ consists of the rotor RPMs and tilt angles. The control allocation problem specific to the airship can therefore be expressed as:
\begin{equation}
    \dot{\pseudo} = \Matrix{B} \dot{\Vector{u}}
\end{equation}

The derivation of the control effectiveness matrix $\Matrix{B}$ is approached by the extended \gls{INDI} inversion law introduced in the inner loop section. For convenience, the inversion law is restated here:

\begin{equation}
    \Vector{u}_{cmd} = \Vector{u} + \Matrix{K}_{act}^{-1} inv\left( \frac{\partial \Vector{F}(\Vector{x}, \Vector{u})}{\partial \Vector{u}} \right) \dot{\pseudo}_{cmd}
\end{equation}

As shown, the inversion law requires the Jacobian of the pseudo control vector with respect to the inputs, $\frac{\partial \Vector{F}(\Vector{x}, \Vector{u})}{\partial \Vector{u}}$, to compute the commanded actuator inputs.

The function $\Vector{F}(\Vector{x}, \Vector{u})$ represents the relationship between the system state and input vector and the pseudo controls. This mapping was established earlier using the equations governing the propulsion forces and moments, detailed in \cref{subsec:propulsion}.

For the airship with four rotors indexed by $i = 1, 2, 3, 4$, the pseudo control vector can be expressed as a summation of the contributions from each rotor:
\begin{equation}
    \pseudo = \Vector{F}(\Vector{x}, \Vector{u}) = \begin{bmatrix}
        \sum_{i=1}^{4} -\sin{\gamma_i} \rho k_N \sigma_{dir,i} \Omega_i^2 - y^{RP_i} \cos{\gamma_i} \rho k_T \Omega_i^2 \\
        \sum_{i=1}^{4} \left(z^{RP_i} \sin{\gamma_i} + x^{RP_i} \cos{\gamma_i} \right) \rho k_T \Omega_i^2              \\
        \sum_{i=1}^{4} \cos{\gamma_i} \rho k_N \sigma_{dir,i} \Omega_i^2 - y^{RP_i} \sin{\gamma_i} \rho k_T \Omega_i^2  \\
        \sum_{i=1}^{4} sin{\gamma_i} \rho k_T \Omega_i^2                                                                \\
        \sum_{i=1}^{4} -\cos{\gamma_i} \rho k_T \Omega_i^2                                                              \\
    \end{bmatrix}
\end{equation}
To form the control effectiveness matrix, each component of the function $\Vector{F}$ must be differentiated with respect to each element of $\Vector{u}$.
The input vector $\Vector{u}$ consists of the RPMs and the tilt angles of the respective propellers:
\begin{equation}
    \Vector{u} = \left[ \Omega_1 \quad \Omega_2 \quad \Omega_3 \quad \Omega_4 \quad \gamma_1 \quad \gamma_2 \quad \gamma_3 \quad \gamma_4 \right]^T
\end{equation}
The general expressions for these derivatives are as follows:
\begin{align*}
    b_{\Omega_i,1} = \frac{\partial}{\partial \Omega_i}\left( -\sgamma \rho k_N \sigma_{dir,i} \Omega_i^2 - y^{RP_i} \cgamma \rho k_T \Omega_i^2 \right)               & = -2 \rho \left( \sgamma k_N \sigma_{dir,i} + \cgamma k_T y^{RP_i} \right) \Omega_i    \\
    b_{\gamma_i,1} = \frac{\partial}{\partial \gamma_i}\left( -\sgamma \rho k_N \sigma_{dir,i} \Omega_i^2 - y^{RP_i} \cgamma \rho k_T \Omega_i^2 \right)               & =    \rho \left( \sgamma k_T y^{RP_i} - \cgamma k_N \sigma_{dir,i} \right) \Omega_i^2  \\
    b_{\Omega_i,2} = \frac{\partial}{\partial \Omega_i} \left( \left(z^{RP_i} \sgamma + x^{RP_i} \cos{\gamma_i} \right) \rho k_T \Omega_i^2 \right)                    & =
    2 \rho k_T \left( z^{RP_i} \sgamma + x^{RP_i} \cgamma \right) \Omega_i                                                                                                                                                                                      \\
    b_{\gamma_i,2} = \frac{\partial}{\partial \gamma_i} \left( \left(z^{RP_i} \sgamma + x^{RP_i} \cos{\gamma_i} \right) \rho k_T \Omega_i^2 \right)                    & = \rho k_T \left( z^{RP_i} \cgamma - x^{RP_i} \sgamma \right) \Omega_i^2               \\
    b_{\Omega_i,3} = \frac{\partial}{\partial \Omega_i} \left( \cos{\gamma_i} \rho k_N \sigma_{dir,i} \Omega_i^2 - y^{RP_i} \sin{\gamma_i} \rho k_T \Omega_i^2 \right) & =
    2 \rho \left( \cgamma k_N \sigma_{dir,i} - \sgamma k_T y^{RP_i} \right) \Omega_i                                                                                                                                                                            \\
    b_{\gamma_i,3} = \frac{\partial}{\partial \gamma_i} \left( \cos{\gamma_i} \rho k_N \sigma_{dir,i} \Omega_i^2 - y^{RP_i} \sin{\gamma_i} \rho k_T \Omega_i^2 \right) & =    -\rho \left( \sgamma k_N \sigma_{dir,i} + \cgamma k_T y^{RP_i} \right) \Omega_i^2 \\
    b_{\Omega_i,4} = \frac{\partial}{\partial \Omega_i} \left( \sin{\gamma_i} \rho k_T \Omega_i^2 \right)                                                              & =
    2 \sgamma \rho k_T \Omega_i                                                                                                                                                                                                                                 \\
    b_{\gamma_i,4} = \frac{\partial}{\partial \gamma_i} \left( \sin{\gamma_i} \rho k_T \Omega_i^2 \right)                                                              & = \cgamma \rho k_T \Omega_i^2                                                          \\
    b_{\Omega_i,5} = \frac{\partial}{\partial \Omega_i} \left( -\cos{\gamma_i} \rho k_T \Omega_i^2 \right)                                                             & =
    -2 \cgamma \rho k_T \Omega_i                                                                                                                                                                                                                                \\
    b_{\gamma_i,5} = \frac{\partial}{\partial \Omega_i} \left( -\cos{\gamma_i} \rho k_T \Omega_i^2 \right)                                                             & = \sgamma \rho k_T \Omega_i^2
\end{align*}
Using the defined placeholder expressions for the partial derivatives, the control effectiveness matrix can now be assembled. This allows the linear relationship between the rate of change of the pseudo control vector$\pseudodot$ and the rate of change of the input vector $\udot$ to be written:

\begin{equation}
    \dot{\pseudo} = \frac{\partial \Vector{F}(\Vector{x}, \Vector{u})}{\partial \Vector{u}} \dot{\Vector{u}} = \underbrace{\begin{bmatrix}
            b_{\Omega_1,1} & ... & b_{\Omega_4,1} & b_{\gamma_1,1} & ... & b_{\gamma_4,1} \\
            b_{\Omega_1,2} & ... & b_{\Omega_4,2} & b_{\gamma_1,2} & ... & b_{\gamma_4,2} \\
            b_{\Omega_1,3} & ... & b_{\Omega_4,3} & b_{\gamma_1,3} & ... & b_{\gamma_4,3} \\
            b_{\Omega_1,4} & ... & b_{\Omega_4,4} & b_{\gamma_1,4} & ... & b_{\gamma_4,4} \\
            b_{\Omega_1,5} & ... & b_{\Omega_4,5} & b_{\gamma_1,5} & ... & b_{\gamma_4,5}
        \end{bmatrix}}_{\Matrix{B}} \begin{bmatrix}
        \dot{\Omega}_1 \\
        \dot{\Omega}_2 \\
        \dot{\Omega}_3 \\
        \dot{\Omega}_4 \\
        \dot{\gamma}_1 \\
        \dot{\gamma}_2 \\
        \dot{\gamma}_3 \\
        \dot{\gamma}_4 \\
    \end{bmatrix}
\end{equation}
This defines the mapping from the input vector time derivative to the rate of change of the pseudo controls, effectively yielding the control effectiveness matrix $\bm{B}$, providing the foundation for the control allocation algorithm.

To implement this matrix, a MATLAB script was created using the Symbolic Math Toolbox. This method allows all mathematical operations, including differentiation, to be carried out symbolically, improving accuracy compared to purely numerical approaches. Furthermore, the toolbox supports exporting symbolic expressions as MATLAB functions, enabling numerical evaluation. These functions can then be easily integrated the Simulink environment.

\subsection{Comparison of Control Allocation Methods}
A variety of control allocation algorithms are available, each offering different strengths and trade-offs. Some are capable of handling actuator constraints and nonlinearities in the mapping from pseudocontrols to actuator commands, while others prioritize ease of implementation and computational efficiency \cite{JOHANSEN20131087}. The following sections introduce a selection of these methods, including the one chosen for the airship application.
\subsubsection{Pseudoinverse}
If the control effectiveness matrix $\bm{B}$ is square and invertible, i.e. the number of physical inputs equals the number of pseudocontrols, a direct solution to the control allocation problem is simply:

\begin{equation}
    \udot = \bm{B}^{-1} \pseudodot
\end{equation}

However, in the airship case under consideration, the system is overactuated; that is, the number of physical inputs exceeds the number of pseudocontrols. This leads to an underdetermined system with infinitely many possible input solutions. A common strategy to resolve this is to select the input vector $\udot$ that minimizes the control effort, typically defined by the 2-norm:

\begin{equation}\label{eq:argmin}
    \argmin_{\udot} \udot^T \udot \quad \text{subject to} \ \pseudodot = \bm{B} \udot
\end{equation}

This optimization problem can be solved using Lagrange multipliers \cite{WANG20191718}, leading to the following cost function:

\begin{equation}
    H(\udot, \lambda) = \frac{1}{2} \udot^T \udot + \lambda^T (\pseudodot - \bm{B} \udot)
\end{equation}

To find the optimal solution, take the partial derivatives of $H$ with respect to both $\udot$ and the Lagrange multiplier $\lambda$:

\begin{equation}\label{eq:H_derivative}
    \begin{split}
        \frac{\partial H}{\partial \udot} & = \udot^T - \lambda^T \bm{B} = 0 \Rightarrow \udot^T = \lambda^T \bm{B}
    \end{split}
    \quad\quad
    \begin{split}
        \frac{\partial H}{\partial \lambda} & = \pseudodot - \bm{B} \udot = 0 \Rightarrow \pseudodot = \bm{B} \udot
    \end{split}
\end{equation}

Combining both results gives:

\begin{equation}
    \pseudodot = \bm{B} \udot = \bm{B} \bm{B}^T \lambda
\end{equation}

Assuming $\bm{B}$ has full rank, $\bm{B} \bm{B}^T$ is invertible, which allows to solve for $\lambda$:

\begin{equation}
    \lambda = \left( \bm{B} \bm{B}^T \right)^{-1} \pseudodot
\end{equation}

Substituting this back into the earlier expression for $\udot$ yields the final solution:

\begin{equation}\label{eq:pseudo_inverse}
    \udot = \bm{B}^T \left( \bm{B} \bm{B}^T \right)^{-1} \pseudodot
\end{equation}

This expression is known as the Moore-Penrose pseudoinverse, often denoted as $\bm{B}^{+}$. While this solution is easy to compute and guarantees the minimum control effort in terms of the input norm, it has a major drawback: it does not account for actuator constraints. As a result, the computed input vector $\udot$ may not be feasible for the actual system.

\subsubsection{Redistributed Pseudoinverse}
As previously mentioned, one key limitation of using the pseudoinverse for control allocation is its inability to account for actuator constraints. A method designed to address this shortcoming is the \gls{RPI}, also known as \gls{CGI}. The following offers a brief overview of the algorithm without delving into full technical detail.

The process begins by computing the standard pseudoinverse solution, as outlined in \cref{eq:pseudo_inverse}. If the resulting actuator command vector $\udot$ falls entirely within the specified actuator limits, the problem is solved directly. However, if any components of $\udot$ violate these limits, those values are clipped to their respective bounds and held constant for the remainder of the allocation process \cite{ERP}.

Next, a reduced control effectiveness matrix is formed by zeroing the columns associated with the now-saturated actuators. The pseudoinverse of this reduced matrix is then applied to the remaining control demand, after subtracting the contribution from the saturated actuators. This iterative procedure continues until all actuator constraints are satisfied or until the reduced pseudoinverse becomes undefined (e.g., due to a loss of rank) \cite{ERP}.

While this method improves on the naive pseudoinverse by enforcing actuator limits, it has a significant drawback: angle divergence \cite{doi:10.2514/6.2018-3480}. This occurs when the direction of the achieved pseudocontrol deviates from the desired direction. In practical terms, a command intended to generate pure roll might inadvertently result in a combined roll and pitch response, undermining the controller's effectiveness.

\subsubsection{RSPI}

The \gls{RSPI} is a variant of the RPI method, developed at TUM \cite{doi:10.2514/6.2018-3480}. It was specifically designed to address the angle divergence issue inherent in RPI. Rather than clipping individual actuator commands, RSPI scales the entire input vector $\udot$ uniformly to ensure the direction of the desired pseudocontrol is preserved \cite{doi:10.2514/6.2018-3480}.

Because it maintains the direction of the command, RSPI generally yields better results than RPI, particularly in applications where directional accuracy is critical. The tradeoff is a slightly increased computational complexity compared to RPI.

\subsubsection{ERP}
The \textit{\gls{ERP}} method, developed by Johannes Stephan and Walter Fichter \cite{ERP}, is a control allocation algorithm designed to ensure that the computed input vector remains within actuator limits. In contrast to other limit-respecting approaches like the previously introduced \gls{RPI}, the \gls{ERP} algorithm stands out by making full use of the available control authority and handling unachievable control demands in a direction-preserving way \cite{ERP}. Its core concept is similar to that of the \gls{RSPI} method, relying on scaling rather than simply clipping input components. The \gls{ERP} algorithm has also been applied to the airship studied in this thesis. A brief explanation of its working principle applied to the airship is presented here, while further details and derivations can be found in \cite{ERP}.

At first, the commanded rate of change of the pseudo control vector is split up into two components, namely the so called trim vector $\pseudodot_0$ and the acceleration command $\pseudodot_\Delta$
\begin{equation}
    \pseudodot_{cmd} = \pseudodot_0 + \pseudodot_\Delta
\end{equation}
with the requirement that $\pseudodot_0$ must always be applied fully, meaning it must lie entirely within the allowable actuator range. For the airship, this trim command is zero.

Furthermore, the acceleration command $\pseudodot_\Delta$ is scaled by a factor $c$ to ensure that actuator limits are not exceeded. The resulting pseudo control rate of change vector is then expressed as:
\begin{equation}
    \label{eq:nu0_and_nu_delta}
    \pseudodot = \pseudodot_0 + c \pseudodot_\Delta
\end{equation}
The objective is to determine the largest possible value of $c$ and the corresponding actuator input rate of change vector $\dot{\Vector{u}}$, such that all elements of $\dot{\Vector{u}}$ remain within their respective limits. The value of $c$ is therefore constrained between 0 and 1, 0 representing the case where all actuators are already at their limit and no additionaly command can be fulfilled and 1 representing the case where the full demand can be fulfilled.

The algorithm proceeds iteratively. In each step, $c$ is increased until one or more actuator commands reach their saturation limits. These saturated actuators are then excluded from the solution by zeroing out their corresponding columns in the control effectiveness matrix, which is then referred to as the reduced matrix $\Matrix{B}_\epsilon$. Once removed, these actuators no longer influence the allocation process.

An initial solution is found by applying the pseudoinverse of $\bm{B}$ to the trim vector $\pseudodot_0$, resulting in the input vector $\dot{\Vector{u}}_0$. Since $\dot{\Vector{u}}_0$ is fully contained within the actuator limits, this input vector is guaranteed to remain within bounds:

\begin{equation}\label{eq:u0}
    \udot(0) = \udot_0 = \bm{B^+} \pseudodot_0
\end{equation}

Following this, the solution is refined through a recursive process. In each iteration, the additional input resulting from the scaled acceleration command $\pseudodot_\Delta$ is calculated as:

\begin{equation}\label{eq:delta_u}
    \udot_{\Delta}(k) = \bm{B_\epsilon^+}(k) \pseudodot_\Delta
\end{equation}

Here, $k$ denotes the current iteration index, and $\bm{B_\epsilon^+}(k)$ represents the pseudoinverse of the reduced control effectiveness matrix at step $k$. Initially, this matrix is identical to the original $\bm{B}$, since no actuators have been saturated or excluded before the first iteration.

The additional input $\bm{u_{\Delta}}(k)$ cannot simply be added to the previous solution, as doing so may result in actuator saturation. To prevent this, a scaling factor is introduced, yielding the following update rule:

\begin{equation}
    \label{eq:u(k)}
    \udot(k) = \udot(k-1) + d(k) \udot_{\Delta}(k)
\end{equation}

Here, $d(k)$ is a scaling factor computed at each iteration to ensure that actuator limits are not violated. Determining $d(k)$ requires distinguishing between two cases, to guarantee that the accumulated scaling factor $c$ always satisfies $c \leq 1$. The two cases are defined as:

\begin{equation}\label{eq:d(k)}
    d(k) =
    \begin{dcases}
        1 - c(k-1), \quad & \text{if } d_{max} > 1 - c(k-1) \\
        d_{max},          & \text{else}
    \end{dcases}
\end{equation}

Here, $d_{max}$ is the maximum allowable scaling before any of the remaining unsaturated actuators reach their limit during iteration $k$.

The first case ensures that $c$ does not exceed one, even if $d_{max}$ is too large. The second case uses the maximum feasible $d_{max}$ to increase $c$ as much as possible without violating actuator constraints.

After computing $d(k)$, the cumulative scaling factor $c$ introduced in \cref{eq:nu0_and_nu_delta} is updated by summing all previously applied increments:
\begin{equation}\label{eq:update_c}
    c(k) = \sum_{j=1}^{k} d(j)
\end{equation}

To determine the value of $d(k)$, the maximum allowable scaling factor $d_{max}$ must first be computed, which requires taking into account the actuator constraints. Each actuator is subject to a minimum limit $\underline{\dot{u}}_i$ and a maximum limit $\overline{\dot{u}}_i$.

By combining these bounds with the expression for the input at iteration $k$ from \cref{eq:u(k)}, the following condition must be satisfied:
\begin{equation}\label{eq:d_derivation}
    \begin{aligned}
        \underline{\dot{u}}_i                  & \leq \dot{u}_i(k) \leq \overline{\dot{u}}_i                                \\
        \underline{\dot{u}}_i                  & \leq \dot{u}_i(k-1) + d(k) \dot{u}_{\Delta,i}(k) \leq \overline{\dot{u}}_i \\
        \underline{\dot{u}}_i - \dot{u}_i(k-1) & \leq d(k) u_{\Delta,i}(k) \leq \overline{\dot{u}}_i - u_i(k-1).
    \end{aligned}
\end{equation}

The bounds on $d(k)$ depend on the sign of the corresponding entry in $\udot(k)$. Three scenarios must be considered: whether the value is positive, negative, or zero.

From this, the individual upper bound $\overline{d}_i(k)$ for each actuator input can be derived as:

\begin{equation}\label{eq:d_bar}
    \overline{d}_i(k) =
    \begin{dcases}
        (\overline{\dot{u}}_i - \dot{u}_i(k-1)) / \dot{u}_{\Delta,i}(k), \quad & \text{if } \dot{u}_{\Delta,i}(k) > 0 \\
        (\underline{\dot{u}}_i - \dot{u}_i(k-1)) / \dot{u}_{\Delta,i}(k),      & \text{if } \dot{u}_{\Delta,i}(k) < 0 \\
        \text{not defined},                                                    & \text{else}.
    \end{dcases}
\end{equation}
In this context, $\overline{\dot{u}}_i$ and $\underline{\dot{u}}_i$ denote the upper and lower rate limits of the i-th actuator, respectively. To derive these bounds, two sources must be taken into account:

First, the actual rate limitations of the actuator. These are given for both the propellers and the tilt actuators in \cref{tab:actuator_limits}.

Second, since the allocation is performed on the rate level, a phase-plane-based approach is employed to ensure that the absolute actuator limits are also respected. The corresponding absolute constraints are likewise provided in \cref{tab:actuator_limits}.

Both contributions are combined to define the final rate bounds used in the control allocation process:
\begin{align}
    \overline{\dot{\Omega}_i}   & = \min \left[  \dot{\Omega}_{max}, \omega_{\Omega_i} \left(\Omega_{max} - \Omega_i\right) \right]  \\
    \underline{ \dot{\Omega}_i} & = \max \left[  \dot{\Omega}_{min}, \omega_{\Omega_i} \left(\Omega_{min} - \Omega_i\right) \right]  \\
    \overline{ \dot{\gamma}_i}  & = \min \left[  \dot{\gamma}_{max},  \omega_{\gamma_i} \left(\gamma_{max} - \gamma_i\right) \right] \\
    \underline{ \dot{\gamma}_i} & = \max \left[  \dot{\gamma}_{min},  \omega_{\gamma_i} \left(\gamma_{min} - \gamma_i\right) \right]
\end{align}
Here, $\dot{\Omega}_{max}$ and $\dot{\gamma}_{max}$ denote the upper rate limits, while $\dot{\Omega}_{min}$ and $\dot{\gamma}_{min}$ are the corresponding lower rate limits. Similarly, $\Omega_{max}$, $\gamma_{max}$, $\Omega_{min}$, and $\gamma_{min}$ define the absolute upper and lower bounds for the actuator values. The parameters $\omega_{\Omega_i}$ and $\omega_{\gamma_i}$ represent the natural frequencies or bandwidths of the respective actuators, as introduced in \cref{chap:system_description}.

These bounds are then used to identify the smallest $\overline{d}_i(k)$ across all actuators, which defines $d_{max}$ for the current iteration.

Finally, among the actuators still participating in the allocation, the one associated with the smallest value of $\overline{d}_i(k)$ becomes the limiting factor for that iteration. This value defines the maximum allowable scaling factor $d_{max}(k)$ at step $k$.

The limiting actuator is then considered saturated and removed from further optimization by modifying the reduced control effectiveness matrix $\bm{B_\epsilon}$. Specifically, the column corresponding to this actuator is set to zero, ensuring it no longer contributes to future solutions. This update prepares the algorithm for the next iteration.

Before proceeding to the next step, a termination condition must be checked. The algorithm continues iterating until one of the following criteria is met:
\begin{enumerate}
    \item $c(k) = 1$: The full control demand $\udot_{cmd}$ has been successfully allocated.
    \item $\text{rank}(\bm{B_\epsilon}(k)) < m$: The reduced control effectiveness matrix becomes rank deficient, where $m$ is the number of pseudo controls.
\end{enumerate}

In the first case, the allocation is complete and successful. In the second, allocation is no longer possible because the pseudo-inverse $\bm{B_\epsilon}^+(k)$ cannot be computed due to insufficient rank.

Since each iteration removes one actuator from the solution space, the algorithm is guaranteed to terminate after at most $p - m + 1$ steps, where $p$ is the total number of actuators. However, it may terminate earlier if linear dependence among the remaining actuators causes an earlier rank drop.

To demonstrate that the outcome of the algorithm matches the intended form defined in \cref{eq:nu0_and_nu_delta}, consider expressing the achieved pseudo control as a function of the computed actuator input:

\begin{equation}
    \pseudodot_{achieved} = \bm{B} \udot(k)
\end{equation}

This expression can be expanded by recursively substituting the update rule from \cref{eq:u(k)}:
\begin{align}
    \pseudodot_{achieved} & = \bm{B} \left( \udot(k-1) + d(k) \udot_\Delta(k) \right)                                                       \\
    \pseudodot_{achieved} & = \bm{B} \left( \udot(k-2) + d(k-1) \udot_\Delta(k-1) + d(k) \udot_\Delta(k) \right)                            \\
    \pseudodot_{achieved} & = \bm{B} \left( \udot(k-3) + d(k-2) \udot_\Delta(k-2) + d(k-1) \udot_\Delta(k-1) + d(k) \udot_\Delta(k) \right) \\
                          & \vdots                                                                                                          \\
    \pseudodot_{achieved} & = \bm{B} \left( \udot(0) + d(1) \udot_\Delta(1) + ... + d(k) \udot_\Delta(k) \right)
\end{align}

The term $\bm{u}_\Delta(i)$ can be replaced by \cref{eq:delta_u} and the bracket can be expanded:
\begin{align}
    \pseudodot_{achieved} & = \bm{B} \left( \bm{u}(0) + d(1) \bm{B}_\epsilon(1) \pseudodot_\Delta + ... + d(k) \bm{B}_\epsilon(k) \pseudodot_\Delta \right)                                    \\
    \pseudodot_{achieved} & = \bm{B} \bm{u}(0) + d(1) \underbrace{\bm{B} \bm{B}_\epsilon(1)}_{=I} \pseudodot_\Delta + ... + d(k) \underbrace{\bm{B} \bm{B}_\epsilon(k)}_{=I} \pseudodot_\Delta \\
    \pseudodot_{achieved} & = \bm{B} \bm{u}(0) + \left(d(1) + ... + d(k) \right) \pseudodot_\Delta
\end{align}

Finally, to obtain the desired result, apply \cref{eq:u0} and \cref{eq:update_c}:
\begin{align}
    \pseudodot_{achieved} & = \underbrace{\bm{B} \bm{u}(0)}_{\pseudodot_0} + \underbrace{\left(d(1) + ... + d(k) \right)}_{c} \pseudodot_\Delta \\
    \pseudodot_{achieved} & = \bm{\nu_0} + c \pseudodot_\Delta
\end{align}
Therefore, it is confirmed that the actual result matches the desired result.

To conclude this section, in \cref{fig:erp_flow} the \gls{ERP} algorithm is depicted in a flow diagram.

\begin{figure}
    \centering

    \begin{tikzpicture}[node distance=2cm]

        \tikzstyle{startstop} = [rectangle, rounded corners, minimum width=3cm, minimum height=1cm,text centered, draw=black, fill=red!30]
        \tikzstyle{process} = [rectangle, minimum width=3cm, minimum height=1cm, text centered, draw=black, fill=orange!30]
        \tikzstyle{decision} = [diamond, aspect=2, minimum width=3cm, minimum height=1cm, text centered, draw=black, fill=green!30]
        \tikzstyle{arrow} = [thick,->,>=stealth]

        \node (start) [startstop] {Start};
        \node (init) [process, below of=start] {Initialize: \( k=1,\ c(0)=0,\ \bm{B}_\epsilon(1)=\bm{B} \)};
        \node (base) [process, below of=init] {Base step: \( \udot(0) \) using \( \bm{B}^+ \) and \( \pseudodot_0 \) \small{(\cref{eq:u0})}};
        \node (dec1) [decision, below of=base, yshift=-2cm,align=center] {\( c(k) < 1 \) and \\ \( \text{rank}(\bm{B}_\epsilon(k)) = \text{Number of pseudocontrols} \)};
        \node (proc1) [process, below of=dec1, yshift=-2cm,align=center] {Calculate: \( \bm{B}^+(k), \udot_\Delta(k), \overline{d}_i(k), d(k) \)\\ \small{(\cref{eq:delta_u}, \cref{eq:d_bar}, \cref{eq:d(k)})}};
        \node (update) [process, below of=proc1] {Update \( \udot(k), c(k) \) \small{(\cref{eq:u(k)}, \cref{eq:update_c})}};
        \node (incr) [process, below of=update] {Increment \( k \)};
        \node (updN) [process, below of=incr] {Update \( \bm{B}(k) \) by setting column of saturated actuator to zero};
        \node (end) [startstop, below of=updN] {End, \( \udot = \udot(k) \)};

        \draw [arrow] (start) -- (init);
        \draw [arrow] (init) -- (base);
        \draw [arrow] (base) -- (dec1);
        \draw [arrow] (dec1) -- node[anchor=east] {yes} (proc1);
        \draw [arrow] (proc1) -- (update);
        \draw [arrow] (update) -- (incr);
        \draw [arrow] (incr) -- (updN);
        \draw [arrow] (updN) -- ++(-6,0) |- (dec1);
        \draw [arrow] (dec1.east) -- ++(2,0) |- node[near start,right] {no} (end); 
    \end{tikzpicture}
    \caption{Flow diagram of the \gls{ERP} algorithm (adapted from pseudocode from \cite{ERP})}
    \label{fig:erp_flow}
\end{figure}
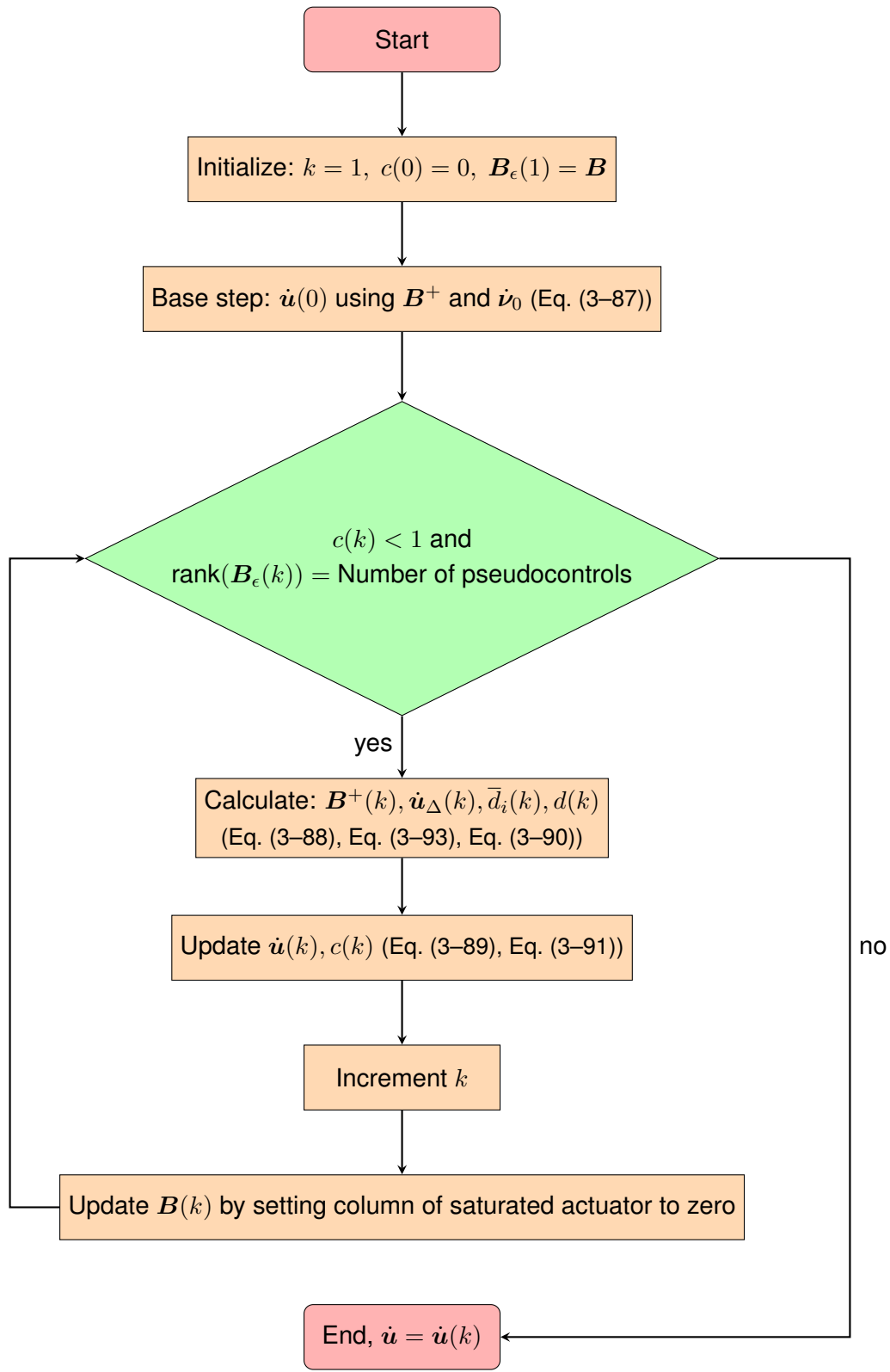

\subsection{Control Surface Allocation}\label{subsec:surface_allocation}
For the control surface allocation, the \gls{ERP} algorithm is also used. The relationship between the control surface deflections and the produced moments was derived in \cref{subsec:control_surfaces}, specifically in \cref{eq:b_matrix_surf}. It is repeated here for convenience:
\begin{equation}
    \begin{bmatrix}
        L_{F} \\
        M_{F} \\
        N_{F}
    \end{bmatrix} = \underbrace{ \frac{\rho}{2} V^2 S_{ref} l_{ref} \begin{bmatrix}
            c_{l \eta}                 & c_{l \eta}                 & c_{l \eta}                 \\
            \cos{\varphi_1} c_{m \eta} & \cos{\varphi_2} c_{m \eta} & 0                          \\
            \sin{\varphi_1} c_{m \eta} & \sin{\varphi_2} c_{m \eta} & \sin{\varphi_3} c_{m \eta}
        \end{bmatrix}}_{\Bsurf} \underbrace{\begin{bmatrix}
            \eta_1 \\
            \eta_2 \\
            \eta_3
        \end{bmatrix}}_{\usurf}
\end{equation}
As seen above, the surface control effectiveness matrix $\Bsurf \in \mathrm{R}^{3 \times 3}$ is square, indicating that the system is not overactuated. Therefore, a unique solution exists for any feasible moment command.
In this case, applying the \gls{ERP} algorithm becomes straightforward and involves only a single allocation step.
Moreover, since $\Bsurf$ depends only on the dynamic pressure and is otherwise constant, its inverse can be computed in advance. This significantly simplifies the allocation process. The inverse of $\Bsurf$ is given by:
\begin{equation}
    \Bsurf^{-1} = \frac{1}{\frac{1}{2} V^2 S_{ref} l_{ref}} \begin{bmatrix}
        c_{l \eta}                 & c_{l \eta}                 & c_{l \eta}                 \\
        \cos{\varphi_1} c_{m \eta} & \cos{\varphi_2} c_{m \eta} & 0                          \\
        \sin{\varphi_1} c_{m \eta} & \sin{\varphi_2} c_{m \eta} & \sin{\varphi_3} c_{m \eta}
    \end{bmatrix}^{-1}
\end{equation}
The matrix within the equation is constant, and its inverse can be calculated offline. During runtime, it only needs to be scaled by the inverse of the dynamic pressure. The allocation itself is then described as:
\begin{align}
    \Vector{u}_{\Delta, surf} & = \Bsurf^{-1} \Vector{M}_{F,cmd}                                                 \\
    \usurf                    & = d \Vector{u}_{\Delta, surf} = d \Bsurf^{-1} \Vector{M}_{F,cmd}\label{eq:usurf}
\end{align}
Here, $d$ is again the scalar scale factor that ensures the computed input remains within actuator limits.
This factor is determined similarly to how it's calculated in the rotor allocation.
First, for each control surface, an individual scaling factor $\overline{d}_i$ is computed:
\begin{equation}
    \overline{d}_i =
    \begin{dcases}
        \overline{u}_{i,surf} / u_{\Delta,i,surf}, \quad & \text{if } u_{\Delta,i,surf} > 0 \\
        \underline{u}_{i,surf} / u_{\Delta,i,surf},      & \text{if } u_{\Delta,i,surf} < 0 \\
        \text{not defined},                              & \text{else}.
    \end{dcases}
\end{equation}
Note that the iteration index $k$ is omitted here, as only a single step is required.
The limits for the control surface deflections are set to \ang{\pm 35}. While the physical actuator limits are given as \ang{\pm 40} in \cref{tab:actuator_limits}, a more conservative bound is used. As illustrated in \cref{fig:control_surfaces_coefficients}, the effectiveness of the control surfaces begins to decline beyond \ang{\pm 35}. To avoid operating in this less effective region, the maximum allowable deflection is intentionally restricted to \ang{\pm 35}.

Among the computed $\overline{d}_i$, the smallest is selected and defined as $d_{max} = \min(\overline{d}_1, \overline{d}_2, \overline{d}_3)$.

Depending on the value of $d_{max}$, two cases are distinguished to decide the final scale factor $d$. If $d_{max} > 1$, this indicates that no actuator is saturated, and the solution of the inverse can be used directly by setting $d = 1$. Otherwise, $d_{max}$ is used as the scale factor:
\begin{equation}
    d =
    \begin{dcases}
        1, \quad & \text{if } d_{max} > 1 \\
        d_{max}, & \text{else}
    \end{dcases}
\end{equation}
Once $d$ is determined, the control surface input vector $\usurf$ can be computed using \cref{eq:usurf}.
Specifically, the desired moment from \cref{eq:pseudo_cmd} $\Vector{M}_{des}$ is used as the control surface deflections.
With the control surface deflections computed, the real, achieved control surface moment is calculated with the allocated control surface deflections:
\begin{equation}
    \Vector{M}_F = \Bsurf \usurf
\end{equation}
This moment $\Vector{M}_F$ reduces the moment demand allocated to the propulsion system, as outlined in \cref{eq:pseudo_cmd}. Since the effectiveness of the control surfaces is directly proportional to the dynamic pressure, their influence becomes negligible during hover or near-hover conditions, where airspeed and thus dynamic pressure is very low.
For this reason, the control surface allocation is only activated when the airspeed exceeds \qty{3}{\meter\per\second}.
This logic is depicted in \cref{fig:surface_allocation_diagram}.
\begin{figure}
    \centering
    \includegraphics{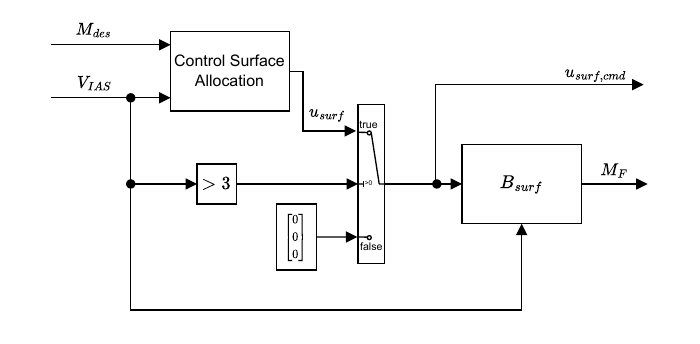}
    \caption{Block diagram of the surface allocation}
    \label{fig:surface_allocation_diagram}
\end{figure}

\subsection{Nullspace Transition}
Due to the incremental nature of the \gls{INDI} control law, which commands actuator increments rather than absolute positions, a phenomenon known as path dependency arises \cite{rupprecht_indi_2024, zhang_modeling_2019}. This becomes problematic when actuators reach their physical limits: further increments cannot be executed, and instead, the actuators must be explicitly “pulled away” from the limit.

Moreover, once a maneuver concludes and steady-state flight is achieved, the desired forces and moments remain constant. Hence, the commanded actuator increments approach zero, as the pseudo control rate vector is zero. Consequently, actuators remain fixed at their current positions, which may be suboptimal. For instance, the front propellers might remain fully tilted forward while the rear propellers are angled backward, leading to unnecessarily high power consumption.

To address this, a nullspace transition strategy is introduced. Since the airship is an overactuated system, i.e. it has more actuators than elements in the pseudo control vector, a nullspace of the control effectiveness matrix $\Matrix{B}$ exists.
The nullspace of a linear mapping are all vectors that are mapped to the zero vector.
This means, if there exists a input rate vector $\dot{\Vector{u}}_{ns}$ that is in the nullspace of $\Matrix{B}$, it has no effect on the pseudo control rate vector $\pseudodot$.
This allows for the fulfillment of secondary objectives, such as actuator position optimization, in addition to the primary control allocation task.

The objective for the airship is to pull the actuators toward predefined optimum actuator positions. For instance, in hover flight, these positions might maximize the attainable moment set, while during cruise, the optimal configuration could involve all propellers pointing straight forward by setting the tilt angle accordingly.

It is important to note that although the term optimum is used here, this does not necessarily imply maximization of a formal performance criterion. The choice of actuator positions may be heuristic or based on empirical reasoning.
In particular, the desired tilt angles during hover are set to \ang{45} for the front propellers and \ang{-45} for the aft propellers, as this maximizes the attainable moment set as already investigated in \cite{schimmer}.
For high speed flight above \qty{5}{\meter\per\second}, the tilts are set to \ang{85}. Between hover and \qty{5}{\meter\per\second} a linear blending is used.
For the RPMs on the other hand, $0$ is used as the desired value in order to pull the rotors away from their upper limit and to ensure lower power consumption by reducing the RPMs.

To guide the actuators toward these desired positions, a proportional controller can be used to compute a desired input rate:
\begin{equation}
    \Vector{\dot{u}}_{opt,des} = \Matrix{K}_{act,opt} \left( \Vector{u}_{opt} - \Vector{u} \right)
\end{equation}
Here, $\Vector{u}$ denotes the current actuator positions, $\Vector{u}_{opt}$ he target positions and $\Matrix{K}_{act,opt}$ a diagonal gain matrix.

However, this desired input rate generally lies outside the nullspace of the control effectiveness matrix and would therefore interfere with the primary control objective. To preserve control performance, the desired input rate must be projected onto the nullspace of the control effectiveness matrix $\Matrix{B}$.
In other words: the goal is to find the input rate vector $\udot_{opt,ns}$ that is in the nullspace of $\Matrix{B}$ and minimizes the distance to the desired input rate vector $\Vector{\dot{u}}_{opt,des}$.
This can be expressed as an optimization problem:
\begin{equation}
    \argmin_{\udot_{opt,ns}} \lVert \udot_{opt,ns} - \udot_{opt,des} \rVert^2_2, \quad \text{subject to } \Matrix{B} \udot_{opt,ns} = \Matrix{0}
\end{equation}
where $\lVert . \rVert_2$ denotes the 2-norm of a vector and $\Matrix{B} \udot_{opt,ns} = \Matrix{0}$ denotes the necessary condition that the $\udot_{opt,ns}$ is in the nullspace, i.e. is mapped to the zero vector.

The optimization problem is reformulated in terms of Lagragian multipliers:
\begin{align}
    H(\udot_{opt,ns}, \Vector{\lambda}) & = \lVert \udot_{opt,ns} - \udot_{opt,des} \rVert^2_2 + \Vector{\lambda}^T \Matrix{B} \udot_{opt,ns}                      \\
    H(\udot_{opt,ns}, \Vector{\lambda}) & = (\udot_{opt,ns} - \udot_{opt,des})^T (\udot_{opt,ns} - \udot_{opt,des}) + \Vector{\lambda}^T \Matrix{B} \udot_{opt,ns}
\end{align}
In order to solve the optimization problem, the partial derivatives with respect to $\udot_{opt,ns}$ and $\Vector{\lambda}$ are determined:
\begin{align}
    \pdv{H(\udot_{opt,ns}, \Vector{\lambda})}{\udot_{opt,ns}}   & = 2 \udot_{opt,ns}^T - 2 \udot_{opt,des}^T + \Vector{\lambda}^T \Matrix{B} \overset{!}{=} 0 \label{eq:nullspace_opt1}              \\
    \pdv{H(\udot_{opt,ns}, \Vector{\lambda})}{\Vector{\lambda}} & = \Matrix{B} \udot_{opt,ns} \overset{!}{=} 0                                                             \label{eq:nullspace_opt2}
\end{align}
The first equation can be solved for $\udot_{opt,ns}$:
\begin{align}
    \udot_{opt,ns}^T & = \udot_{opt,des}^T - \frac{1}{2} \Vector{\lambda}^T \Matrix{B}                        \\
    \udot_{opt,ns}   & = \udot_{opt,des} - \frac{1}{2} \Matrix{B}^T \Vector{\lambda} \label{eq:udot_lagrange}
\end{align}
and the result can be inserted into \cref{eq:nullspace_opt2}:
\begin{align}
    \Matrix{B} \udot_{opt,ns}                                                         & \overset{!}{=} 0                                               \\
    \Matrix{B} (\udot_{opt,des} - \frac{1}{2} \Matrix{B}^T \Vector{\lambda} )         & = 0                                                            \\
    \Matrix{B} \udot_{opt,des} - \frac{1}{2} \Matrix{B} \Matrix{B}^T \Vector{\lambda} & = 0                                                            \\
    \Vector{\lambda}                                                                  & = 2  (\Matrix{B} \Matrix{B}^T)^{-1} \Matrix{B} \udot_{opt,des}
\end{align}
The expression for $\Vector{\lambda}$ can be inserted into \cref{eq:udot_lagrange}:
\begin{align}
    \udot_{opt,ns} & = \udot_{opt,des} - \frac{1}{2} \Matrix{B}^T \Vector{\lambda} \nonumber                                             \\
    \udot_{opt,ns} & = \udot_{opt,des} - \Matrix{B}^T ((\Matrix{B} \Matrix{B}^T)^{-1} \Matrix{B} \udot_{opt,des})                        \\
    \udot_{opt,ns} & = \udot_{opt,des} - \Matrix{B}^T(\Matrix{B} \Matrix{B}^T)^{-1} \Matrix{B} \udot_{opt,des}                           \\
    \udot_{opt,ns} & = (\Matrix{I} - \underbrace{\Matrix{B}^T(\Matrix{B} \Matrix{B}^T)^{-1}}_{\Matrix{B}^{+}} \Matrix{B})\udot_{opt,des}
\end{align}
where $\Matrix{I}$ is the identity matrix and $\Matrix{B}^{+}$ the Moore-Penrose pseudo inverse of the control effectiveness matrix.
It can be shown that this projected input rate $\dot{\Vector{u}}_{opt,ns}$ does not affect the achieved pseudo control rate $\pseudodot$, thereby solving the primary control task while enabling secondary objectives like actuator optimization:
\begin{align}
    \pseudodot & = \Matrix{B} \dot{\Vector{u}}_{total}                                                                                                                                       \\
    \pseudodot & = \Matrix{B} ( \dot{\Vector{u}}_{ERP} + \Vector{\dot{u}}_{opt,ns} )                                                                                                         \\
    \pseudodot & = \Matrix{B} ( \dot{\Vector{u}}_{ERP} + (I - \Matrix{B}^T(\Matrix{B} \Matrix{B}^T)^{-1} \Matrix{B})\udot_{opt,des} )                                                        \\
    \pseudodot & = \Matrix{B} \dot{\Vector{u}}_{ERP} + (\Matrix{B} - \underbrace{\Matrix{B} \Matrix{B}^T(\Matrix{B} \Matrix{B}^T)^{-1}}_{=\Matrix{I}} \Matrix{B}) \Vector{\dot{u}}_{opt,des} \\
    \pseudodot & = \Matrix{B} \dot{\Vector{u}}_{ERP} + (\Matrix{B} - \Matrix{B}) \Vector{\dot{u}}_{opt,des}                                                                                  \\
    \pseudodot & = \Matrix{B} \dot{\Vector{u}}_{ERP}
\end{align}

The effect of the nullspace transition is illustrated with a simple test case.
In \cref{fig:plant_without_ns,fig:actuators_without_ns,fig:plant_with_ns,fig:actuators_with_ns} the results of this test case can be seen.
For both cases, with and without nullspace transition, first a \qty{9}{\meter\per\second} step command in the forward velocity is commanded. After \qty{30}{\second}, the step command is removed and the airship is commanded into hover.
\cref{fig:plant_without_ns} shows the plant response without the nullspace transition, \cref{fig:plant_with_ns} on the other hand shows the plant reaction with the nullspace transition enabled. Both responses are practically indistinguisable from each other, proofing that the nullspace transition does indeed not interfere with the primary control task.

In \cref{fig:actuators_without_ns}, the actuator response without the nullspace transition is shown, while \cref{fig:actuators_with_ns} shows the actuator position with the nullspace transition enabled.
When the nullspace transition is enabled, during cruise all four propellers have the same tilt angle of around \ang{65}. Without nullspace transition, the front propellers are tilted around \ang{75}, while the aft propellers are only tilted around \ang{45}. The desired tilt angles are not reached for the case with nullspace transition, because some vertical force is necessary to compensate for the part of the airship's weight that is not covered by the buoyancy force.

After the airship has come to a stop and reaches hover, the influence of the nullspace transition becomes even more evident as for the case with nullspace transition, the propellers have the desired tilt angles of \ang{\pm 45} while for the case without nullspace transition, the front propellers are stuck at around \ang{75} while the aft propellers stay at around \ang{-60}.
Additonally, for the case with nullspace transition all four rotor speeds reach their trim value earlier, while for the case without nullspace transition, they converge only slowly.

\begin{figure}[H]
    \centering
    \includegraphics{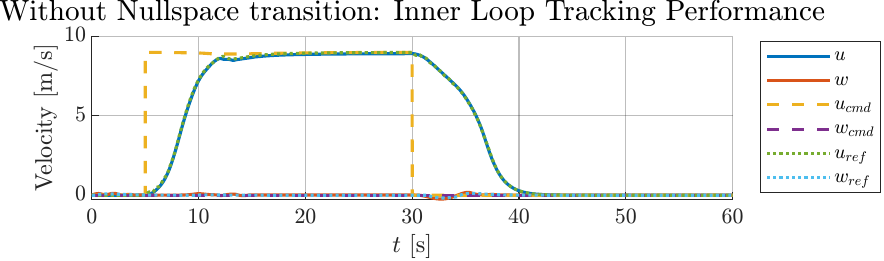}
    \caption{Plant response without nullspace transition}
    \label{fig:plant_without_ns}
\end{figure}

\begin{figure}[H]
    \centering
    \includegraphics{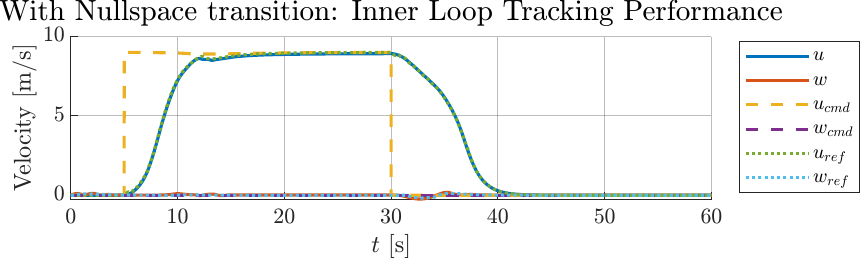}
    \caption{Plant response with nullspace transition}
    \label{fig:plant_with_ns}
\end{figure}

\begin{figure}[H]
    \centering
    \includegraphics{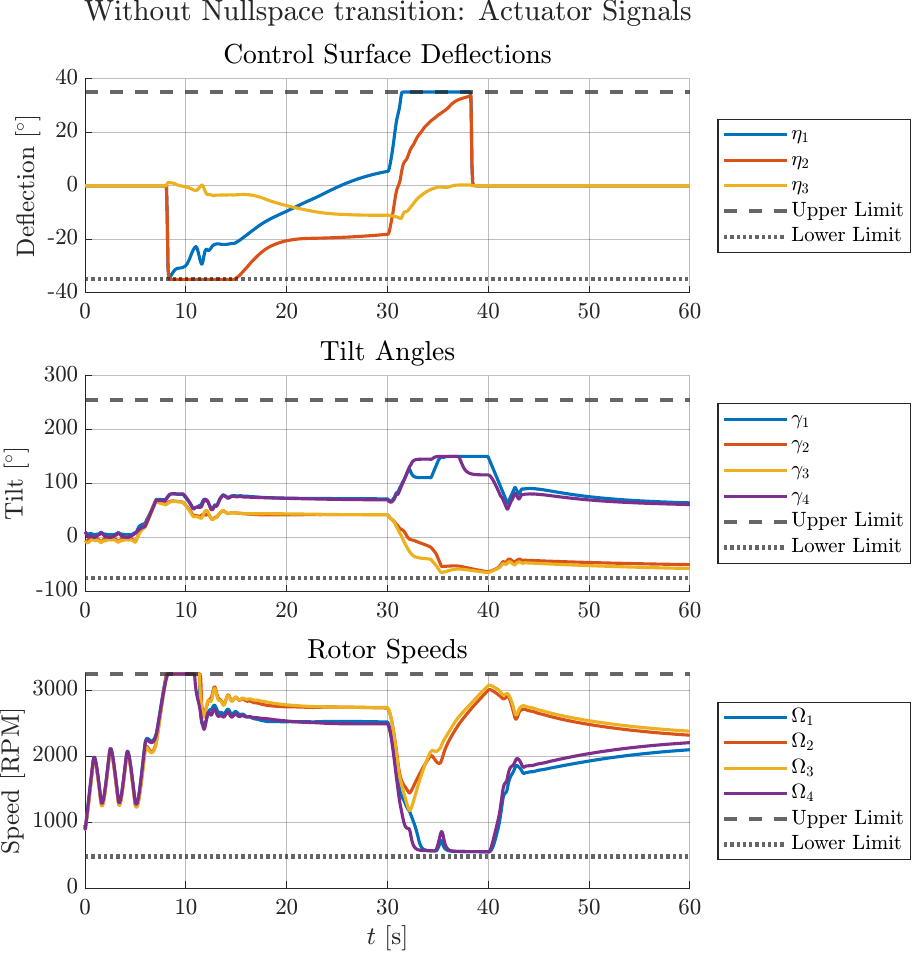}
    \caption{Actuator positions without nullspace transition}
    \label{fig:actuators_without_ns}
\end{figure}

\begin{figure}[H]
    \centering
    \includegraphics{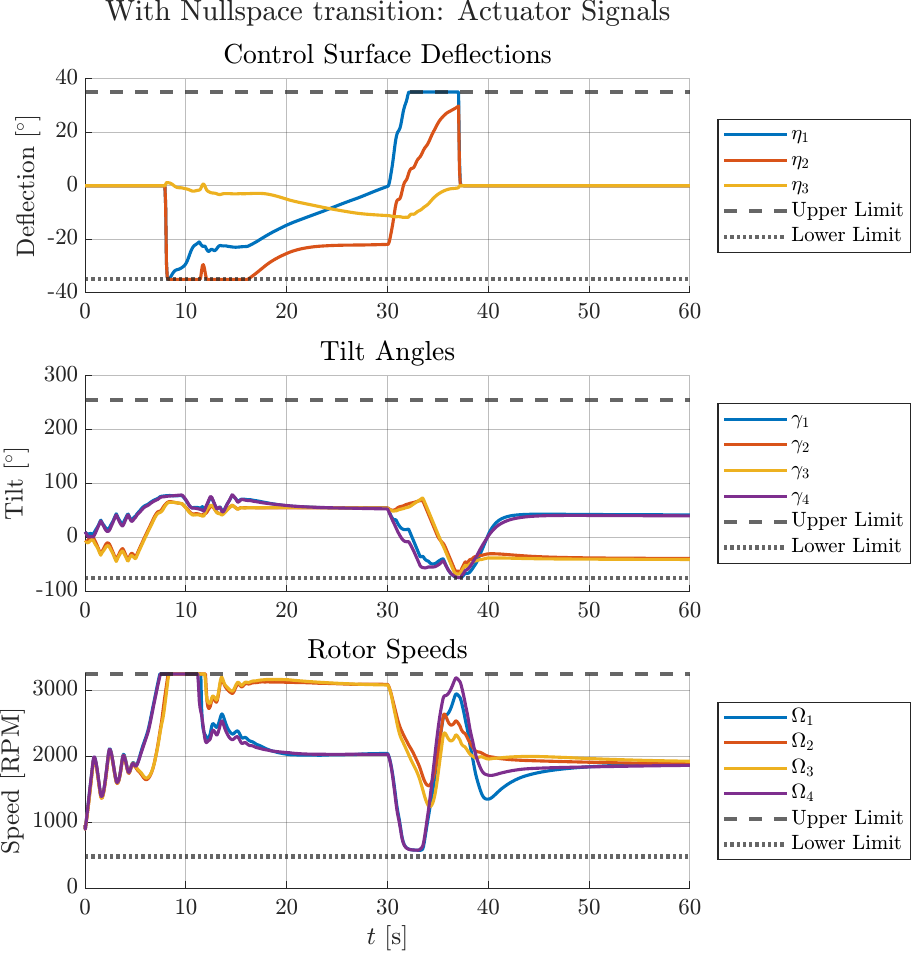}
    \caption{Actuator positions with nullspace transition}
    \label{fig:actuators_with_ns}
\end{figure}
  \chapter{Simulation Results}
In this chapter, the performance of the proposed controller is evaluated using several test harnesses implemented in Simulink. These test scenarios are designed not only to demonstrate the capabilities of the controller presented in this thesis, but also to perform a direct comparison with another controller that has been developed at the institute.

The structure of this chapter is as follows: first, the key working principles of the other control strategy are outlined to provide necessary context. This is followed by a comparative evaluation of both controllers across four representative test cases. Each test case is designed to examine the controllers under different operational conditions and control demands. For every case and controller, a comprehensive presentation of the simulation results is given.

\section{Description of the Second Controller}
For the remainder of this chapter, the term \emph{controller 1} refers to the control strategy developed and presented in this thesis, while \emph{controller 2} designates the baseline controller used for comparison, which was developed independently at the institute.

The primary conceptual difference between controller 1 and controller 2 lies in the way the command for the rate of change of the pseudocontrol vector, $\pseudodot$, is generated. In controller 1, $\pseudodot_{cmd}$ is derived by applying a first-order reference model to the pseudocontrol vector $\pseudo$, as outlined in \cref{subsec:pseudo_ref_model}. In contrast, controller 2 bypasses this reference model entirely and instead computes $\pseudodot_{cmd}$ directly through dynamic inversion of the system's jerk-level equations of motion.

This means that while controller 1 performs dynamic inversion at the acceleration level (i.e., forces and moments), controller 2 performs inversion one dynamic level lower, at the jerk level (i.e., the rate of change of forces and moments). To do this, the equations of motion must be differentiated with respect to time to express $\pseudodot$ explicitly in terms of the system state and inputs.
Another key difference is that the inner loop of controller 2 doesn't control the body-fixed velocities directly.
Instead, it controls the body-fixed specific forces, and therefore acts on the acceleration level.
Considering the rotational dynamics, also the angular velocity is controlled in the inner loop, like in controller 1.
Controlling of the velocity is hence done in the outer loop of controller 2.
The outer loop also has slightly different inputs: instead of the forward velocity in the C-frame, the forward velocity in the body-fixed frame is controlled. The vertical velocity component is epxressed in the C-frame though, like for controller 1.
Additionally, not the heading rate of change $\dot{\varPsi}$ is commanded to the outer loop, but the yaw rate $r$.
They are effecitvely equal during level flight, but differ during climbs or descents, as the pitch angle is different from zero.
The control allocation module is exactly the same as used in controller 1.
The key differences of the both controllers are summarized in \cref{tab:controllers_difference}

A high level block diagram of controller 2 can be seen in \cref{fig:block_diagram_ctrl_2}.
For comparison, the high level block diagram of controller 1 is shown in \cref{fig:block_diagram_ctrl} again.

\begin{figure}[H]
    \centering
    \includegraphics[width=\linewidth]{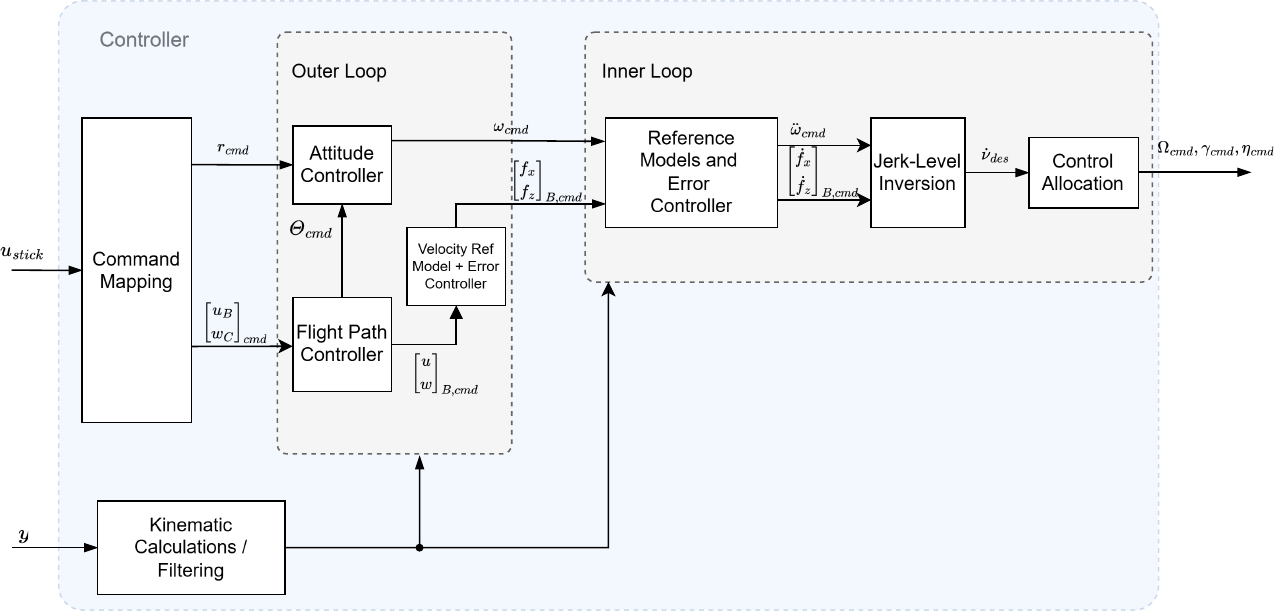}
    \caption{High Level Block Diagram of Controller 2}
    \label{fig:block_diagram_ctrl_2}
\end{figure}

\begin{figure}[H]
    \centering
    \includegraphics[width=\linewidth]{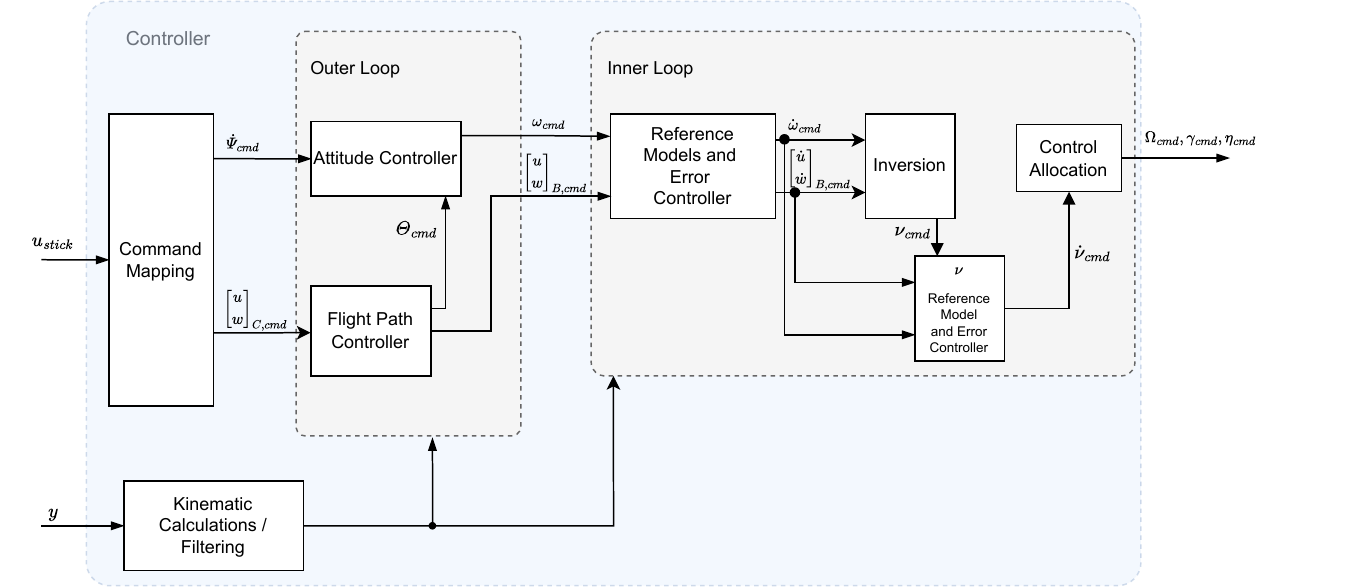}
    \caption{High Level Block Diagram of Controller}
    \label{fig:block_diagram_ctrl}
\end{figure}

\begin{table}
    \centering
    \begin{tabular}{c|c|c}
                            & Controller 1                      & Controller 2               \\
        \hline
        Inversion           & Acceleration Level                & Jerk Level                 \\
        Outer Loop Commands & $\dot{\varPsi}$, $u_{C}$, $w_{C}$ & $r$, $u_{B}$, $w_{C}$      \\
        Inner Loop Commands & $\omega$, $u_{B}$, $w_{B}$        & $\omega$, $f_{x}$, $f_{z}$
    \end{tabular}
    \caption{Key differences between the two controllers}
    \label{tab:controllers_difference}
\end{table}

As mentioned, the inner loop directly calculates $\pseudodot$, which consists of the time derivatives of the propulsion forces in the x- and z-direction as well as the time derivatives of the propulsion moments.
To achieve this, the equations of motion must be differentiated once more, yielding expressions for linear and angular jerks on one side, and the time derivatives of forces and moments on the other.

Since the required rates of change of forces and moments are defined in the body-fixed frame, the time derivative must also be taken with respect to this frame. Assuming constant mass and constant mass distribution, all terms related to mass and inertia remain time-invariant and thus do not require differentiation. Furthermore, under the rigid body assumption, the vector $\bm{r}^{RG}$ remains unchanged in the body frame and can be treated as constant.

As a result, the differentiation only needs to be applied to the linear and angular velocities $\bm{V}$ and $\bm{\omega}$, their respective time derivatives, and the corresponding forces and moments.
For convenience, the equations of motion from \cref{chap:system_description} are repeated here:
\begin{equation}
    \begin{bmatrix}
        \dot{\bm{V}} \\
        \dot{\bm{\omega}}
    \end{bmatrix} = \massmatrix^{-1} \begin{bmatrix}
        \bm{F}_B + \bm{F}_G + \bm{F}_P + \bm{F}_F -m \bm{\omega} \times \left( \bm{\omega} \times \bm{r}^{RG} \right) - \bm{\omega} \times \bm{M}_a \bm{V} \\
        \bm{M}_B + \bm{M}_G + \bm{M}_P + \bm{M}_F -\bm{\omega} \times \bm{J}_a \bm{\omega} + \bm{r}^{RG} \times \left(m \bm{V} \times \bm{\omega}\right)
    \end{bmatrix} \tag{\ref{eq:eom_simple}}
\end{equation}

Applying the time derivative with respect to the B-frame to \cref{eq:eom_simple} yields:
\begin{equation}\label{eq:eom_jerk}
    \begin{bmatrix}
        \ddot{\bm{V}} \\
        \ddot{\bm{\omega}}
    \end{bmatrix}  = \massmatrix^{-1} \begin{bmatrix}
        \left(\dot{\bm{F}}_T\right)^B -m \dot{\bm{\omega}} \times \left( \bm{\omega} \times \bm{r}^{RG} \right) -m \bm{\omega} \times \left( \dot{\bm{\omega}} \times \bm{r}^{RG} \right) - \dot{\bm{\omega}} \times \bm{M}_a \bm{V} - \bm{\omega} \times \bm{M}_a \dot{\bm{V}} \\
        \left(\dot{\bm{M}}_T\right)^B -\dot{\bm{\omega}} \times \bm{J}_a \bm{\omega} -\bm{\omega} \times \bm{J}_a \dot{\bm{\omega}} + \bm{r}^{RG} \times \left(m \dot{\bm{V}} \times \bm{\omega}\right) + \bm{r}^{RG} \times \left(m \bm{V} \times \dot{\bm{\omega}}\right)
    \end{bmatrix}.
\end{equation}
where $\ddot{\Vector{V}} = \left( \ddot{\Vector{V}}_K^R \right)^{EBB}_B$ and $\ddot{\Vector{\omega}} = \left( \ddot{\Vector{\omega}}^{OB}_K \right)^{BB}_B$.

The time derivative of the total force and total moment can be distributed across their individual components, allowing each term to be differentiated separately:
\begin{align}
    \left(\dot{\Vector{F}}_T\right)^B & = \left(\dot{\Vector{F}}_B\right)^B + \left(\dot{\Vector{F}}_G\right)^B + \left(\dot{\Vector{F}}_P\right)^B                                     \\
    \left(\dot{\Vector{M}}_T\right)^B & = \left(\dot{\Vector{M}}_B\right)^B + \left(\dot{\Vector{M}}_G\right)^B + \left(\dot{\Vector{M}}_P\right)^B + \left(\dot{\Vector{M}}_F\right)^B
\end{align}
Special attention must be paid to the treatment of buoyancy and gravitational forces and moments. Although these forces remain constant in the NED frame, they vary in the body-fixed frame due to its rotation relative to the NED frame. Let $\Matrix{M}_{BO}$ denote the transformation matrix from the NED frame to the body-fixed frame. The time derivative of the gravitational force in the body-fixed frame can then be written as:
\begin{align}
    \left(\bm{F}_G\right)_B         & = \bm{M}_{BO} \left(\bm{F}_G\right)_O                                                                                          \\
    \left(\dot{\bm{F}}_G\right)_B^B & = \dot{\bm{M}}_{BO}^B \left(\bm{F}_G\right)_O + \bm{M}_{BO} \overbrace{\left(\dot{\bm{F}}_G\right)_O^O}^{=0}\label{eq:F_G_dot}
\end{align}
Here, the last term vanishes because the gravity force is constant when expressed in the NED frame, therefore the time derivative with respect to this frame is zero.

$\dot{\bm{M}}_{BO}^B$ can be rewritten using the strapdown equation \cite{FSD2}:
\begin{align}
    \left(\bm{\Omega}^{BO}\right)_{BB}              & = \dot{\bm{M}}_{BO}^B \bm{M}_{OB}                               \\
    \left(\bm{\Omega}^{BO}\right)_{BB} \bm{M}_{BO}  & = \dot{\bm{M}}_{BO}^B \underbrace{\bm{M}_{OB} \bm{M}_{BO}}_{=1} \\
    -\left(\bm{\Omega}^{OB}\right)_{BB} \bm{M}_{BO} & = \dot{\bm{M}}_{BO}^B\label{eq:M_BO_dot}
\end{align}
where the term $\left(\bm{\Omega}^{OB}\right)_{BB}$ denotes the skew symmetric matrix which contains the individual components of the angular velocity of the body fixed frame with respect to the NED frame:
\begin{equation}
    \left(\bm{\Omega}^{OB}\right)_{BB} = \begin{bmatrix}
        0  & -r & q  \\
        r  & 0  & -p \\
        -q & p  & 0
    \end{bmatrix}_B
\end{equation}
This matrix vector product can also be represented as the cross product of the angular velocity vector with said vector \cite{FSD2}.
Inserting \cref{eq:M_BO_dot} into \cref{eq:F_G_dot} therefore yields:
\begin{align}
    \left(\dot{\bm{F}}_G\right)_B^B & = -\left(\bm{\Omega}^{OB}\right)_{BB} \left(\bm{F}_G\right)_B    \\
    \left(\dot{\bm{F}}_G\right)_B^B & = -\bm{\omega} \times \left(\bm{F}_G\right)_B \label{eq:f_g_dot}
\end{align}
The same approach applies to the buoyancy force. First, the buoyancy force must be transformed from the NED frame into the body-fixed frame, after which its time derivative is taken with respect to the body-fixed frame:
\begin{align}
    \left(\bm{F}_B\right)_B         & = \bm{M}_{BO} \left(\bm{F}_B\right)_O                                                                                          \\
    \left(\dot{\bm{F}}_B\right)_B^B & = \dot{\bm{M}}_{BO}^B \left(\bm{F}_B\right)_O + \bm{M}_{BO} \overbrace{\left(\dot{\bm{F}}_B\right)_O^O}^{=0}\label{eq:F_B_dot}
\end{align}
The last term is again zero because the buoyancy force is constant in the NED frame, therefore performing the time derivative with respect to this frame yields zero.
Using again the relationship of the strapdown matrix (\cref{eq:M_BO_dot}), this result can be expressed in terms of the angular velocity:
\begin{align}
    \left(\dot{\bm{F}}_B\right)_B^B & = -\left(\bm{\Omega}^{OB}\right)_{BB} \left(\bm{F}_B\right)_B \\
    \left(\dot{\bm{F}}_B\right)_B^B & = -\bm{\omega} \times \left(\bm{F}_B\right)_B
\end{align}
With the expressions for $\left(\dot{\bm{F}}_G\right)_B^B$ and $\left(\dot{\bm{F}}_B\right)_B^B$ available, the corresponding results for $\left(\bm{\dot{M}}_B\right)^B$ and $\left(\bm{\dot{M}}_G\right)^B$ follow directly. This is due to the fact that, under the rigid body assumption, the time derivatives of both $\bm{r}^{RG}$ and $\bm{r}^{RB}$ within the body-fixed frame are zero:
{
\small
\begin{equation}
    \begin{split}
        \bm{M}_G                      & = \bm{r}^{RG} \times \bm{F}_G                                                                                              \\
        \left(\bm{\dot{M}}_G\right)^B & = \overbrace{\left( \bm{\dot{r}}^{RG} \right)^B}^{=0} \times \bm{F}_G + \bm{r}^{RG} \times \left( \bm{\dot{F}}_G \right)^B \\
        \left(\bm{\dot{M}}_G\right)^B & = - \bm{r}^{RG} \times \left( \bm{\omega} \times \bm{F}_G \right)
    \end{split}
    \quad
    \begin{split}
        \bm{M}_B                      & = \bm{r}^{RB} \times \bm{F}_B                                                                                            \\
        \left(\bm{\dot{M}}_B\right)^B & = \overbrace{\left( \bm{\dot{r}}^{RB} \right)^B}^{=0} \times \bm{F}_B + \bm{r}^{RB} \times \left(\bm{\dot{F}}_B\right)^B \\
        \left(\bm{\dot{M}}_B\right)^B & = - \bm{r}^{RB} \times \left( \bm{\omega} \times \bm{F}_B \right)
    \end{split}
\end{equation}
}
Because the reference point is coincident with the center of buoyancy, i.e. \(\bm{r}^{RB} = [0 \ 0 \ 0]^T\) the moment and time derivative of the moment due to buoyancy can be discarded.

Finally, \cref{eq:eom_jerk} can be rearranged to solve for $\left( \dot{\bm{F}}_P \right)^B$ and $\left( \dot{\bm{M}}_P \right)^B$, yielding the inversion law that maps linear and angular jerks to the corresponding time derivatives of forces and moments:
\begin{flalign}\label{eq:eom_jerk_inverted}
     & \begin{bmatrix}
           \dot{\bm{F}}_P \\
           \dot{\bm{M}}_P
       \end{bmatrix}^B = \massmatrix \begin{bmatrix}
                                         \ddot{\bm{V}} \\
                                         \ddot{\bm{\omega}}
                                     \end{bmatrix} + \nonumber \\ &\begin{bmatrix}
        \dot{\bm{\omega}} \times \left( m \bm{\omega} \times \bm{r}^{RG} + \bm{M}_a \bm{V} \right) + \bm{\omega} \times \left( m \dot{\bm{\omega}} \times \bm{r}^{RG} + \bm{M}_a \dot{\bm{V}} + \left(\bm{F}_B\right) + \left(\bm{F}_G\right) \right) \\
        \dot{\bm{\omega}} \times \bm{J}_a \bm{\omega} +\bm{\omega} \times \bm{J}_a \dot{\bm{\omega}} - \bm{r}^{RG} \times \left(m\dot{\bm{V}} \times \bm{\omega} + m\bm{V} \times \dot{\bm{\omega}} - \bm{\omega} \times \left(\bm{F}_G\right)\right) - \left(\dot{\Vector{M}}_F\right)^B
    \end{bmatrix}.
\end{flalign}
However, the equation still contains the linear jerk $\ddot{\Vector{V}}$. As previously discussed, the inner loop operates by controlling the specific forces. Consequently, it is necessary to establish a relationship between the specific forces and the linear jerk. A suitable starting point for this derivation is \cref{eq:sforces}, which was previously used to determine the linear acceleration at the reference point based on the specific force measurements. For clarity, the equation is restated here:
\begin{equation}
    \left( \dot{\Vector{V}}_K^R \right)^{EB} = \Vector{f}^G + \Vector{g} - \left( \dot{\Vector{\omega}}^{OB} \right)^B \times \Vector{r}^{RG} - \Vector{\omega}^{OB} \times \left( \Vector{V}_K^G \right)^E \tag{\ref{eq:sforces}}
\end{equation}
By taking the time derivative of both sides with respect to the body-fixed frame one obtains:
\begin{equation}\label{eq:sforces_dot}
    \left( \ddot{\Vector{V}}_K^R \right)^{EBB} = \left(\dot{\Vector{f}}^G\right)^B + \dot{\Vector{g}}^B - \left( \ddot{\Vector{\omega}}^{OB} \right)^{BB} \times \Vector{r}^{RG} - \left( \dot{\Vector{\omega}}^{OB} \right)^B \times \left( \Vector{V}_K^G \right)^E - \omega^{OB} \times \left( \dot{\Vector{V}}_K^G \right)^{EB}
\end{equation}
$\dot{\Vector{g}}^B$ essentially is just \cref{eq:f_g_dot} divided by mass, therefore the same procedure can be applied to calculate the time derivative:
\begin{equation}
    \dot{\Vector{g}}^B = -\Vector{\omega} \times \Vector{g}_B
\end{equation}
where $\Vector{g}_B$ is the gravity vector expressed in the body-fixed frame.
Inserting this result into \cref{eq:sforces_dot}, the relationship between the linear jerk and the time derivative of the specific forces has been found:
\begin{equation}\label{eq:sforces_dot_final}
    \ddot{\Vector{V}} = \dot{\Vector{f}} -\Vector{\omega} \times \Vector{g}_B - \ddot{\Vector{\omega}} \times \Vector{r}^{RG} - \dot{\Vector{\omega}} \times \Vector{V}^G  - \omega \times \dot{\Vector{V}}^G
\end{equation}
where $\dot{\Vector{f}} = \left(\dot{\Vector{f}}^G\right)^B$, $\Vector{V}^G = \left( \Vector{V}_K^G \right)^E$ and $\dot{\Vector{V}}^G = \left( \dot{\Vector{V}}_K^G \right)^{EB}$.

\cref{eq:sforces_dot_final} can be inserted into \cref{eq:eom_jerk_inverted}.
After a few transformation steps, which are not further outlined here (see \cref{app:jerk} in the appendix for more detail), the final result is:
\begin{flalign}
     & \begin{bmatrix}
           \dot{\bm{F}}_P \\
           \dot{\bm{M}}_P
       \end{bmatrix} = \begin{bmatrix}
                           \Matrix{M}_a            & \Matrix{M}_v\Matrix{r}^{RG\times}                            \\
                           m \Vector{r}^{RG\times} & \Matrix{J}_a + m \Vector{r}^{RG\times} \Vector{r}^{RG\times}
                       \end{bmatrix}
    \begin{bmatrix}
        \dot{\bm{f}} \nonumber \\
        \ddot{\bm{\omega}}
    \end{bmatrix}
    +                                                                                                                                                                                                                                                                                                                                                                                      \\
     & \begin{bmatrix}
           \dot{\Vector{\omega}} \times \Matrix{M}_v \Vector{V} + \Vector{\omega} \times \left(\Matrix{M}_v \dot{\Vector{V}} + \Vector{F}_B + \Vector{F}_G\right) - \Matrix{M}_v (\dot{\Vector{\omega}} \times \Vector{V}^G + \Vector{\omega} \times \dot{\Vector{V}}^G) - (\Matrix{M}_v + m)(\Vector{\omega} \times \Vector{g}_B) \\
           \dot{\Vector{\omega}} \times \Matrix{J}_a \Vector{\omega} + \Vector{\omega} \times \Matrix{J}_a \dot{\Vector{\omega}} - m \Vector{r}^{RG} \times (\dot{\Vector{\omega}} \times (\Vector{\omega} \times \Vector{r}^{RG}) + \Vector{\omega} \times (\dot{\Vector{\omega}} \times \Vector{r}^{RG}) - \Vector{\omega}\times\Vector{g}_B) - \dot{\Vector{M}}_F
       \end{bmatrix}
\end{flalign}
where $\dot{\Vector{M}}_F$ is the rate of change of the control surface moments.
The pseudo control rate of change vector $\pseudodot$ can be constructed by using a selection matrix:
\begin{equation}
    \pseudodot = \begin{bmatrix}
        0 & 0 & 0 & 1 & 0 & 0 \\
        0 & 0 & 0 & 0 & 1 & 0 \\
        0 & 0 & 0 & 0 & 0 & 1 \\
        1 & 0 & 0 & 0 & 0 & 0 \\
        0 & 0 & 1 & 0 & 0 & 0 \\
    \end{bmatrix} \begin{bmatrix}
        \dot{\bm{F}}_P \\
        \dot{\bm{M}}_P
    \end{bmatrix}
\end{equation}

The command values for $\ddot{\omega}$ and $\dot{f}$ are obtained by a second and first order reference system, respectively, in combination with a proportional error controller.

This concludes the description of controller 2.
What follows next is the first test case, comparing the performances of both controllers.

\section{Case 1}
In the first scenario, the forward velocity is commanded to reach its maximum allowable value. Subsequently, the airship is instructed to perform turning maneuvers in both directions using the maximum available turn rate. During these turns, the controller is further challenged by commanding climbs and descents using full stick deflections. This test case is designed to evaluate whether the controller can maintain stable and accurate control of the airship under maximum performance demands, including aggressive changes in heading and altitude. The scenario represents a high-stress condition for the control system and serves to demonstrate its robustness and responsiveness when operating near the limits of the vehicle's flight envelope.

\subsection{Controller 1}
The simulation results clearly demonstrate the strong performance of the full airship control architecture across all control loops. Starting with the outer loop in \cref{fig:ctrl1_case1_outer_loop}, the commanded forward and vertical velocities ($u_{cmd}, w_{cmd}$) are closely followed by the actual velocities ($u, w$), even during step inputs and reversals. The heading rate tracking ($\dot{\psi}$) is also very precise, with the actual yaw rate nearly indistinguishable from the command throughout the steady-state phases of the maneuver.

Looking at the inner loop performance in \cref{fig:ctrl1_case1_inner_loop}, the velocity tracking within the body-fixed frame further confirms the accuracy of the cascaded control structure. The reference signals ($u_{ref}, w_{ref}$) are matched closely by the actual values, validating the ability of the inner loop to realize the desired dynamics imposed by the outer loop. Similarly, the attitude controller exhibits good tracking of all angular rates. The measured rates ($p, q, r$) follow both the commanded and reference rates very closely, with only slight, short-lived discrepancies during rapid changes, particularly in roll. These deviations small, and the system returns to nominal tracking.

Continuing with \cref{fig:ctrl1_case1_nu}, the tracking of the pseudo control vector $\pseudo$, which consists of the propulsion forces in body-fixed x- and z-direction and the propulsion moments in the body frame, reveals that the airship can overall follow the desired pseudo control targets. Throughout the entire simulation, the achieved values ($X_P, Z_P M_P, N_P$) are close to their reference values, including in dynamically demanding segments such as sharp turns and altitude changes. For the rolling moment there is one particular larger deviation at around \qty{80}{\second}, indicating the demanded rolling moment is too high for the airship to achieve.


Next, the performance of the control allocation is investigated in \cref{fig:ctrl1_case1_nu_dot_angle}.
The angle between the commanded and achieved $\dot{\pseudo}$ vectors remains zero for the vast majority of the simulation, however there are some short time periods during which the angle is not zero, indicating a misallocation between the desired and achieved pseudo control rate vector.
Due to time constraints, the root cause of the issue could not be fully identified. However, it was consistently observed that the spikes only occur when one or more rotors reach their absolute upper limit, i.e., $\Omega_i = \Omega_{i,\text{max}}$. It is assumed that, under these conditions, actuator increments in the direction of the saturation can no longer be allocated. Consequently, the corresponding input rate is effectively clamped to zero, i.e., $\overline{\dot{u}}_i = 0$. This constraint appears to interfere with the direction-preserving properties of the allocation algorithm. Although a detailed analysis of the phenomenon was not feasible within the scope of this work, the observation is considered noteworthy and is documented here for completeness.

Lastly, the actuator activity for Controller 1 in Case 1 is depicted in \cref{fig:ctrl1_case1_actuators}. The control surfaces are utilized extensively, with multiple instances of maximum deflection observed throughout the simulation. In contrast, the tilt angles remain relatively constant, predominantly oriented forward, suggesting that tilt mechanisms play a minimal role in this scenario. The rotor speeds, however, show notable variability. Specifically, the aft propellers ($\Omega\_2$ and $\Omega\_3$) reach their maximum RPM several times, maintaining saturation for a few seconds during peak demand phases.

\begin{figure}
    \centering
    \includegraphics{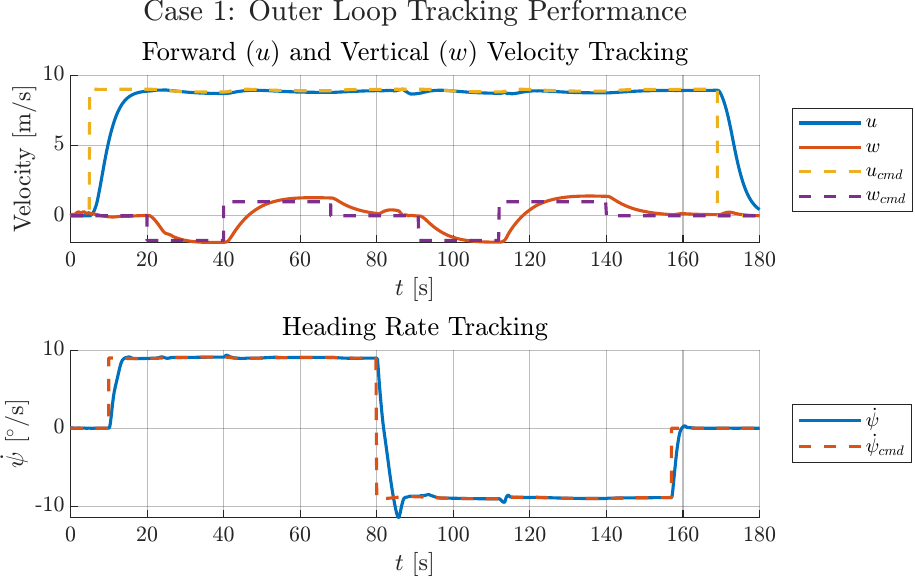}
    \caption{Controller 1, Case 1: Outer Loop Tracking}
    \label{fig:ctrl1_case1_outer_loop}
\end{figure}

\begin{figure}
    \centering
    \includegraphics{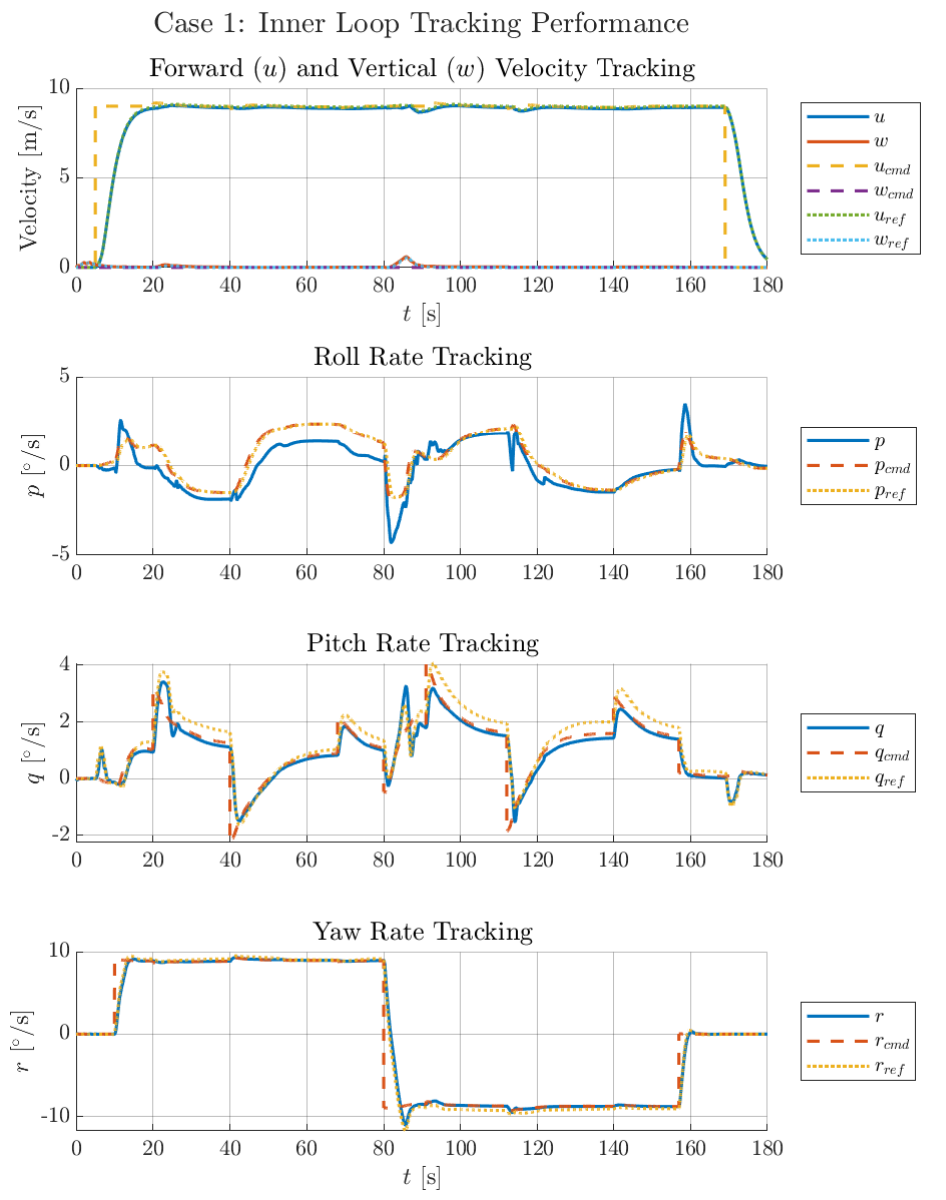}
    \caption{Controller 1, Case 1: Inner Loop Tracking}
    \label{fig:ctrl1_case1_inner_loop}
\end{figure}

\begin{figure}
    \centering
    \includegraphics{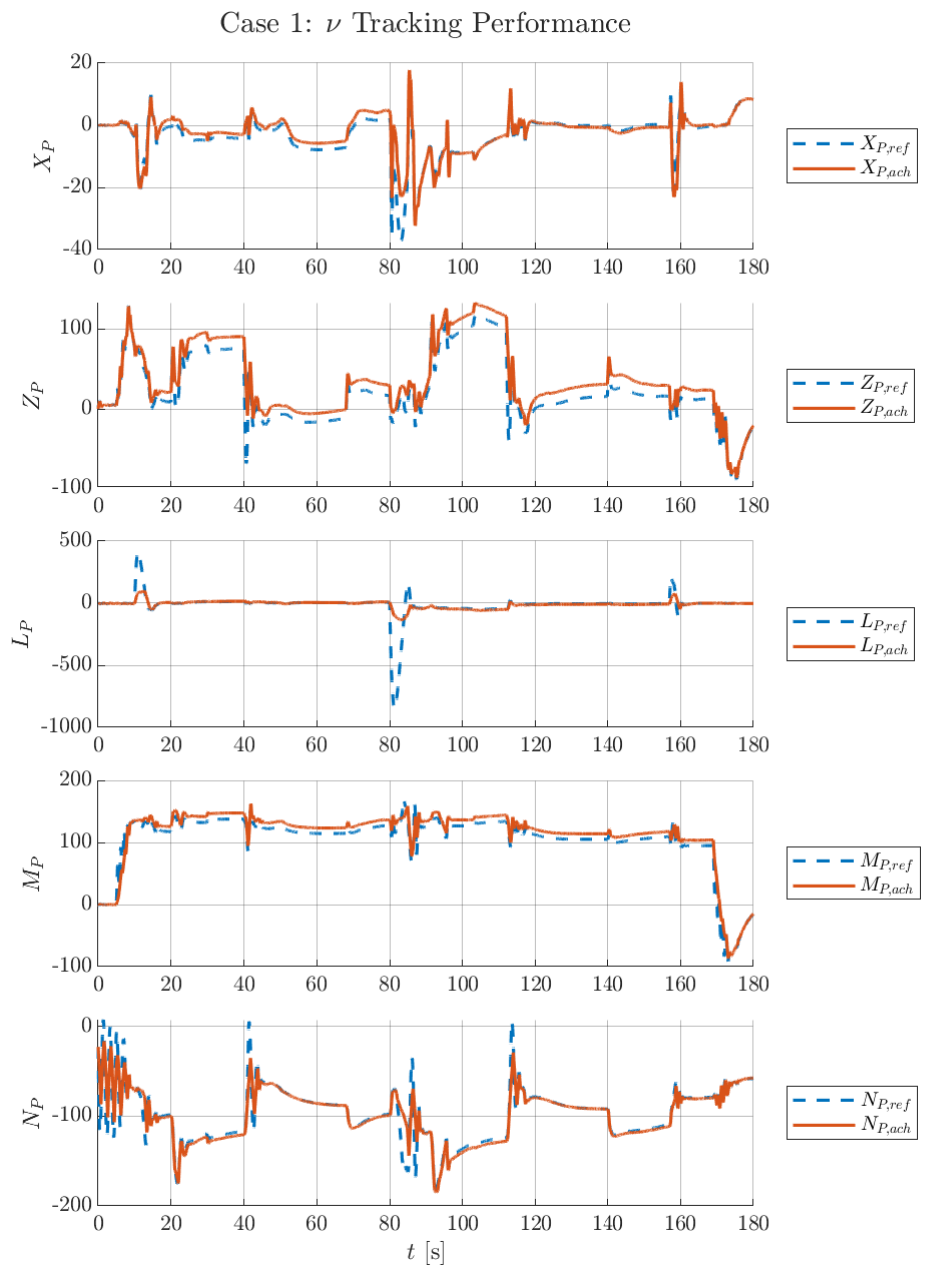}
    \caption{Controller 1, Case 1: $\pseudo$ Tracking}
    \label{fig:ctrl1_case1_nu}
\end{figure}

\begin{figure}
    \centering
    \includegraphics{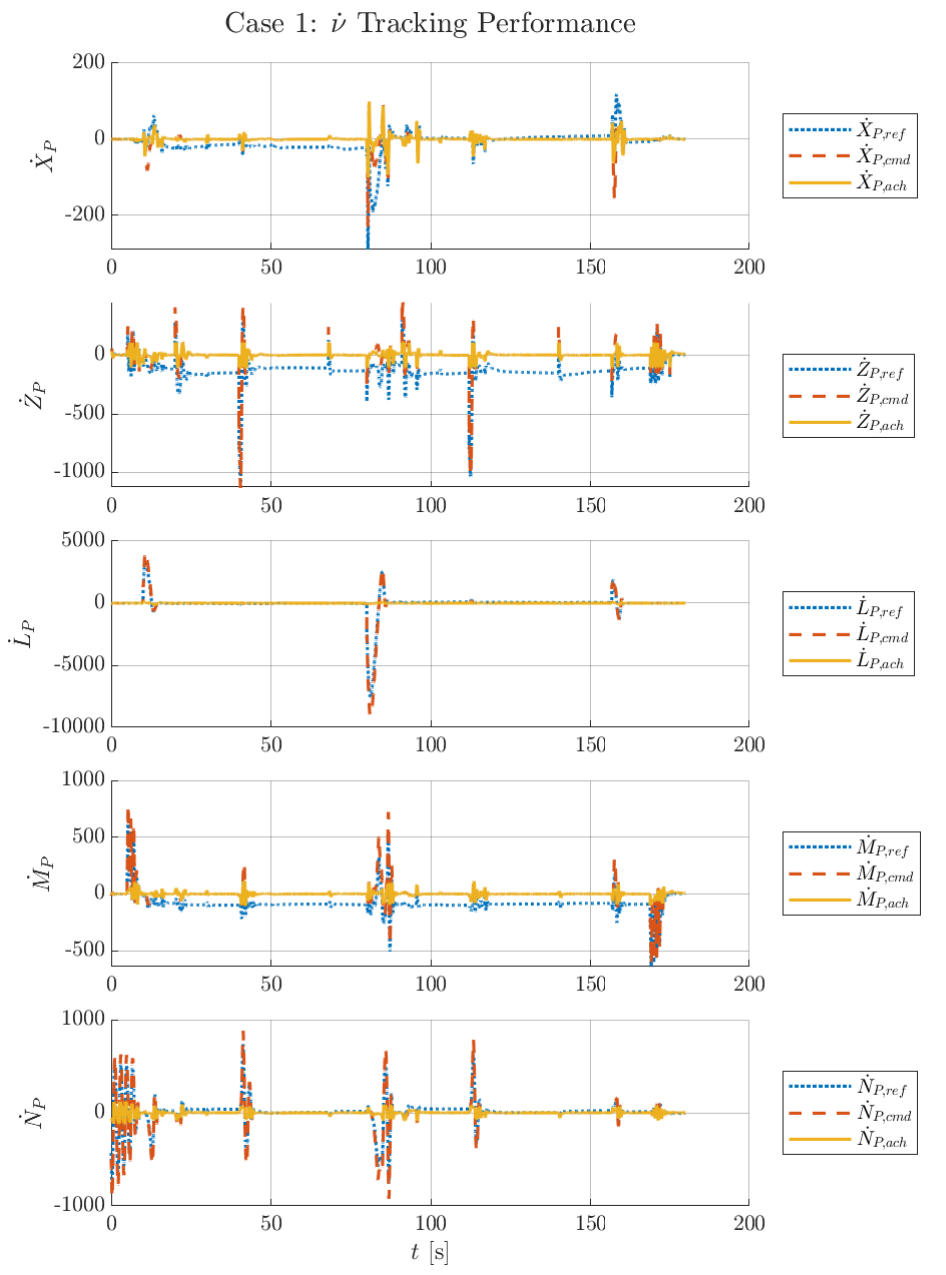}
    \caption{Controller 1, Case 1: $\pseudodot$ Tracking}
    \label{fig:ctrl1_case1_nu_dot}
\end{figure}

\begin{figure}
    \centering
    \includegraphics{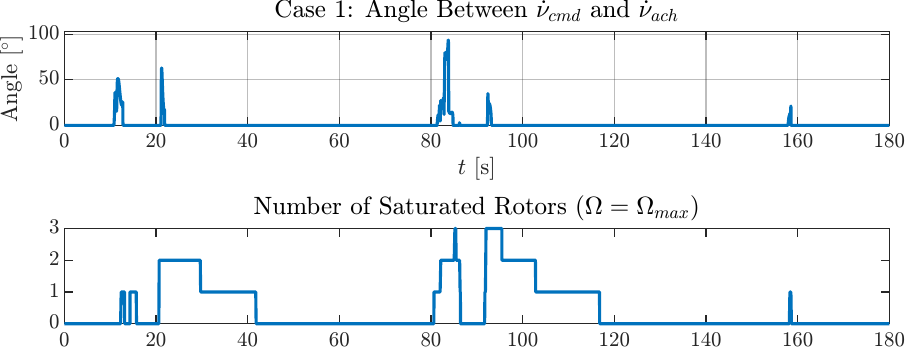}
    \caption{Controller 1, Case 1: Angle between $\pseudodot_{cmd}$ and $\pseudodot_{ach}$}
    \label{fig:ctrl1_case1_nu_dot_angle}
\end{figure}

\begin{figure}
    \centering
    \includegraphics{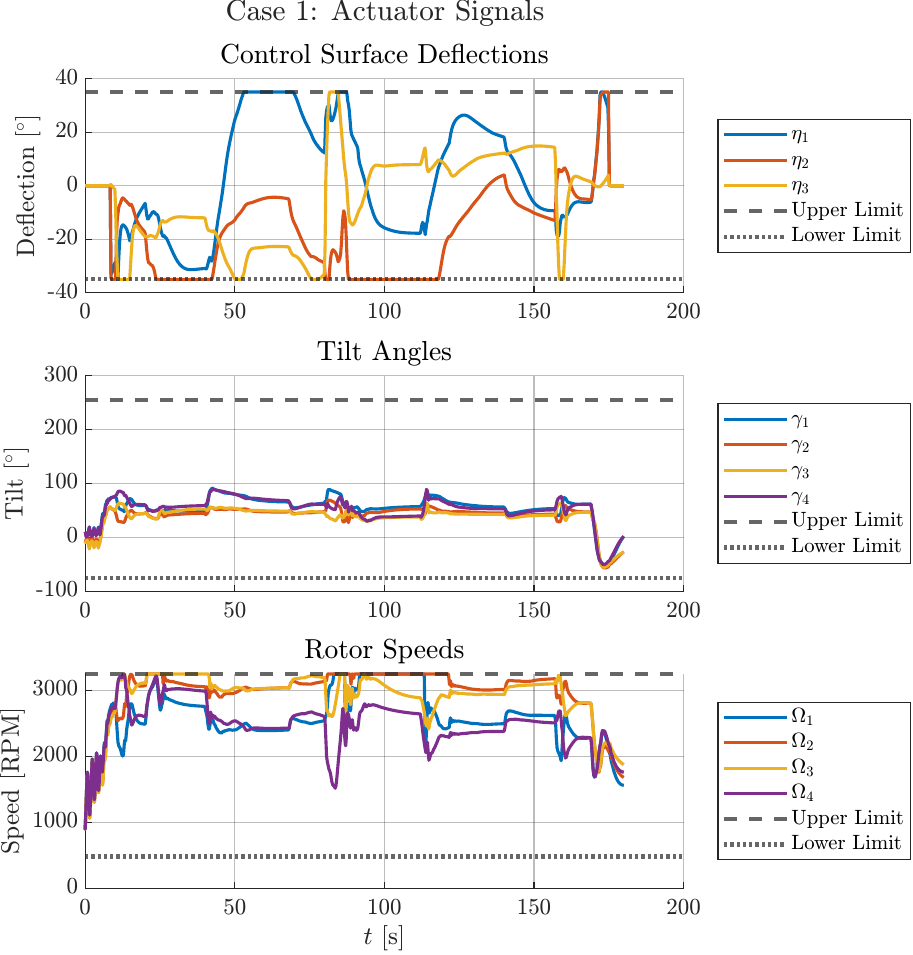}
    \caption{Controller 1, Case 1: Actuator Positions}
    \label{fig:ctrl1_case1_actuators}
\end{figure}

\subsection{Controller 2}

The following figures show the performance of the second controller under the same input conditions as previously used for evaluating the original controller. The same demanding input scenario consisting of maximum forward velocity, maximum climb and descent commands, and turns at maximum heading rate was applied.

Starting with the outer loop performance in \cref{fig:ctrl2_case1_outer_loop}, the top subplot illustrates the tracking of forward and vertical velocities. While the forward velocity $u_B$ closely follows the command $u_{B,cmd}$, it exhibits slightly more deviations compared to controller 1.
The vertical velocity $w_C$ shows good tracking during horizontal and descending flight phases but displays more pronounced deviations during climb phases. Notably, the reference trajectories $u_{B,ref}$ and $w_{C,ref}$ are less accurately followed than before. The yaw rate $r$, shown in the lower subplot, aligns reasonably well with the commanded rate $r_{cmd}$, however minor deviations occur after the desired yaw rate has been achieved.

The inner loop tracking performance is shown in \cref{fig:ctrl2_case1_inner_loop}. The specific force tracking in both the forward direction ($f_x$) and the vertical direction ($f_z$) is very accurate as the achieved values follow the reference values closely.
The roll rate tracking shows greater oscillations and deviations compared to controller 1.
The pitch rate tracking is more accurate than the roll rate tracking, still there is some lag observed between the achieved and the reference values during the climb and descent maneuvers.
Overall this lag is small, and the agreement between the achieved and reference values is very good during the rest of the simulation.
The yaw rate tracking was already described in the outer loop plots.

In the $\pseudodot$ tracking is shown in \cref{fig:ctrl2_case1_nu_dot}.
The commanded values are often too high to be achievable by the plant. This is similar to controller 1. Nevertheless the controller is able to deliver resonable good performance with regards to outer and inner loop variables.

The angle between $\pseudodot_{cmd}$ and $\pseudodot_{ach}$ in \cref{fig:ctrl2_case1_nu_dot_angle} shows a high number of spikes exceeding \qtyrange{60}{100}{\degree}. This suggests a misalignment between commanded and achieved actuation efforts. It is again noted that this almost exclusively appears to happen when the number of saturated rotors is higher than zero, except for one occurance at the beginning of the simulation.

The actuator usage is shown in \cref{fig:ctrl2_case1_actuators}.
The control surfaces and the tilt angles remain well within their limits for the whole simulation.
The rotor speeds on the other hand experience some prolonged saturations, particularly the aft right ($\Omega_2$) propeller stays at its maximum value between \qtyrange{40}{90}{\second}, while the front left ($\Omega_4$) stays at its maximum value between \qtyrange{90}{170}{\second}.


\begin{figure}
    \centering
    \includegraphics{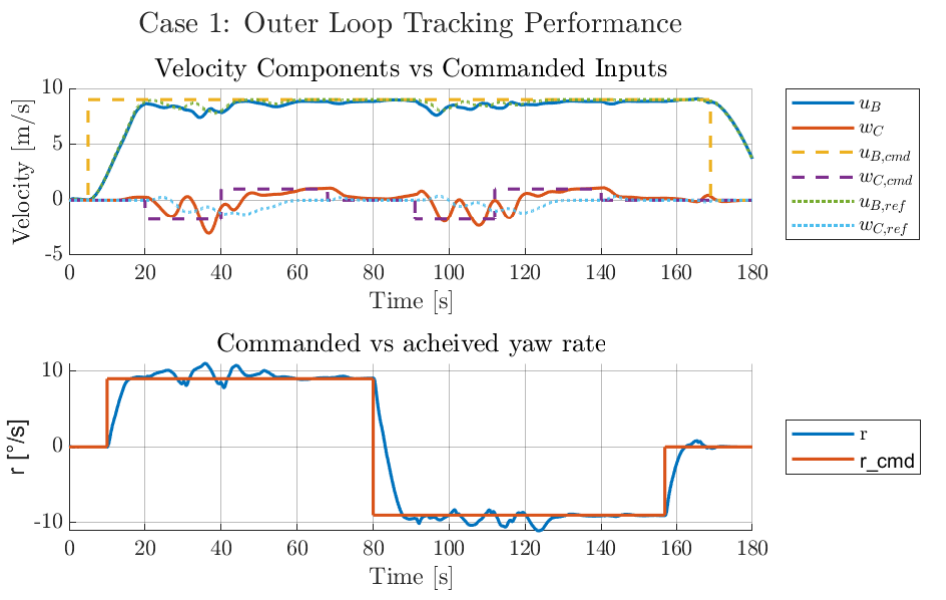}
    \caption{Controller 2, Case 1: Outer Loop Tracking}
    \label{fig:ctrl2_case1_outer_loop}
\end{figure}

\begin{figure}
    \centering
    \includegraphics{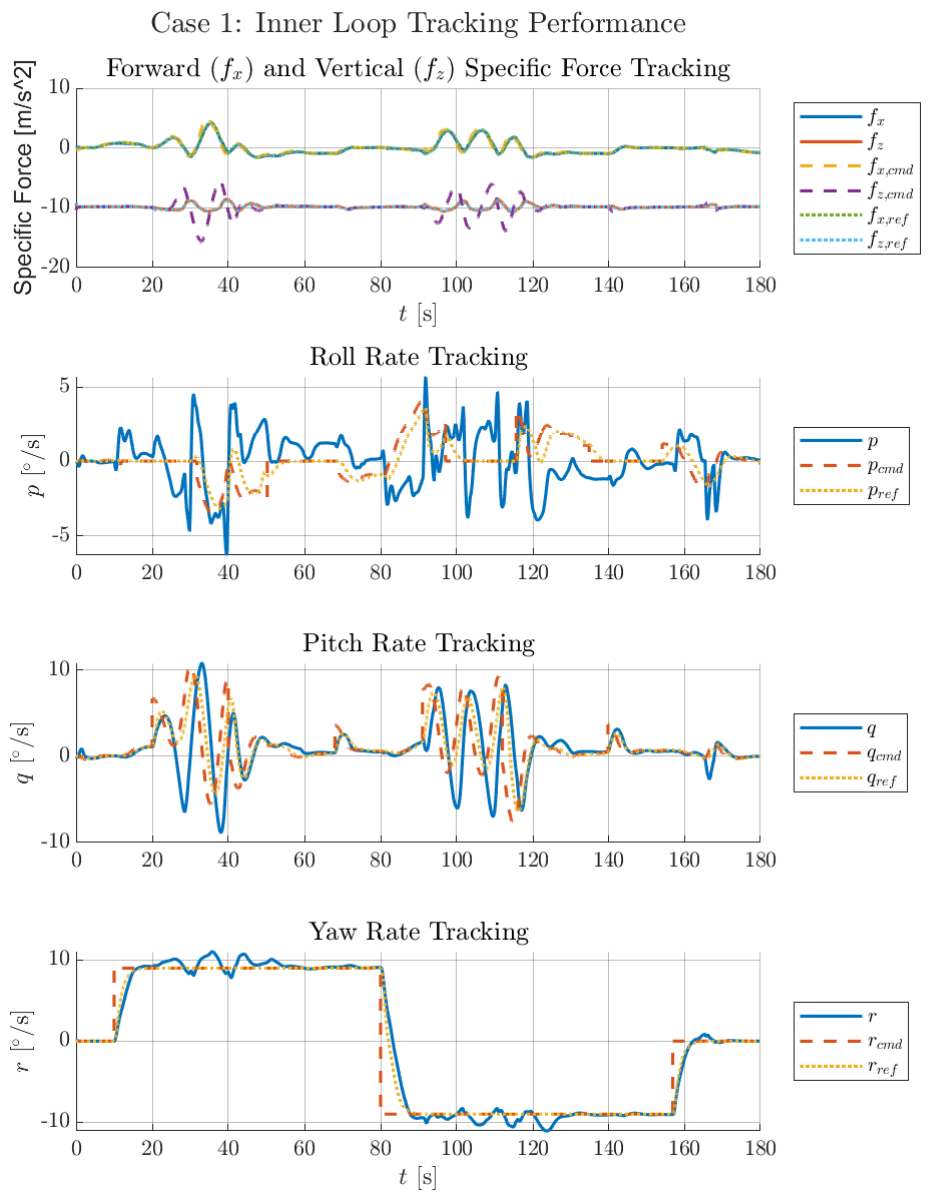}
    \caption{Controller 2, Case 1: Inner Loop Tracking}
    \label{fig:ctrl2_case1_inner_loop}
\end{figure}

\begin{figure}
    \centering
    \includegraphics{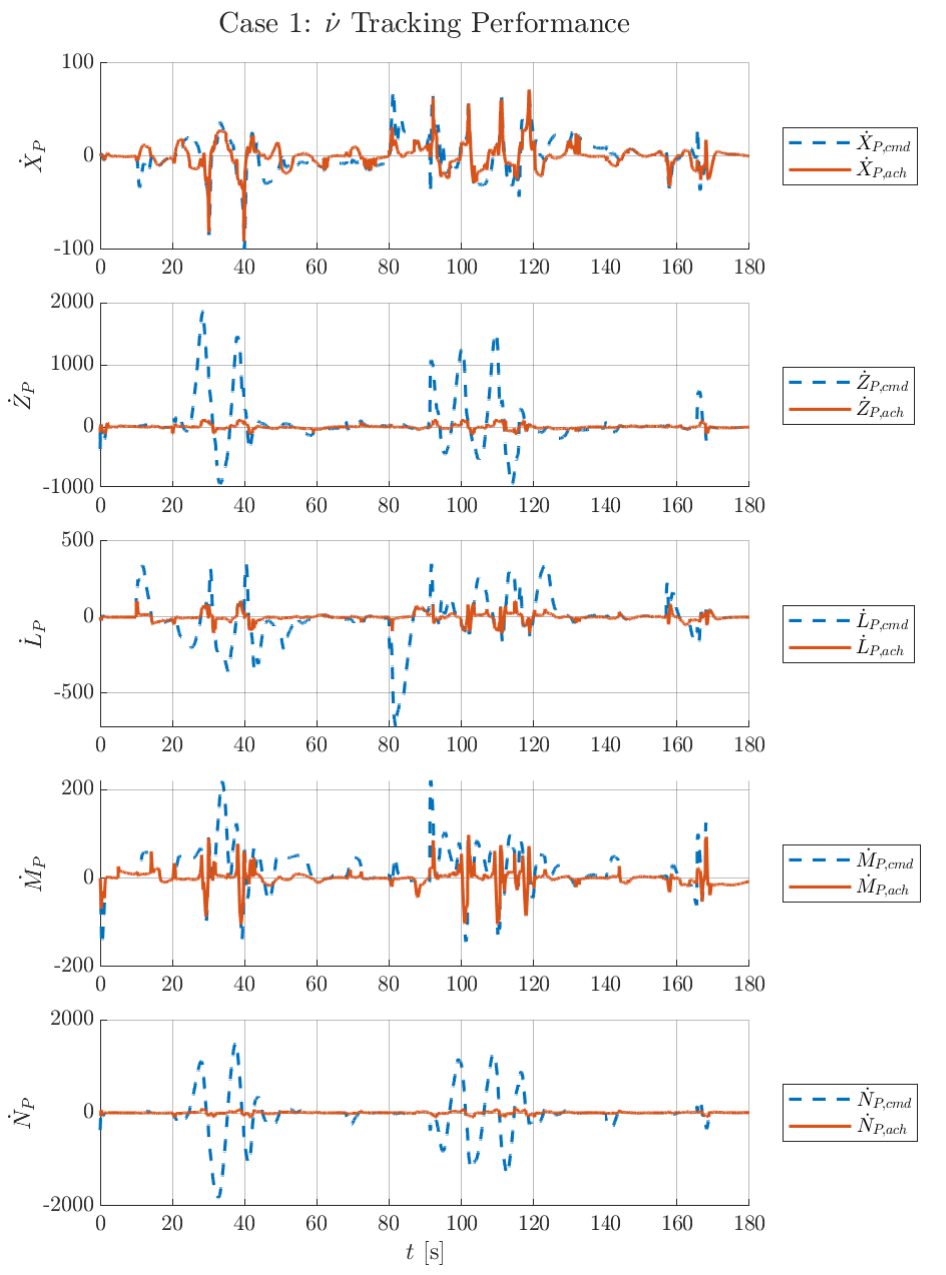}
    \caption{Controller 2, Case 1: $\pseudodot$ Tracking}
    \label{fig:ctrl2_case1_nu_dot}
\end{figure}

\begin{figure}
    \centering
    \includegraphics{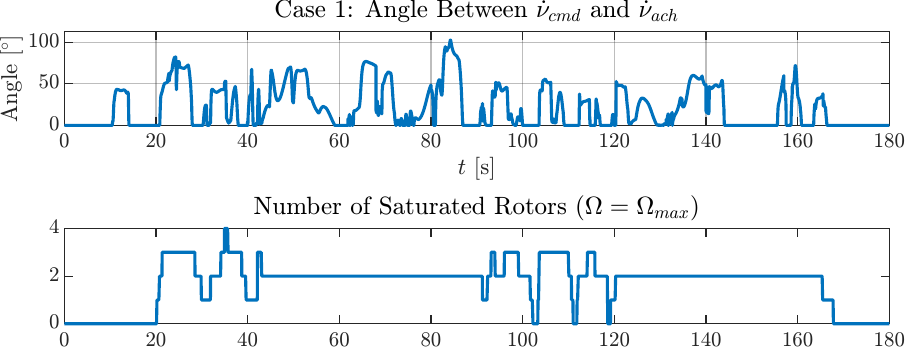}
    \caption{Controller 2, Case 1: Angle between $\pseudodot_{cmd}$ and $\pseudodot_{ach}$}
    \label{fig:ctrl2_case1_nu_dot_angle}
\end{figure}

\begin{figure}
    \centering
    \includegraphics{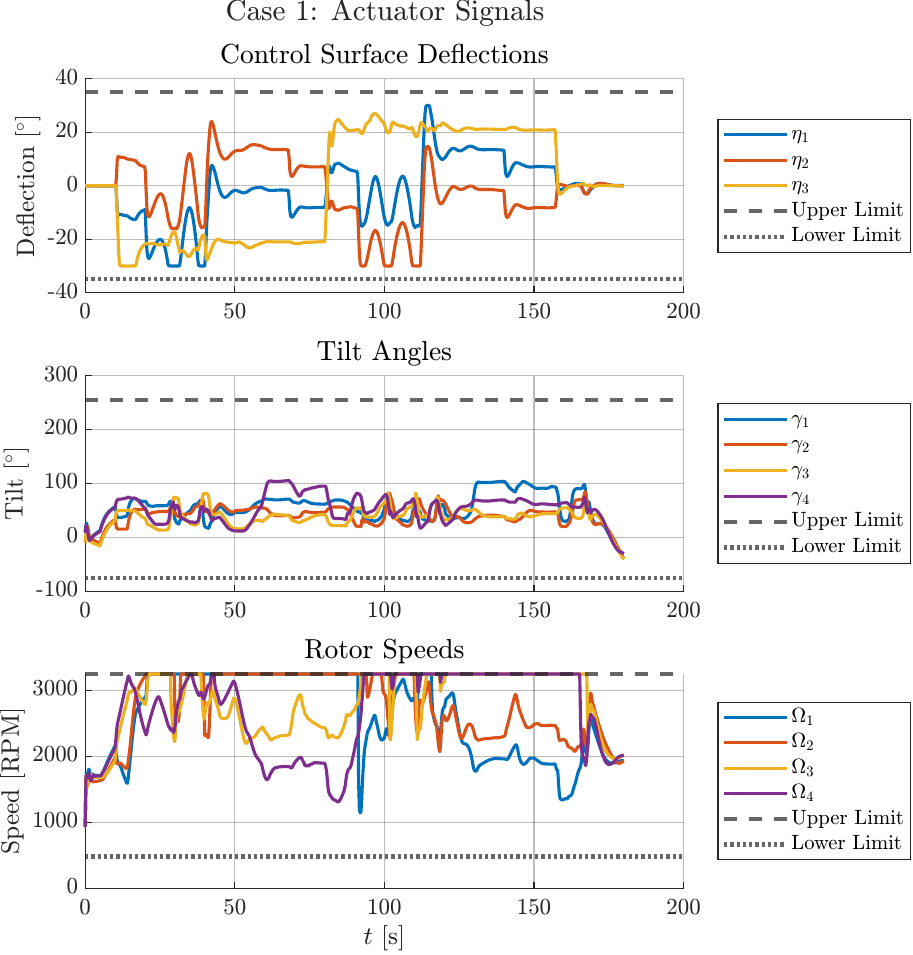}
    \caption{Controller 2, Case 1: Actuator Positions}
    \label{fig:ctrl2_case1_actuators}
\end{figure}

\section{Case 2: Wind Gust}
In this simulation scenario, the airship is subjected to a sudden atmospheric gust at t = \qty{15}{\second}. The gust comes from below and from the left, with an amplitude of \qty{3}{\meter\per\second} and a length of \qty{3}{\meter}, introducing a brief but significant disturbance. Despite this external excitation, the controller demonstrates its ability in restoring stable flight and maintaining trajectory tracking.

\subsection{Controller 1}
Starting with the outer loop tracking performance in \cref{fig:ctrl1_case2_outer_loop}, it can be seen that the forward velocity $u$ tracks its commanded value well throughout the simulation, even in the presence of the gust. After a brief disturbance around \qty{15}{\second}, the velocity stabilizes rapidly back to its commanded level. The vertical velocity $w$, which reacts more strongly to the gust due to the vertical component of the disturbance, also returns to zero shortly after the event, indicating effective rejection of vertical perturbations. The heading rate $\dot{\varPsi}$ shows a short but strong deviation coinciding with the gust impact. The controller quickly counters this though, and the airship regains the commanded heading rate within a couple of seconds.

For comparison, the undisturbed response, i.e. the same simulation without the gust, is shown for the outer loop in \cref{fig:ctrl1_case2_undisturbed_outer_loop}.

The inner loop plots (\cref{fig:ctrl1_case2_inner_loop}) reinforce this observation. Forward and vertical velocity tracking remain robust with only minor deviations during the gust. The angular velocity tracking  for roll, pitch, and yaw respectively, shows short but significant spikes at \qty{15}{\second} in response to the sudden aerodynamic changes. Importantly, the system returns to nominal tracking performance almost immediately, demonstrating fast dynamic response and strong disturbance rejection.

In \cref{fig:ctrl1_case2_nu}, the $\pseudo$-tracking plots show that the required forces and moments during the gust are too high to be achieved by the propulsion system.

The $\pseudodot$ tracking performance shown in \cref{fig:ctrl1_case2_nu_dot} further confirms this, as the commanded values are way higher than the system capabilities, which shows how significant the gust affects the airship.

The angular deviation plot (\cref{fig:ctrl1_case2_nu_dot_angle}) between the commanded and achieved $\pseudodot$ vectors is zero for almost the entire simulation, indicating that the achieved pseudo control vector and the commanded pseudo control vector point in the same direction. There is one spike towards the end of the simulation.
Again it is noted this occurs when more than zero rotors are saturated.
Unitl around \qty{26}{\second}, no rotors are saturated and in this period, the angle between the commanded and achieved $\pseudodot$ vector is zero.

Finally, \cref{fig:ctrl1_case2_actuators} presents the actuator signals. The plots illustrate the system's response across the various control surfaces, rotor tilt angles, and rotor speeds. During the gust, there is a clearly visible spike in actuator activity, especially for the surface deflections. Rotor speeds and tilt angles adapt also visibly but remain well below the control surface response. The control surfaces momentarily deflect to their saturation limits before returning to nominal deflections, indicating maximum use of available control authority.

Overall, this scenario demonstrates that the controller handles a significant atmospheric disturbance good performance. All signals recover quickly after the gust impact, without loss of stability.
\begin{figure}
    \centering
    \includegraphics{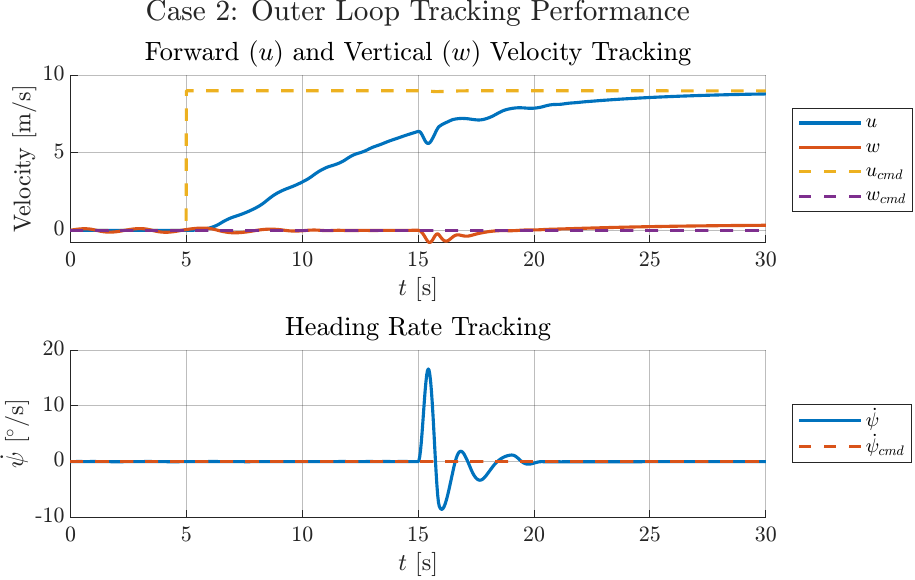}
    \caption{Controller 1, Case 2: Outer Loop Tracking}
    \label{fig:ctrl1_case2_outer_loop}
\end{figure}

\begin{figure}
    \centering
    \includegraphics{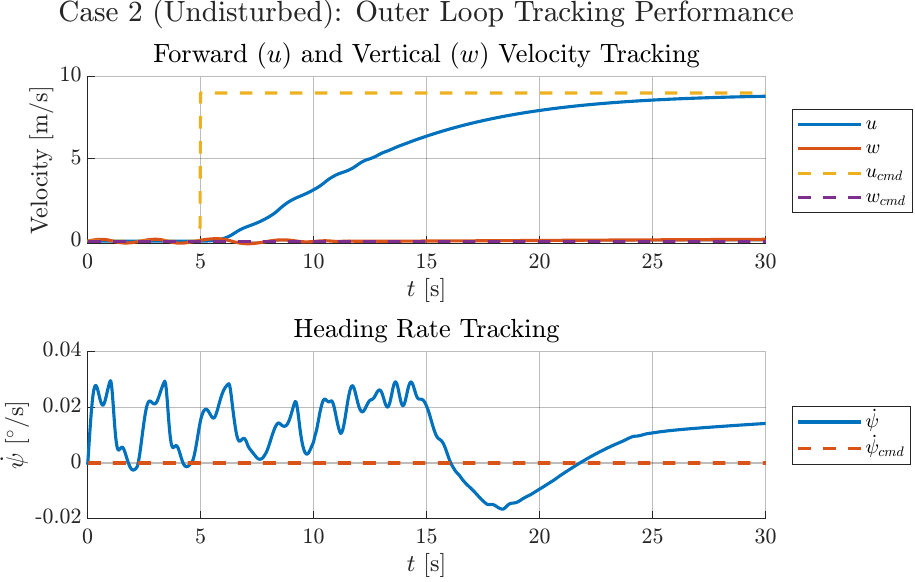}
    \caption{Controller 1, Case 2 (Undisturbed): Outer Loop Tracking}
    \label{fig:ctrl1_case2_undisturbed_outer_loop}
\end{figure}

\begin{figure}
    \centering
    \includegraphics{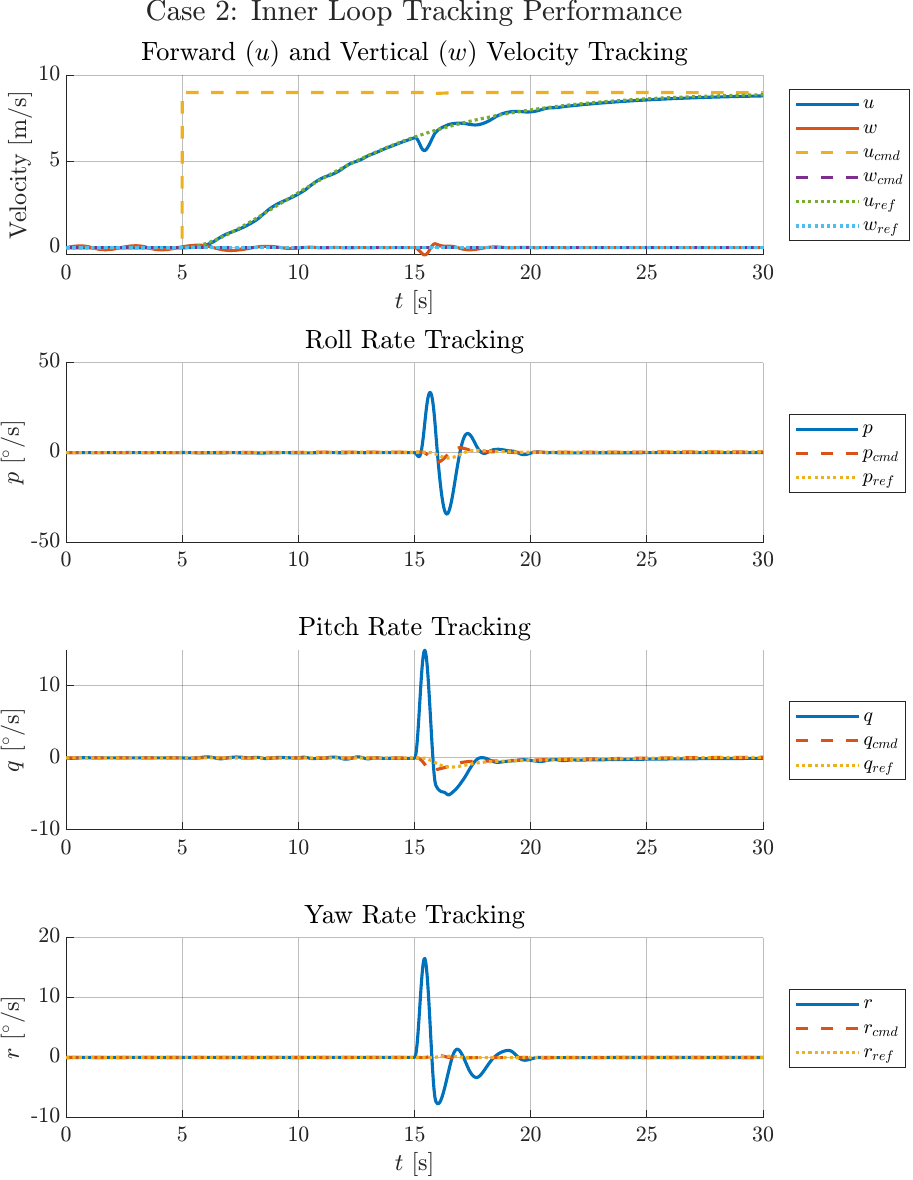}
    \caption{Controller 1, Case 2: Inner Loop Tracking}
    \label{fig:ctrl1_case2_inner_loop}
\end{figure}

\begin{figure}
    \centering
    \includegraphics{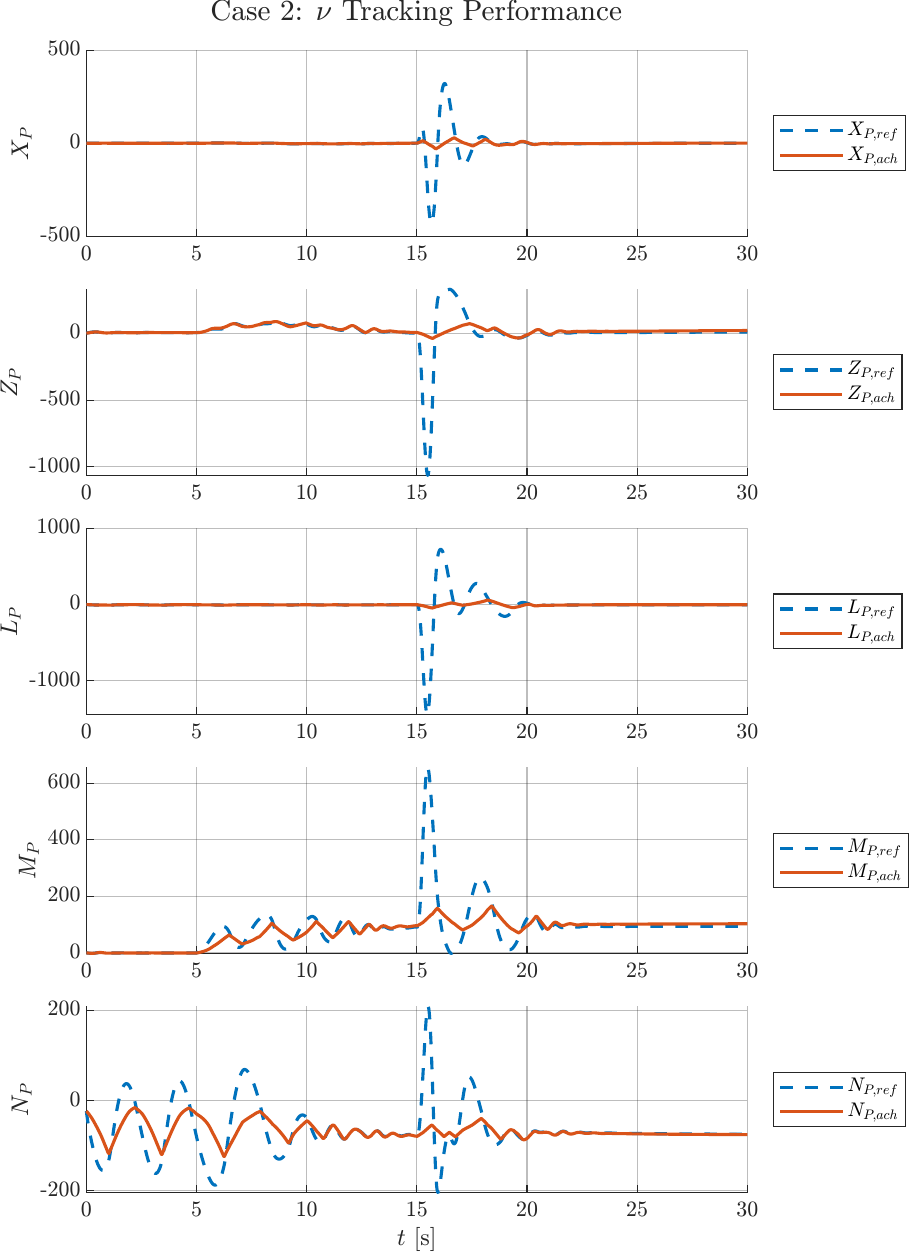}
    \caption{Controller 1, Case 2: $\pseudo$ Tracking}
    \label{fig:ctrl1_case2_nu}
\end{figure}

\begin{figure}
    \centering
    \includegraphics{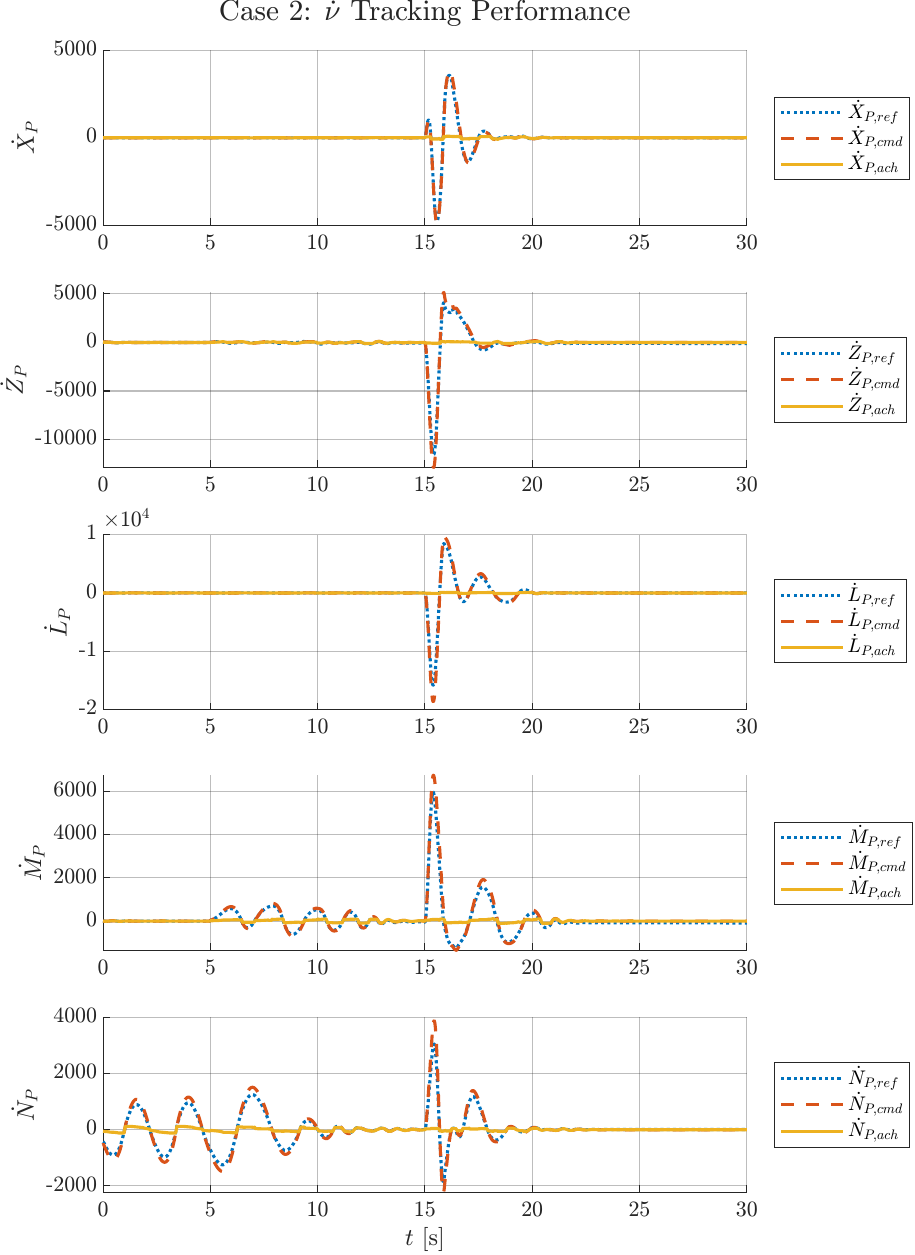}
    \caption{Controller 1, Case 2: $\pseudodot$ Tracking}
    \label{fig:ctrl1_case2_nu_dot}
\end{figure}

\begin{figure}
    \centering
    \includegraphics{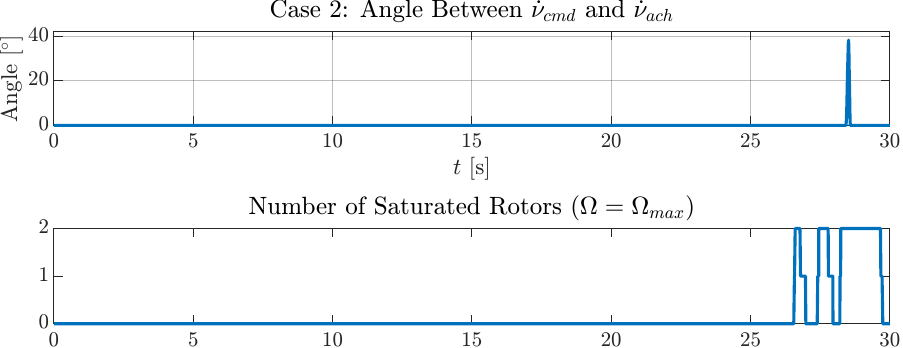}
    \caption{Controller 1, Case 2: Angle between $\pseudodot_{cmd}$ and $\pseudodot_{ach}$}
    \label{fig:ctrl1_case2_nu_dot_angle}
\end{figure}

\begin{figure}
    \centering
    \includegraphics{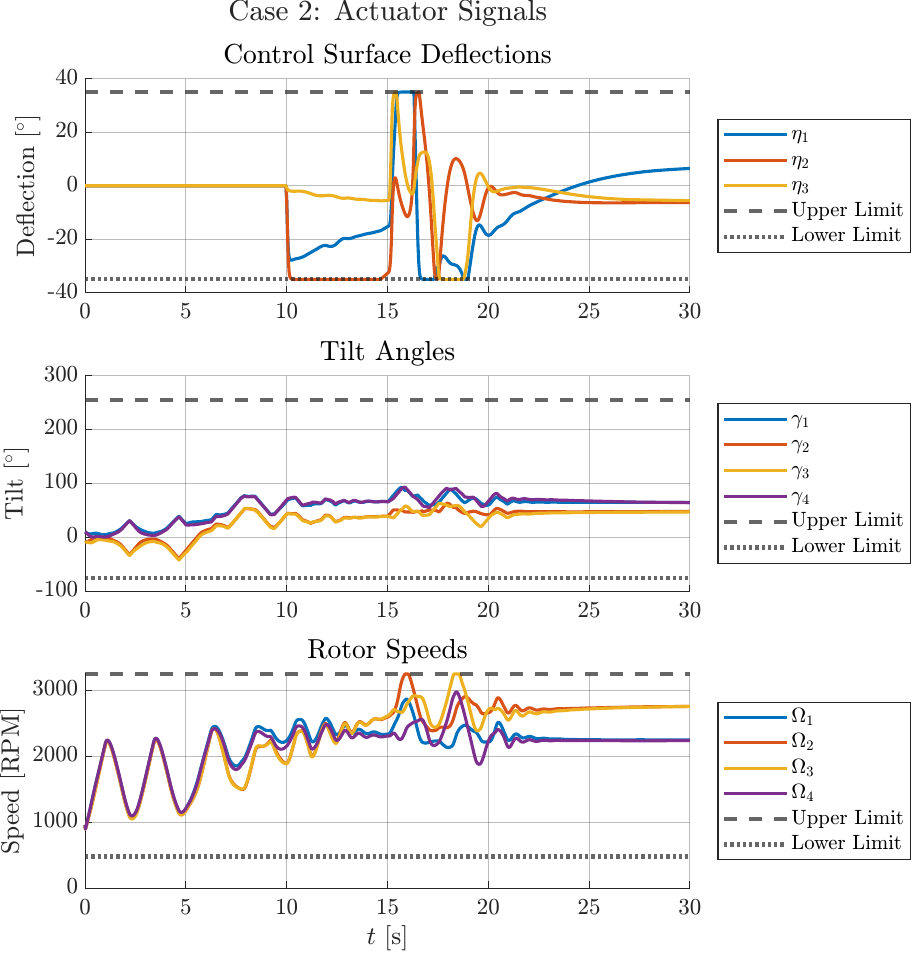}
    \caption{Controller 1, Case 2: Actuator Positions}
    \label{fig:ctrl1_case2_actuators}
\end{figure}

\subsection{Controller 2}
In the outer loop tracking plots (\cref{fig:ctrl2_case2_outer_loop}), the achieved velocity $u_B$ is closely following the reference velocity $u_{B,ref}$ with only a minimal deviation at the moment the gust hits the airship.
The vertical velocity $w_C$ also tracks its reference signal $w_{C,ref}$ very accurately with only minor errors.
The yaw rate $r$ experiences a sudden excursion during the gust with rougly the same maximum amplitude compared to controller 1.
The excursion is recovered after around \qty{5}{\second} with one additional overshoot compared to the other controller.

For comparison, the undisturbed response, i.e. the same simulation without the gust, is shown for the outer loop in \cref{fig:ctrl2_case2_undisturbed_outer_loop}.

The inner loop tracking performance is shown in \cref{fig:ctrl2_case2_inner_loop}.
The horizontal specific force $f_x$ follows the references accurately.
The vertical specific force $f_z$ also exhibits good tracking throughout most of the simulation, however, during the gust the achieved value goes in the opposite direction of the commanded value. This is recovered quickly though, and for the rest of the simulation the achieved and reference value are in perfect agreement.
The angular velocity tracking shows short but significant spikes at the time of the gust, which are quickly recovered by the controller.

In the $\pseudodot$ tracking plots (\cref{fig:ctrl2_case2_nu_dot}), the discrepancy between the commanded and achieved values is clear during the gust. For the remainder of the simulation the commanded values are achieved by the system.

In \cref{fig:ctrl2_case2_nu_dot_angle} the angle between the commanded and achieved $\pseudodot$ is shown. The angle is zero throughout the simulation with a small exception at around $t=\qty{17}{\second}$, where the angle is approximately \ang{2.5}, which is still very small.
This point in time is coincident with the time that 2 rotors are saturated, reinforcing the previous observations.
Overall, this shows that the achieved and commanded pseudo control rate vector point in the same direction.

\Cref{fig:ctrl2_case2_actuators} illustrates the actuator response of controller 2 during the gust rejection scenario in case 2. Compared to controller 1, the control surface deflections are significantly more conservative, reaching only approximately half of their allowable range.
The tilt angles exhibit only minor deviations during the gust event. Instead, the majority of compensation is handled by the rotor speeds, where two rotors temporarily reach saturation.

\begin{figure}
    \centering
    \includegraphics{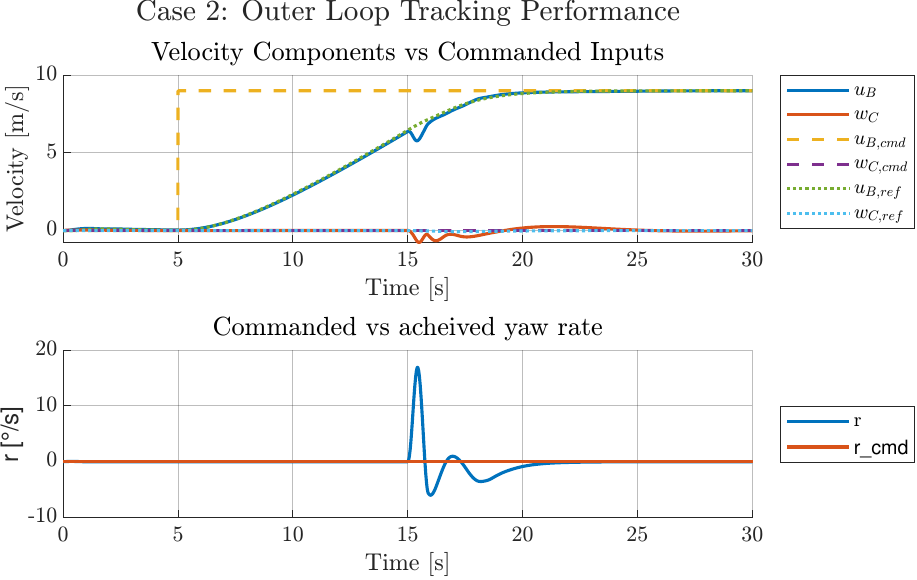}
    \caption{Controller 2, Case 2: Outer Loop Tracking}
    \label{fig:ctrl2_case2_outer_loop}
\end{figure}

\begin{figure}
    \centering
    \includegraphics{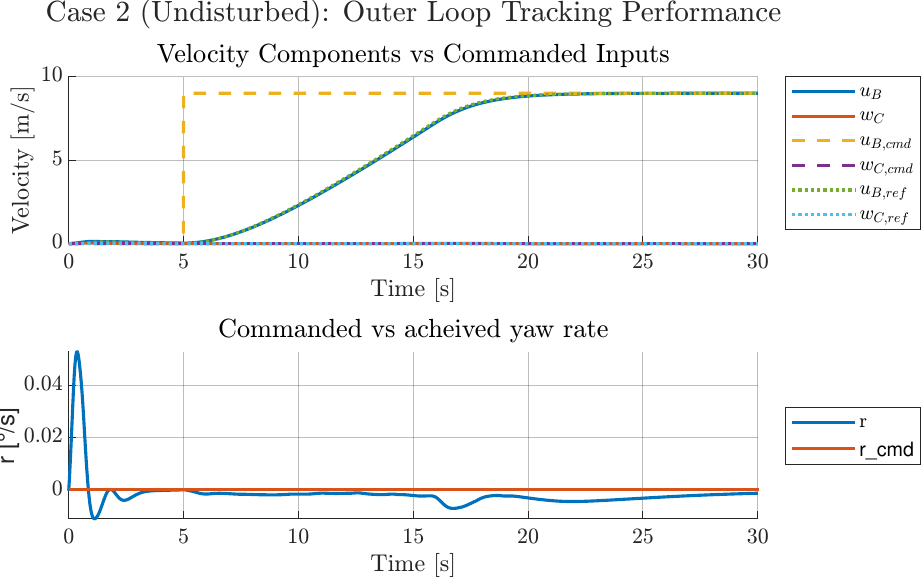}
    \caption{Controller 2, Case 2 (Undisturbed): Outer Loop Tracking}
    \label{fig:ctrl2_case2_undisturbed_outer_loop}
\end{figure}

\begin{figure}
    \centering
    \includegraphics{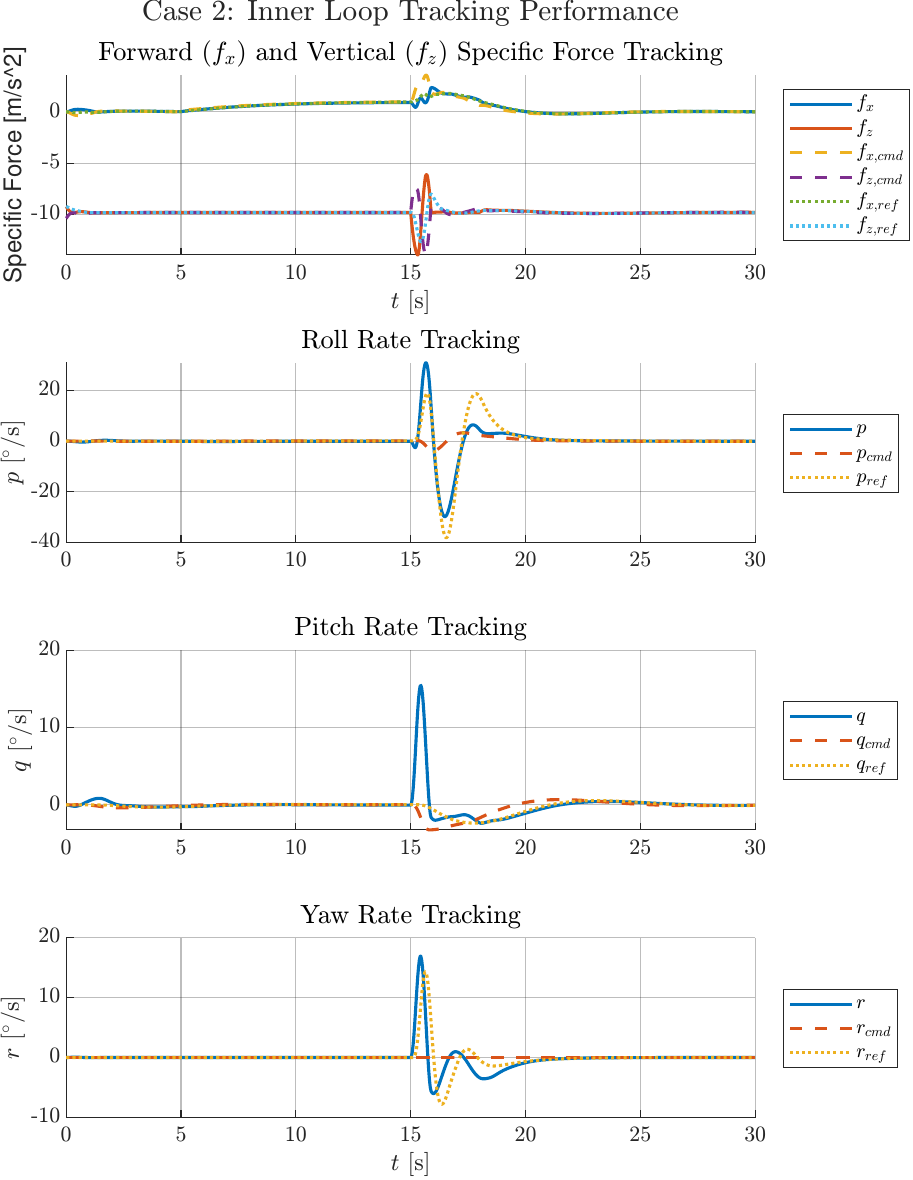}
    \caption{Controller 2, Case 2: Inner Loop Tracking}
    \label{fig:ctrl2_case2_inner_loop}
\end{figure}

\begin{figure}
    \centering
    \includegraphics{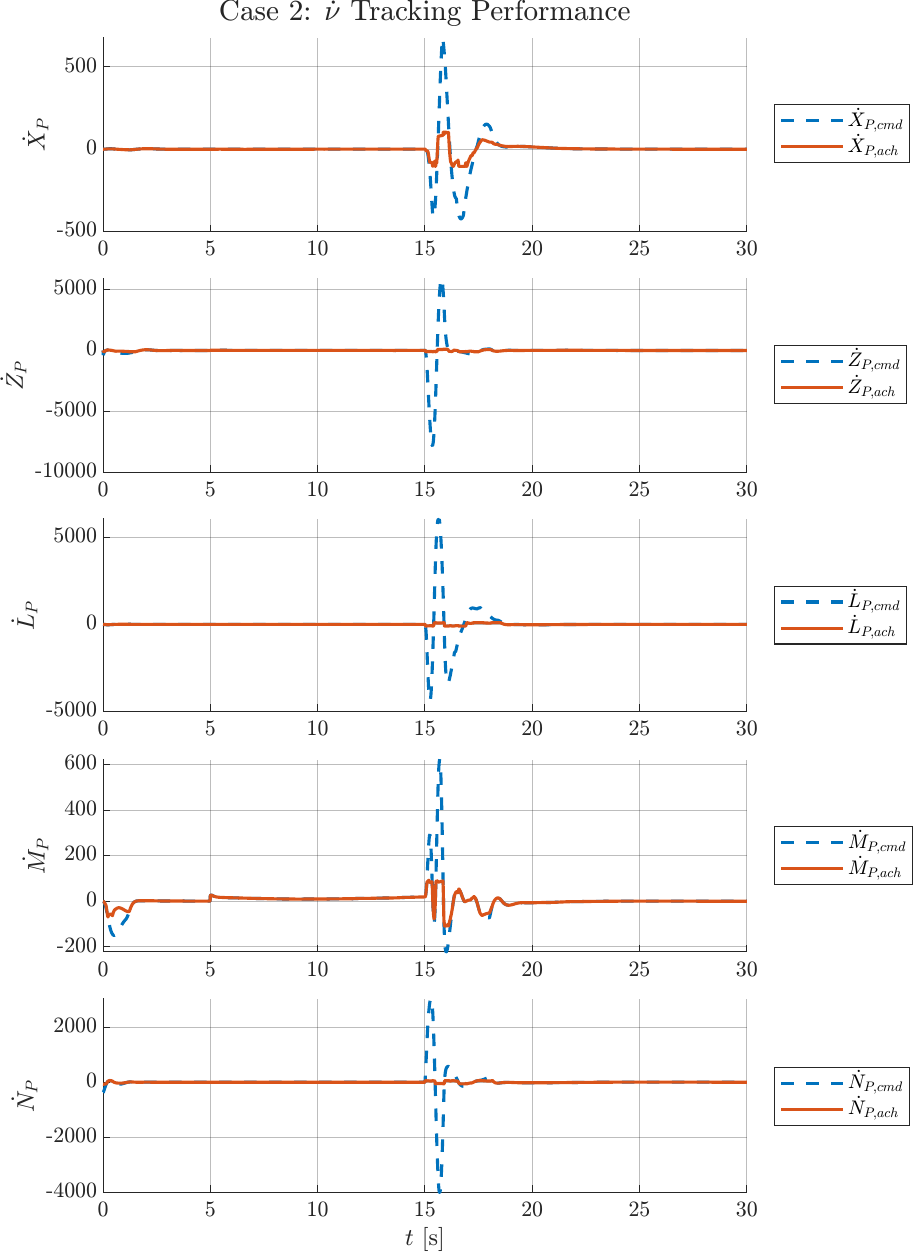}
    \caption{Controller 2, Case 2: $\pseudodot$ Tracking}
    \label{fig:ctrl2_case2_nu_dot}
\end{figure}

\begin{figure}
    \centering
    \includegraphics{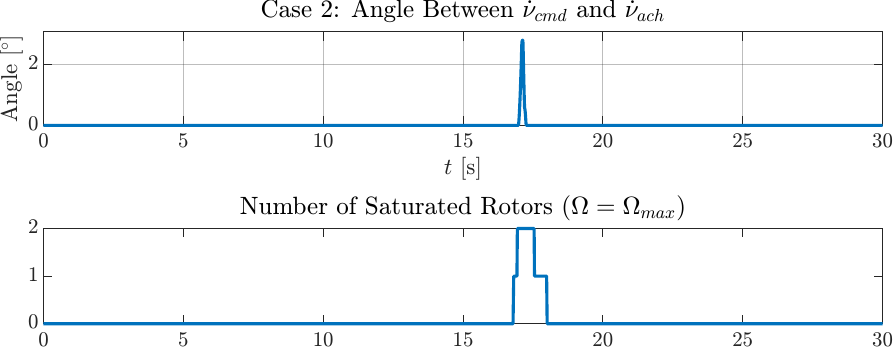}
    \caption{Controller 2, Case 2: Angle between $\pseudodot_{cmd}$ and $\pseudodot_{ach}$}
    \label{fig:ctrl2_case2_nu_dot_angle}
\end{figure}

\begin{figure}
    \centering
    \includegraphics{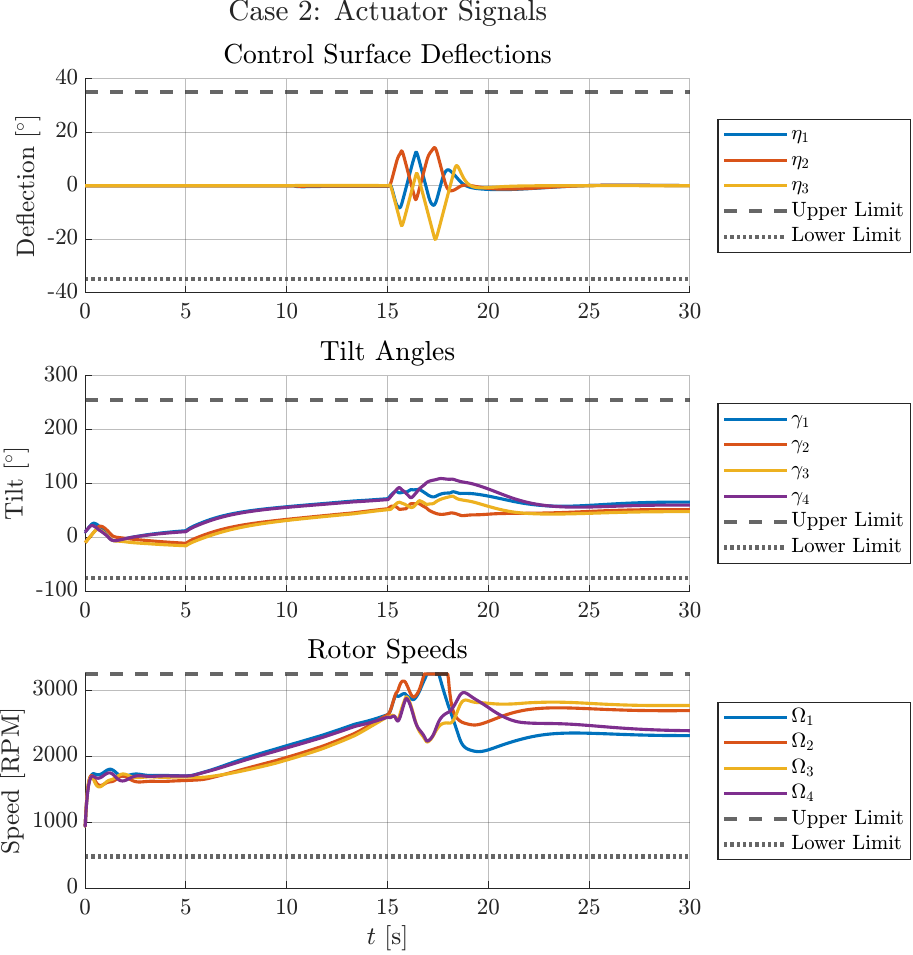}
    \caption{Controller 2, Case 2: Actuator Positions}
    \label{fig:ctrl2_case2_actuators}
\end{figure}

\section{Case 3: Athmospheric Turbulence}
In Case 3, the airship is subjected to the same demanding pilot input sequence as in Case 1, but with the added complexity of atmospheric turbulence modeled using the Dryden turbulence model. This scenario presents a particularly demanding test case, as the controller is required to maintain good tracking despite multiple challenges: maximum commanded forward speed, maximum climb and descent rates, and maximum turn maneuvers at the same time, while additionally being continuously perturbed by stochastic wind disturbances.

\subsection{Controller 1}
\cref{fig:ctrl1_case3_outer_loop} depicts the outer loop tracking performance, remains very stable despite the presence of turbulence. As shown in the first figure, the forward velocity $u$ tracks its commanded value accurately, with only minimal deviations due to wind disturbances. The vertical velocity $w$ also follows the command well, although minor deviations are observed during the left turn climb phase. The heading rate $\dot{\psi}$ tracking is also very accurate with only a minor deviation while initiating the left turn. Overall, there is good alignment with the command even during phases of combined yaw and vertical control, illustrating the controller's robustness.

The inner loop plots in \cref{fig:ctrl1_case3_inner_loop} show a similar picture. The tracking of the body-fixed velocity components remains accurate throughout the simulation with only minimal deviations noticable, particularly at around \qty{95}{\second}.
Roll, pitch, and yaw rates all track their respective commands with only minor deviations during peak maneuvering.
It it observed that the roll rate is noisier than the pitch rate and yaw rate, indicating higher influence of the turbulence due to a lower moment of inertia around the roll axis.

The $\pseudo$-tracking performance plots in \cref{fig:ctrl1_case3_nu} shows that the actual values $\pseudo_{ach}$ closely follow their references $\pseudo_{ref}$ in all axes throughout most of the flight.

In the $\dot{\nu}$-tracking plots (\cref{fig:ctrl1_case3_nu_dot}), the impact of turbulence is more pronounced.
The commanded and reference values are way higher than the capabilities of the plant, resulting in a larger discrepancy between the reference and the achieved values due to physical limitations.
Nevertheless, as it was shown in the previous plot, the $\pseudo$ tracking is still accurate.

\cref{fig:ctrl1_case3_nu_dot_angle} shows the angle between the  $\dot{\pseudo}_{cmd}$ and the  $\dot{\pseudo}_{ach}$ vector. Although the angle between the commanded and achieved rate vectors exhibits several spikes, these only occur for very short amount of times. Again it can be observed that these spikes only occur when the number of saturated rotors is larger than zero.

In \cref{fig:ctrl1_case3_actuators} the actuator signals are presented. The actuators show increased activity, especially for the control surfaces, as expected under turbulent conditions. Rotor speeds operate near their limits during transient events, yet stay within constraints. The control surfaces also show significantly more motion, reacting dynamically to compensate for external forces. The tilt angles move significantly less than the other two actuators, which is expected due to their lower bandwith.

In conclusion, the controller demonstrates good performance even in the case of atmospheric turbulence.

\begin{figure}
    \centering
    \includegraphics{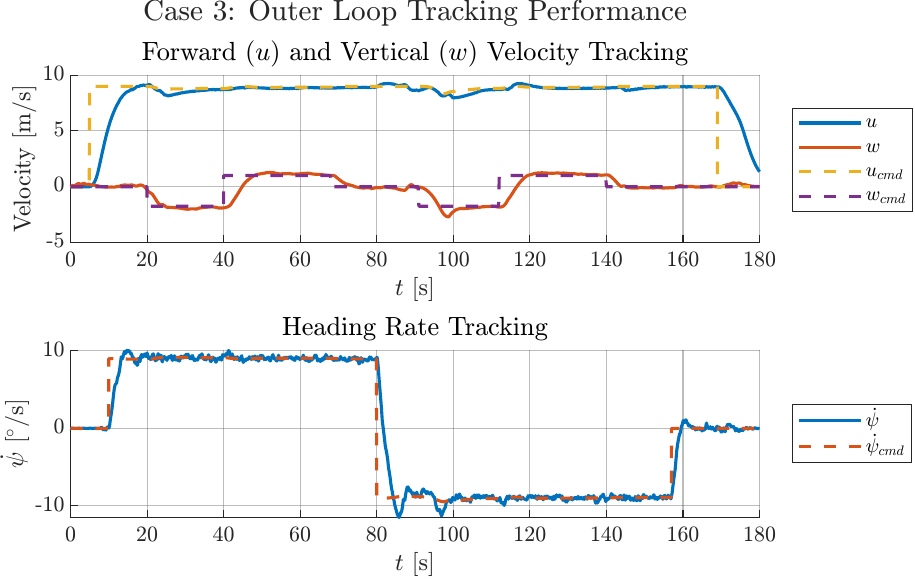}
    \caption{Controller 1, Case 3: Outer Loop Tracking}
    \label{fig:ctrl1_case3_outer_loop}
\end{figure}

\begin{figure}
    \centering
    \includegraphics{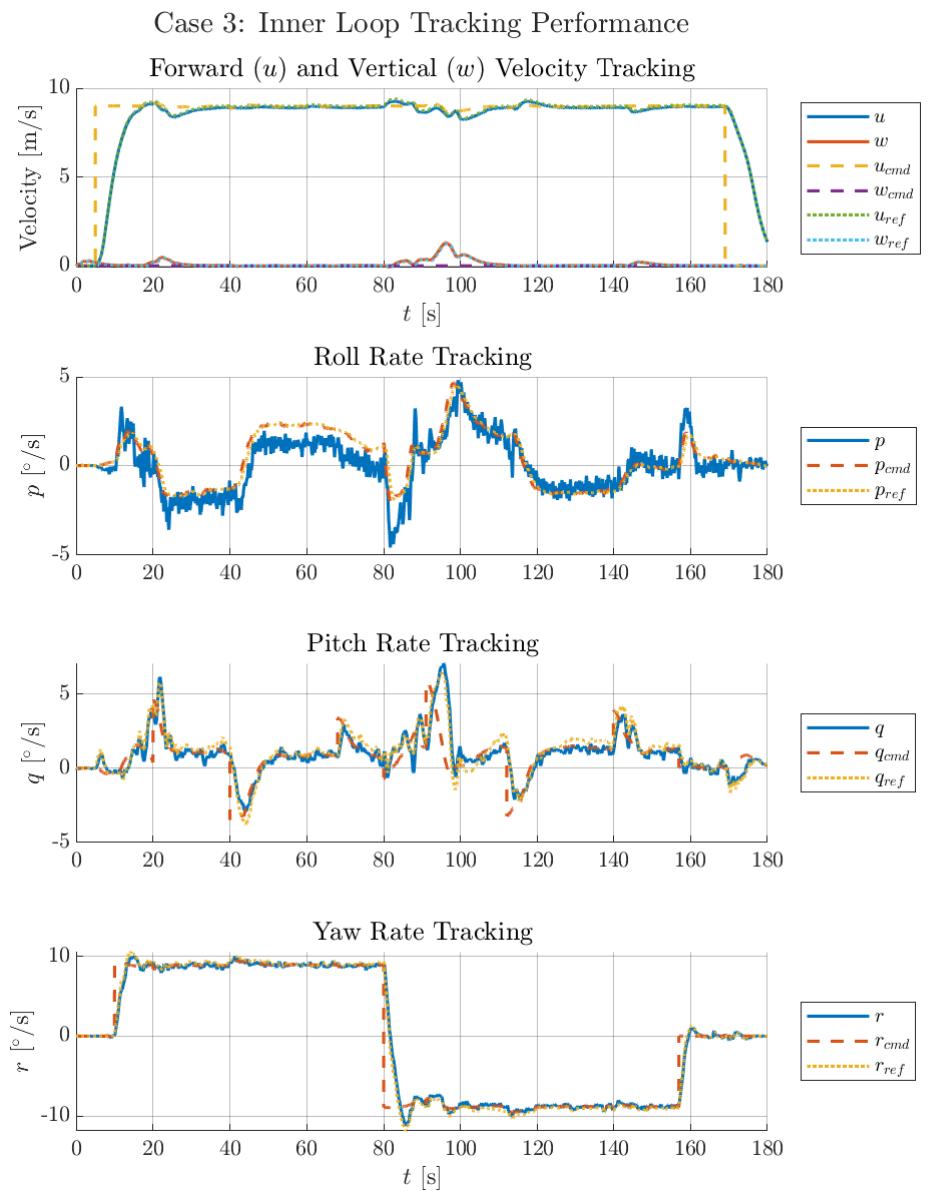}
    \caption{Controller 1, Case 3: Inner Loop Tracking}
    \label{fig:ctrl1_case3_inner_loop}
\end{figure}

\begin{figure}
    \centering
    \includegraphics{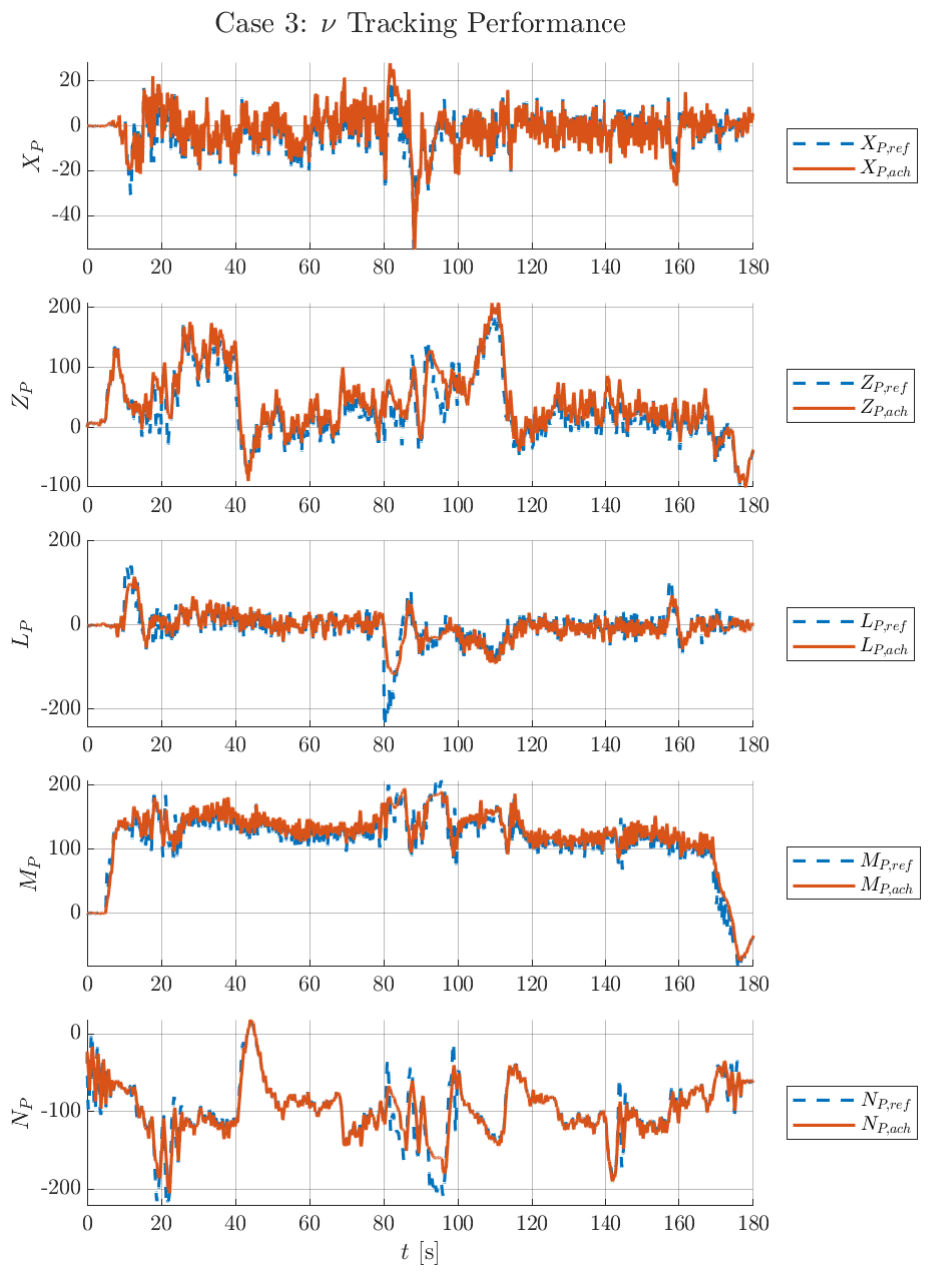}
    \caption{Controller 1, Case 3: $\pseudo$ Tracking}
    \label{fig:ctrl1_case3_nu}
\end{figure}

\begin{figure}
    \centering
    \includegraphics{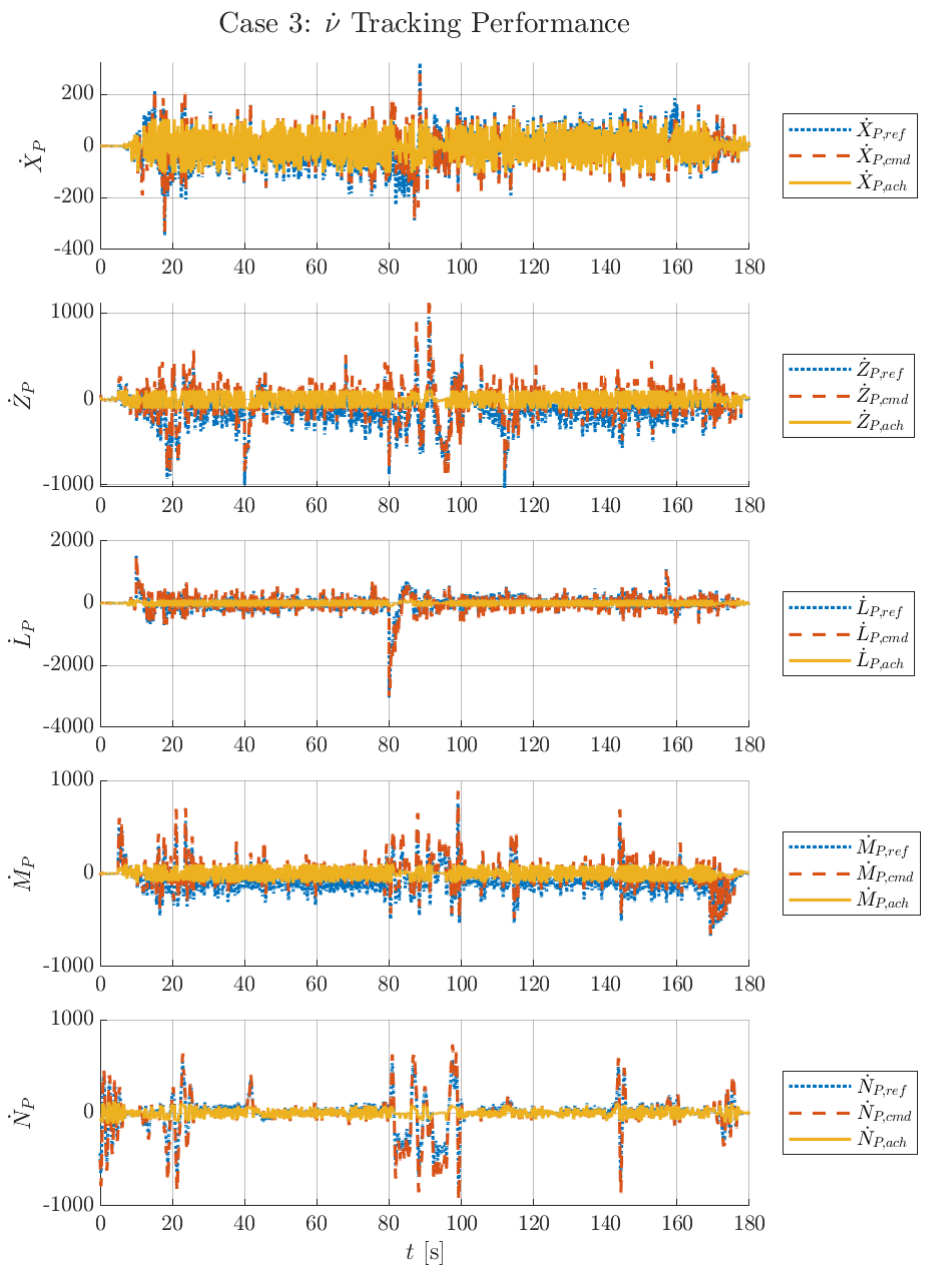}
    \caption{Controller 1, Case 3: $\pseudodot$ Tracking}
    \label{fig:ctrl1_case3_nu_dot}
\end{figure}

\begin{figure}
    \centering
    \includegraphics{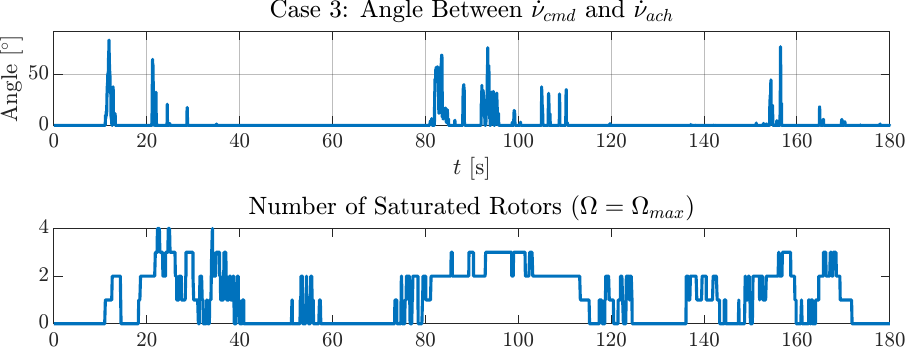}
    \caption{Controller 1, Case 3: Angle between $\pseudodot_{cmd}$ and $\pseudodot_{ach}$}
    \label{fig:ctrl1_case3_nu_dot_angle}
\end{figure}

\begin{figure}
    \centering
    \includegraphics{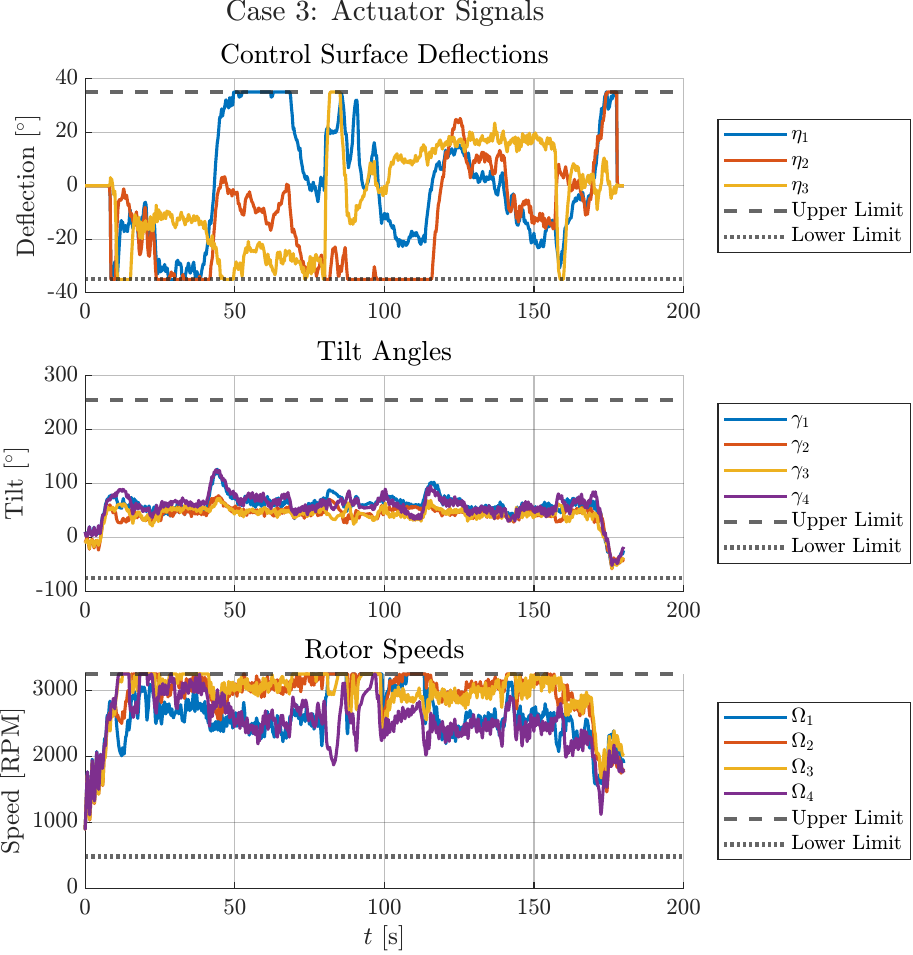}
    \caption{Controller 1, Case 3: Actuator Positions}
    \label{fig:ctrl1_case3_actuators}
\end{figure}

\subsection{Controller 2}
Starting with the outer loop tracking in \cref{fig:ctrl2_case3_outer_loop}, the forward velocity $u_B$ shows a generally good agreement with the reference value throughout the simulation, although with slightly more error than the first controller especially during the middle of the simulation. Nevertheless, it returns to nominal conditions after the short turbulence-induced disturbance at $t=\qty{100}{\second}$.
The vertical velocity $w_C$ demonstrates satisfactory performance during level flight and during descent.
During climb maneuvers larger oscillations around the reference signal are observed.
The yaw rate $r$ overall tracks well, but with slightly higher oscillations and deviations during turns compared to the first controller.

Examining the inner loop in \cref{fig:ctrl2_case3_inner_loop}, the specific force tracking in the $f_x$ and $f_z$ directions still follows the reference fairly well.
The roll rate $p$ is quite noisy and often shows larger deviations to the reference signal.
The pitch rate tracking is more accurate, however again there is some lag between the achieved and the reference signal. Additionally, at around $t=\qty{100}{\second}$, the pitch rate overshoots its reference signal during some oscillations. This is recovered shortly after.
The yaw rate tracking was already discussed in the outer loop.

The $\pseudodot$ tracking performance in \cref{fig:ctrl2_case3_nu_dot} shows more noise compared to the previous case which is explained by the athmospheric turbulence.
Overall, with an exception for the forward force rate of change, for most of the simulation the commanded signals are higher than the airship can achieve, due to physical limitations of the propulsion system. A similar observation was made for controller 1.

In \cref{fig:ctrl2_case3_nu_dot_angle} the angle between the commanded and achieved $\pseudodot$ vector is shown.
The angle is non-zero for most of the simulation. During most of the simulation, one or more rotors are saturated.
There is one angle deviation at around \qty{10}{\second} where no rotor is saturated.

Finally, the actuator behavior is shown in \cref{fig:ctrl2_case3_actuators}. The control surface deflections remain moderate, without reaching their respective limits through the simulation.
Similarly, the tilt angles stay within their allowed bounds and shows less activity than the other two actuators.
The rotor speeds show greater activity, and the aft right ($\Omega_2$) and front left ($\Omega_4$) propellers saturate for prolonged periods of time.
The aft right stays at its maximum value from \qtyrange{20}{100}{\second} while the front left propeller is saturated from \qtyrange{120}{170}{\second}.

\begin{figure}
    \centering
    \includegraphics{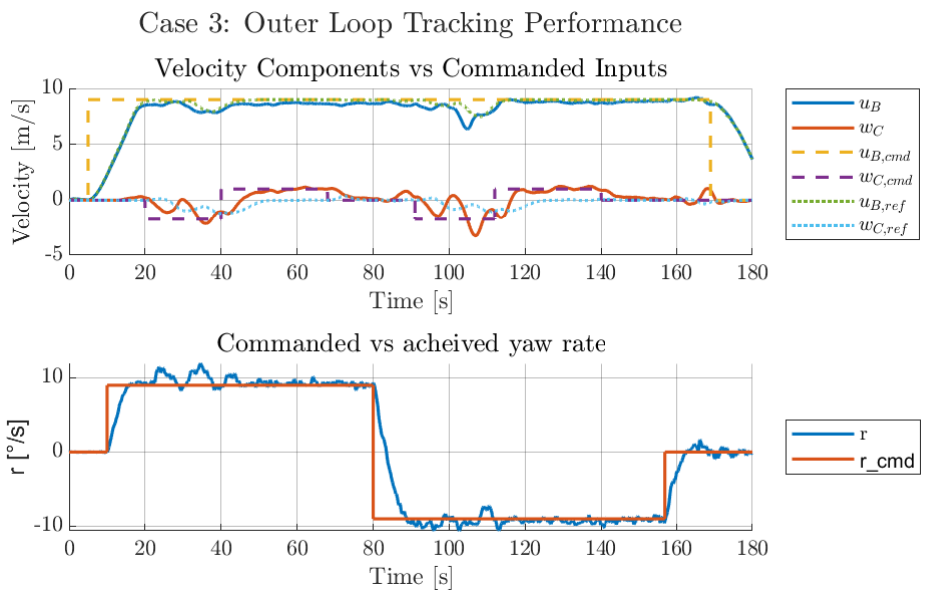}
    \caption{Controller 2, Case 3: Outer Loop Tracking}
    \label{fig:ctrl2_case3_outer_loop}
\end{figure}

\begin{figure}
    \centering
    \includegraphics{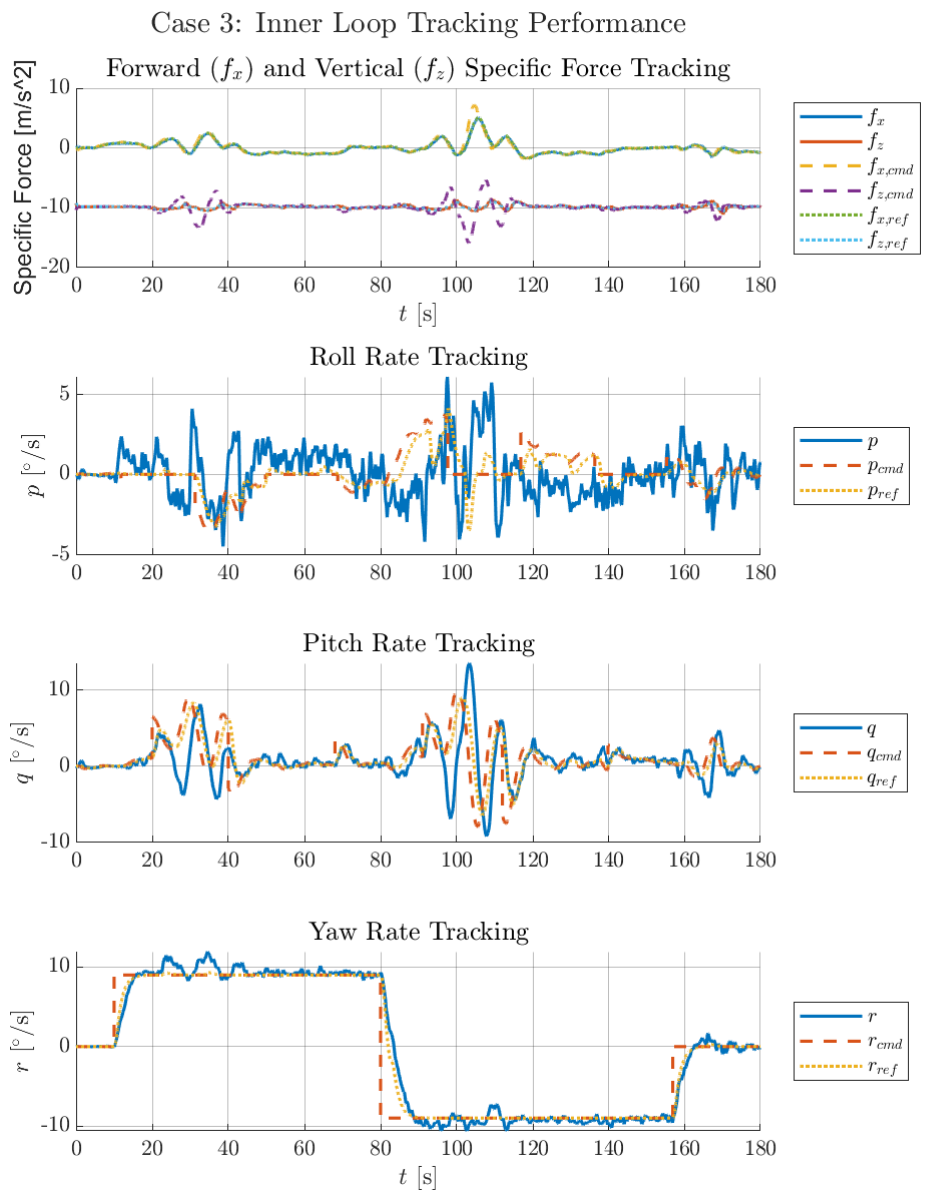}
    \caption{Controller 2, Case 3: Inner Loop Tracking}
    \label{fig:ctrl2_case3_inner_loop}
\end{figure}

\begin{figure}
    \centering
    \includegraphics{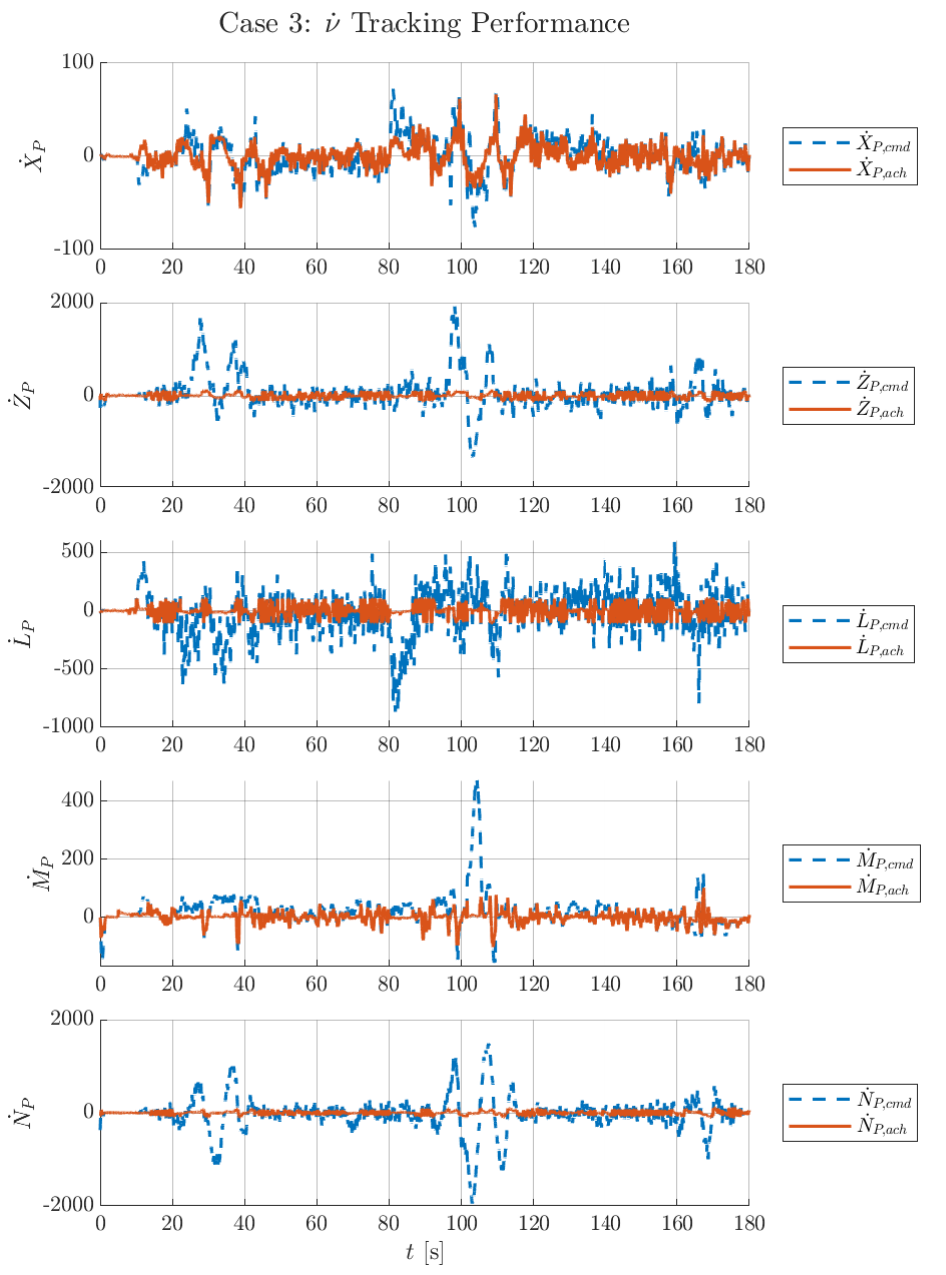}
    \caption{Controller 2, Case 3: $\pseudodot$ Tracking}
    \label{fig:ctrl2_case3_nu_dot}
\end{figure}

\begin{figure}
    \centering
    \includegraphics{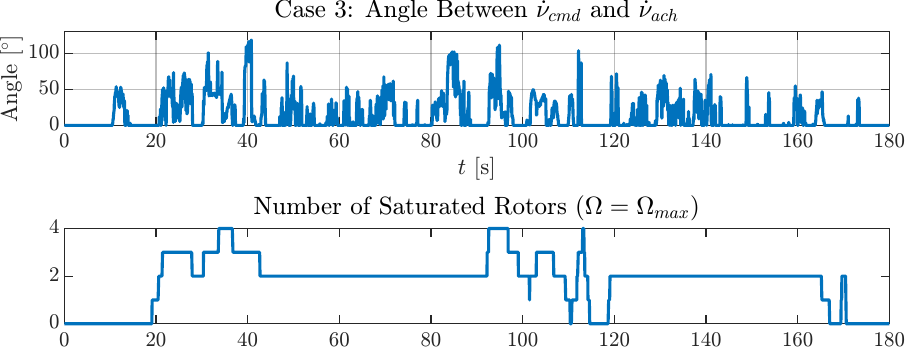}
    \caption{Controller 2, Case 3: Angle between $\pseudodot_{cmd}$ and $\pseudodot_{ach}$}
    \label{fig:ctrl2_case3_nu_dot_angle}
\end{figure}

\begin{figure}
    \centering
    \includegraphics{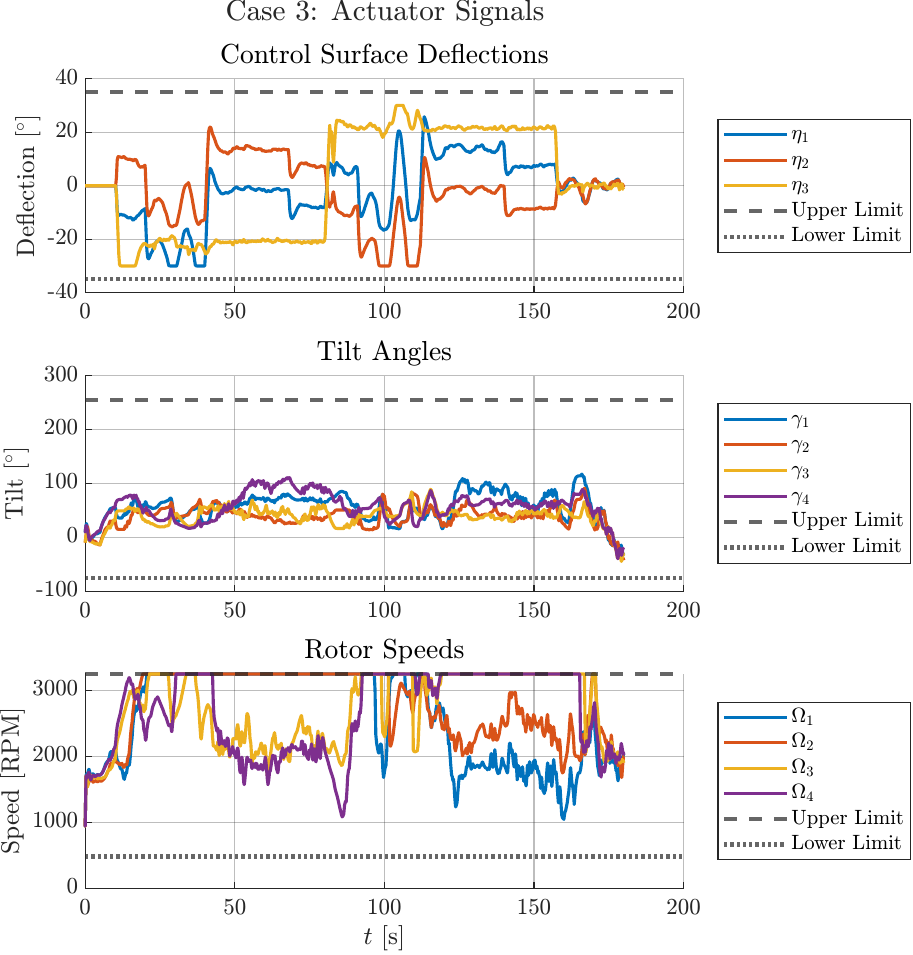}
    \caption{Controller 2, Case 3: Actuator Positions}
    \label{fig:ctrl2_case3_actuators}
\end{figure}

\section{Case 4: Model Uncertainties}
In Case 4, the objective is to evaluate the robustness of the control architecture under substantial model uncertainties.
The mass, inertia matrix, added mass, added inertia, buoyancy forces and control surface effectiveness used for control computations are all deliberately underestimated to a large extent.
Specifically, for case 4 uncertainties in the following variables are investigated (all at the same time):
\begin{itemize}
    \item Mass $m$ of the airship: real value \pmb{\qty{50}{\percent} higher} than assumed
    \item Added mass $\Matrix{M}_v$: real value \pmb{\qty{50}{\percent} higher} than assumed
    \item Inertia $\Matrix{J}$ of airship: real value \pmb{\qty{50}{\percent} higher} than assumed
    \item Added inertia $\Matrix{J}_V$: real value \pmb{\qty{50}{\percent} higher} than assumed
    \item Buoyancy force $\Vector{F}_B$ real value \pmb{\qty{50}{\percent} higher} than assumed
    \item Control surface moment derivatives $c_{l\eta}, c_{m\eta}, c_{n\eta}$: real value \pmb{\qty{50}{\percent} lower} than assumed
\end{itemize}
additionally, uncertainties in the actuators are taken into account.
Concretely, the tilt actuators and the electric motors driving the propellers are assumed faster than they really are.
In numbers, the rate limit $\dot{\gamma}_{i,max}$ is assumed \pmb{\qty{30}{\percent} higher} than it really is, the same is true for the rate limit $\dot{\Omega}_{i,max}$ for the propeller dynamics.

This simulates an almost worst-case mismatch between the plant and the controller model, as important physical properties such as mass and inertia are underestimated while the capabilities of the plant are overestimated.

To put the controllers to the test even more, the athomospheric turbulence introduced in Case 3 is also enabled in this test case.

\subsection{Controller 1}
Despite this aggressive deviation from nominal conditions, controller 1 demonstrates remarkable performance. Starting with the outer loop tracking performance in \cref{fig:ctrl1_case4_outer_loop}, the forward velocity $u$ follows its commanded reference $u_{cmd}$ almost perfectly, even though a slight overshoot occurs around $t = \qty{90}{\second}$. This deviation is handled smoothly, with the velocity returning quickly to its setpoint. The vertical velocity $w$ and heading rate $\dot{\varPsi}$ are also tracked very well throughout the scenario. These results illustrate that the outer loop controller can tolerate significant discrepancies in physical parameters without loss of stability or control authority.

The inner loop tracking also remains highly effective. \cref{fig:ctrl1_case4_inner_loop} shows that the velocity tracking in both the $u$ and $w$ directions aligns closely with their respective references $u_{ref}$ and $w_{ref}$. The yaw rate $r$ follows its respective reference commands almost perfectly.
The roll rate tracking is alsow very accurate, for the pitch rate tracking there are some minro deviations at around \qty{110}{\second}, which are however recovered quickly.
Overall, it is hard to spot differences compared to case 3, where the same situation without any model uncertainties was examined, indicating great robustness.

Moving to the tracking of the pseudo control vector $\pseudo$ in \cref{fig:ctrl1_case4_nu}, the system once again delivers great performance. The achieved forces and moments $\pseudo_{ach}$ track the references $\pseudo_{ref}$ with good accuracy, even during phases of high control demand.

In the $\dot{\pseudo}$-tracking plots (\cref{fig:ctrl1_case4_nu_dot}), a similar situation as in case 3 is present, as the commanded and reference values are higher than the plants capabilities.
Nevertheless, tracking of $\pseudo$ is still possible and accurate, as shown.

Next, \cref{fig:ctrl1_case4_nu_dot_angle} shows that the angle between $\dot{\pseudo}_{cmd}$ and $\dot{\pseudo}_{ach}$ remains close to zero for the majority of the time, indicating that the achieved control directions align well with the intended ones. However, a few brief spikes can be observed. Again, these spikes coincide with periods during which one or more rotors are saturated.

Finally, the actuator signals for controller 1 in case 4 are shown in \cref{fig:ctrl1_case4_actuators}. As evident from the figure, all actuators remain within their respective saturation limits throughout the simulation. The tilt angles exhibit minimal variation, indicating that their contribution is limited in this scenario. Instead, the primary actuation effort is handled by the control surfaces and rotor speeds. The control surfaces reach their maximum deflection only briefly on a few occasions. Notably, the aft rotors operate at their maximum velocity for extended durations, particularly between approximately $t = \qty{80}{\second}$ and $t = \qty{120}{\second}$. Despite this rotor saturation, the control objectives are still successfully achieved by leveraging the remaining available actuators, as shown in both inner and outer loop tracking.

In summary, Case 4 validates the robustness of the airship's control architecture against extreme modeling inaccuracies. The ability to maintain high-quality tracking with such large deviations in physical parameters shows the robustness of the designed controller.

\begin{figure}
    \centering
    \includegraphics{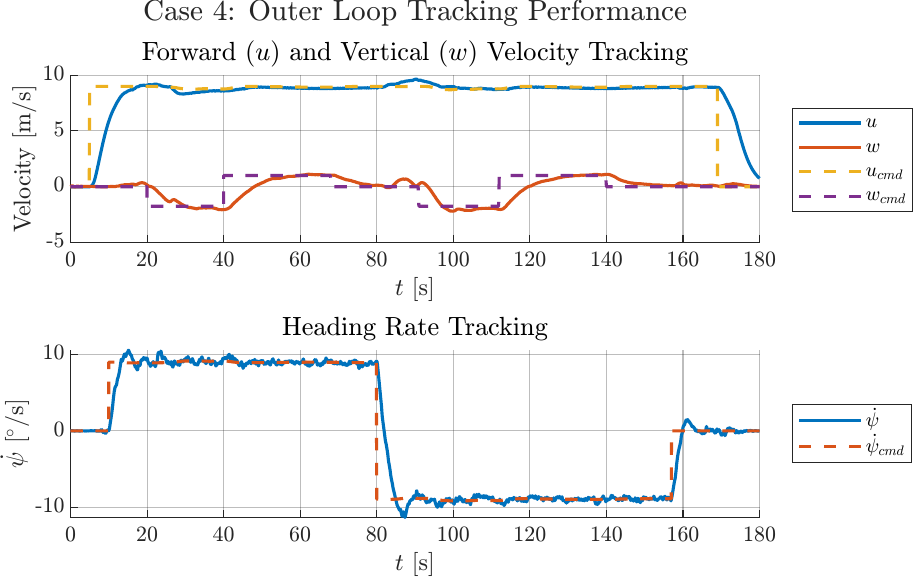}
    \caption{Controller 1, Case 4: Outer Loop Tracking}
    \label{fig:ctrl1_case4_outer_loop}
\end{figure}

\begin{figure}
    \centering
    \includegraphics{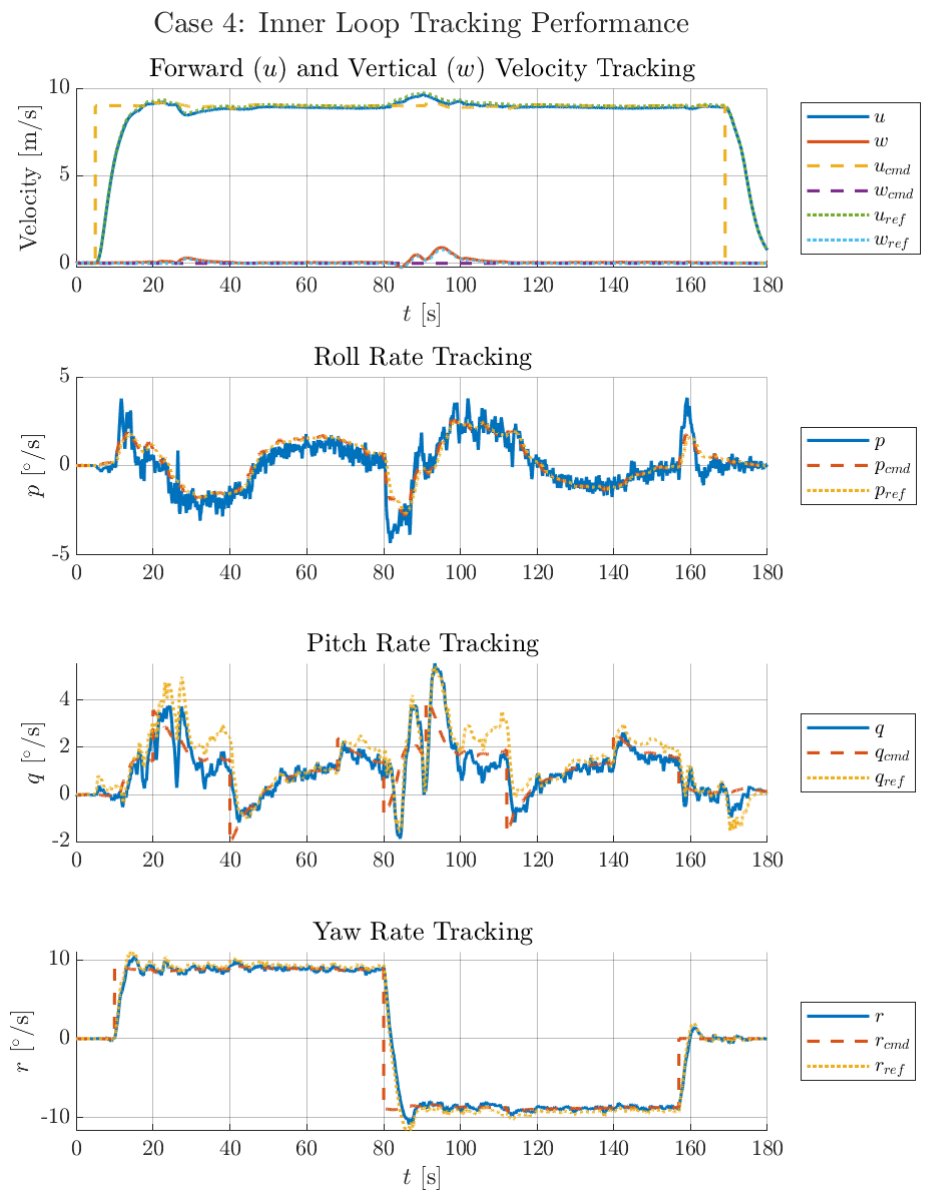}
    \caption{Controller 1, Case 4: Inner Loop Tracking}
    \label{fig:ctrl1_case4_inner_loop}
\end{figure}

\begin{figure}
    \centering
    \includegraphics{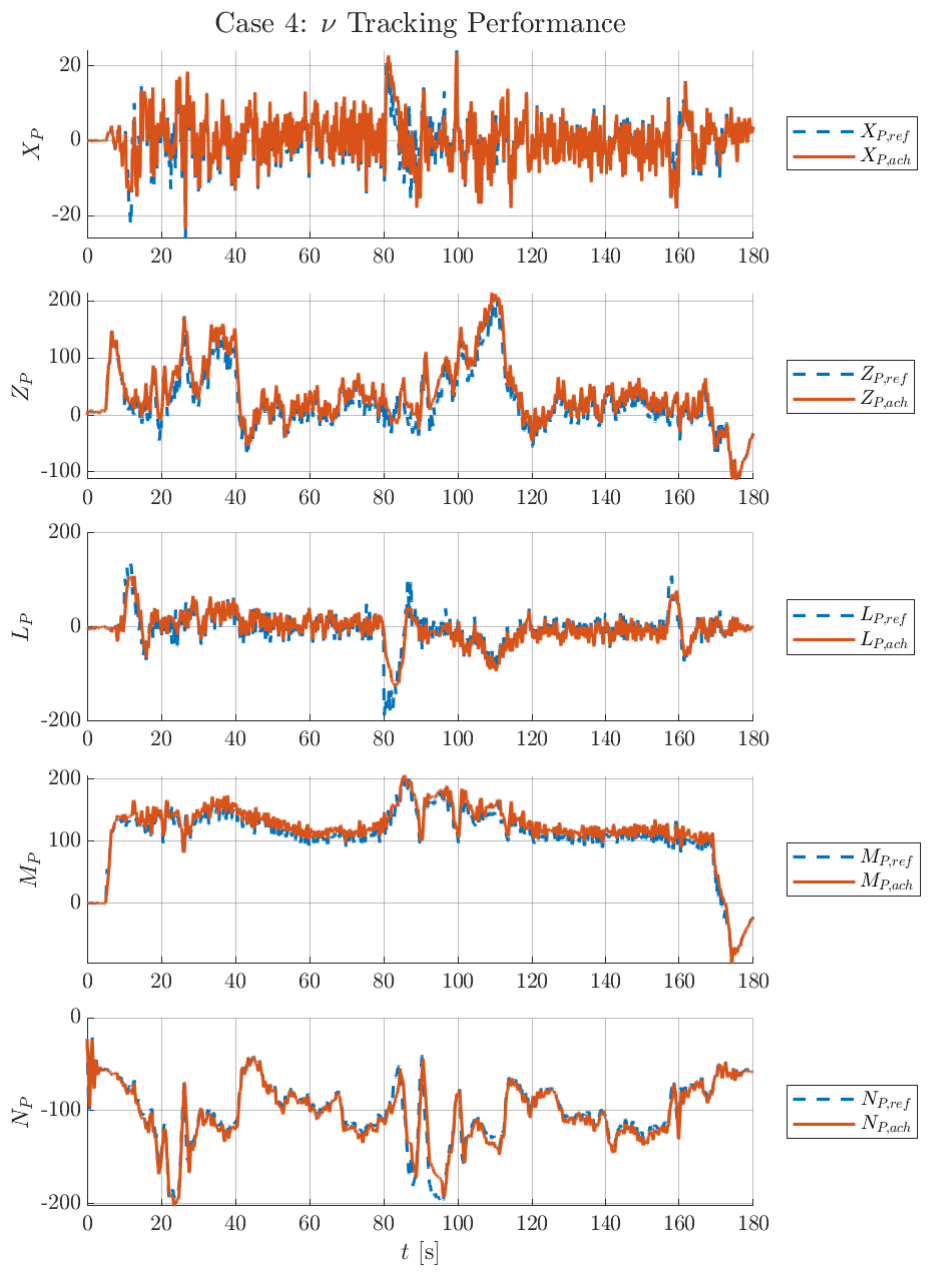}
    \caption{Controller 1, Case 4: $\pseudo$ Tracking}
    \label{fig:ctrl1_case4_nu}
\end{figure}

\begin{figure}
    \centering
    \includegraphics{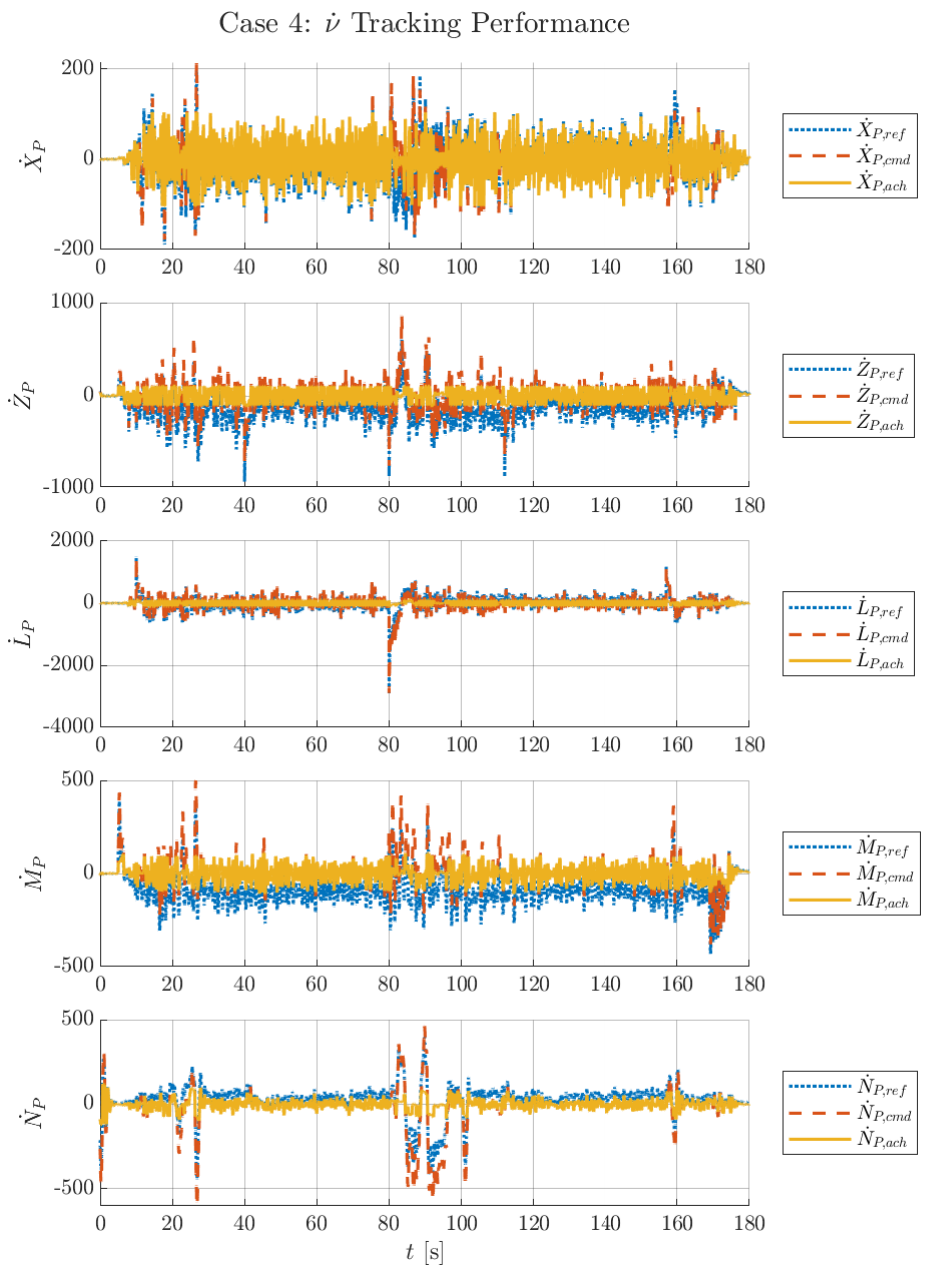}
    \caption{Controller 1, Case 4: $\pseudodot$ Tracking}
    \label{fig:ctrl1_case4_nu_dot}
\end{figure}

\begin{figure}
    \centering
    \includegraphics{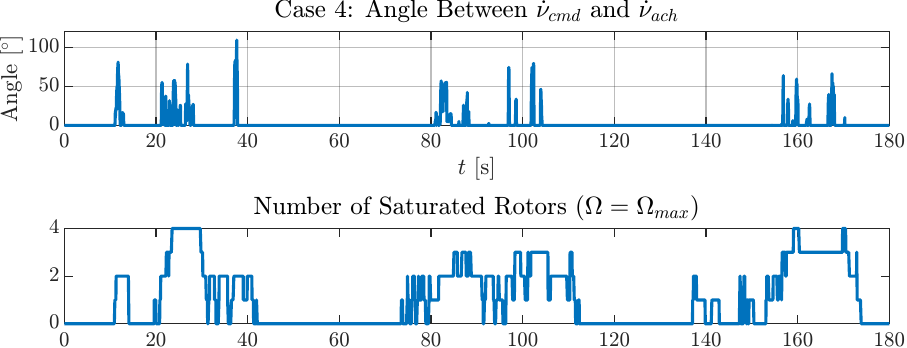}
    \caption{Controller 1, Case 4: Angle between $\pseudodot_{cmd}$ and $\pseudodot_{ach}$}
    \label{fig:ctrl1_case4_nu_dot_angle}
\end{figure}

\begin{figure}
    \centering
    \includegraphics{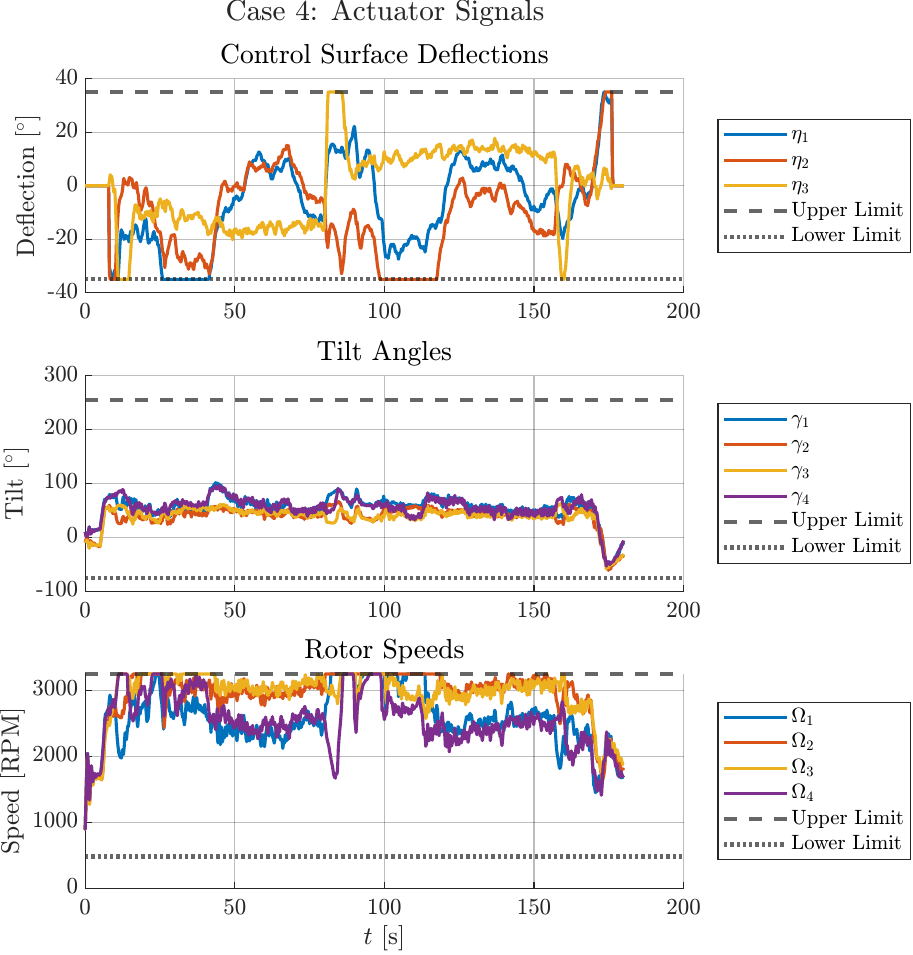}
    \caption{Controller 1, Case 4: Actuator Positions}
    \label{fig:ctrl1_case4_actuators}
\end{figure}

\subsection{Controller 2}
Continuing with controller 2 in \cref{fig:ctrl2_case4_outer_loop}, the forward velocity $u_B$ demonstrates stable and accurate tracking behavior.
Vertical velocity $w_C$ suffers more under these uncertainties. While it roughly follows the commanded changes, it exhibits oscillatory behavior and elevated tracking error throughout the mid and endsection of the simulation. It recovers at the very end, which is a positive sign, but overall tracking in this channel is weaker, although still stable throughout the simulation.

The yaw rate $r$ tracking is reasonable accurate during the first half of the simulation, where a right turn is performed.
During the second half of the simulation, while performing a left turn, the tracking is initally still accurate, however at around \qty{140}{\second} large deviations occur, which are recovered once the commanded signal returns back to \qty{0}{\degree\per\second}.

Continuing with the inner loop plots in \cref{fig:ctrl2_case4_inner_loop}, the forward specific force $f_x$ is in perfect agreement with its reference and command value for the whole simulation.
The vertical component $f_z$ generally follows the reference signal well, however some oscillations develop towards the end of the simulation during $t = \qtyrange{150}{160}{\second}$.

The angular rate tracking plots (for $p$, $q$, and $r$) show a mixed picture. Roll rate $p$ is noisy with notable deviations from the reference, particularly in the later half.
Pitch rate $q$ shows some oscillatory behavior and again lags behind its reference signal. At around \qty{50}{\second} the achieved pitch rate has the opposite sign of the command. After that, the tracking stabilizes.
Yaw rate $r$, already noted in the outer loop discussion, is still quite accurate during the first two thirds of the simulation, only towards the end larger deviations occur.

The commanded and achieved pseudo control rates are shown in \cref{fig:ctrl2_case4_nu_dot}.
In every axis, the commanded values are generally higher than the achieved values, particularly for $\dot{Z}_P$, $\dot{L}_P$, and $\dot{N}_P$.
For $\dot{X}_P$ and $\dot{M}_P$, discrepancies between the commanded and achieved values are the smallest of all 5 axes.

The angle between $\pseudodot_{cmd}$ and $\pseudodot_{ach}$ is depicted in \cref{fig:ctrl2_case4_nu_dot_angle}.
Compared to controller 1, there are large, sustained deviations for the majority of the simulation, often between \qtyrange{40}{85}{\degree}, which indicates a misalignment between the commanded and achieved pseudo control rate vector.
More than one rotor is saturated for most of the simulation. At the beginning and the end there are periods where the angle is larger than zero, although no rotors are saturated. This could not fully be resolved, yet the observation shall be noted.

The actuator plots are shown in \cref{fig:ctrl2_case4_actuators}. The control surface deflections stay well within bounds and avoid saturation.
The tilt angles vary more compared to the previous cases, especially at around $t = \qty{50}{\second}$ as well as at $t = \qty{120}{\second}$ and onward.

The rotors are the most utilized actuator during this simulation, with some rotors frequently hitting the maximum RPM two rotors almost hitting the minimum RPM during $t=\qtyrange{50}{70}{\second}$ and at $t = \qty{120}{\second}$.
The front left ($\Omega_4$) rotor is saturated at its maximum value during $t=\qtyrange{20}{60}{\second}$.
The aft right ($\Omega_2$) rotor is at its maximum RPM during $t=\qtyrange{80}{115}{\second}$.

\begin{figure}
    \centering
    \includegraphics{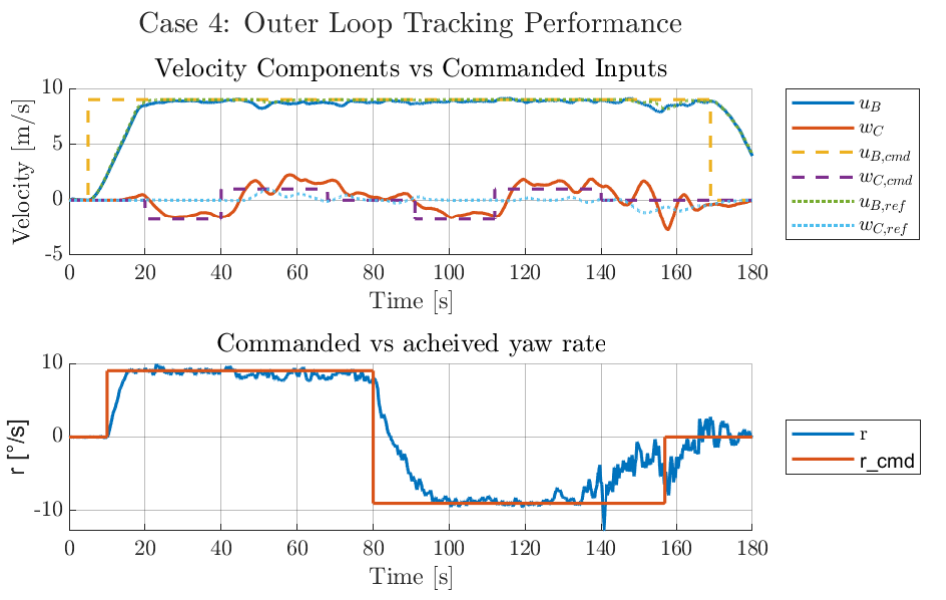}
    \caption{Controller 2, Case 4: Outer Loop Tracking}
    \label{fig:ctrl2_case4_outer_loop}
\end{figure}

\begin{figure}
    \centering
    \includegraphics{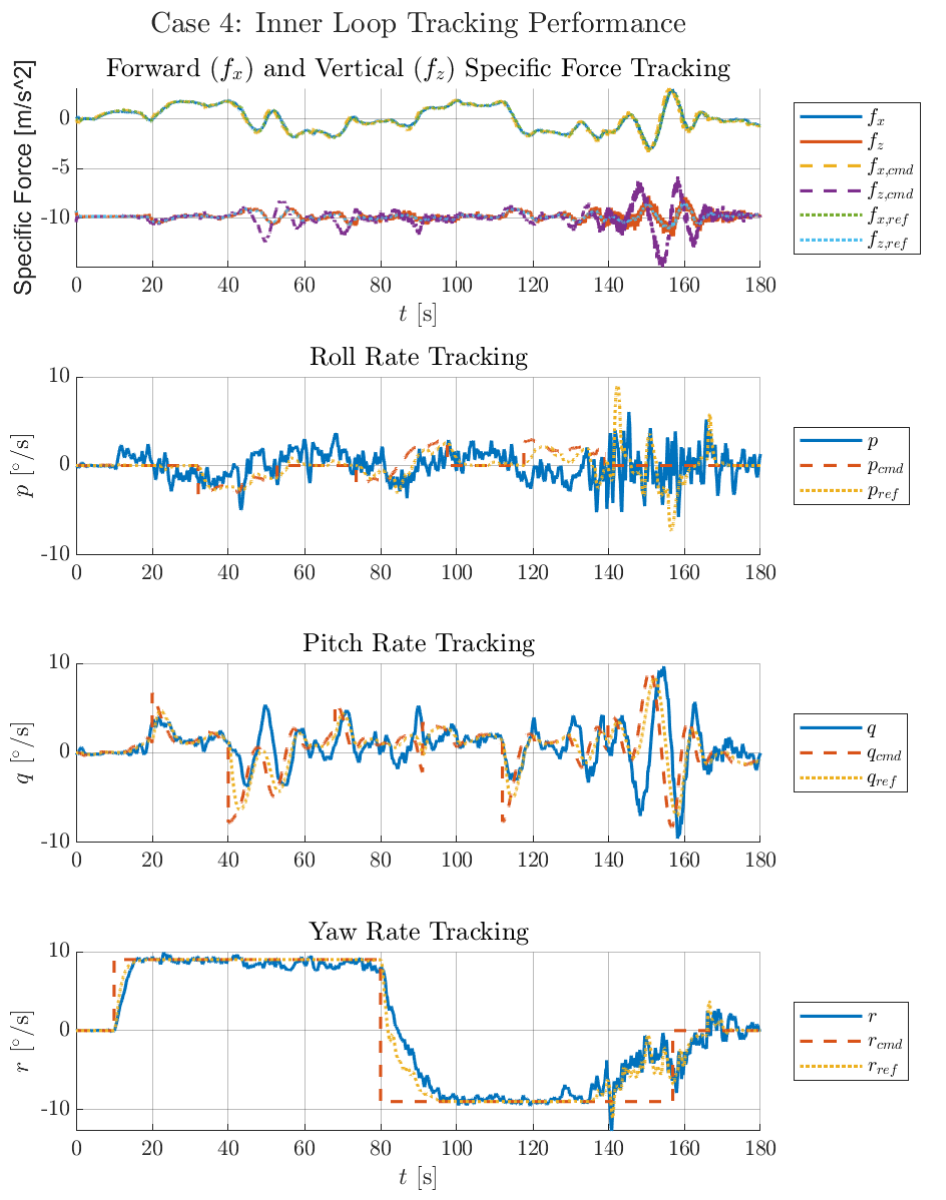}
    \caption{Controller 2, Case 4: Inner Loop Tracking}
    \label{fig:ctrl2_case4_inner_loop}
\end{figure}

\begin{figure}
    \centering
    \includegraphics{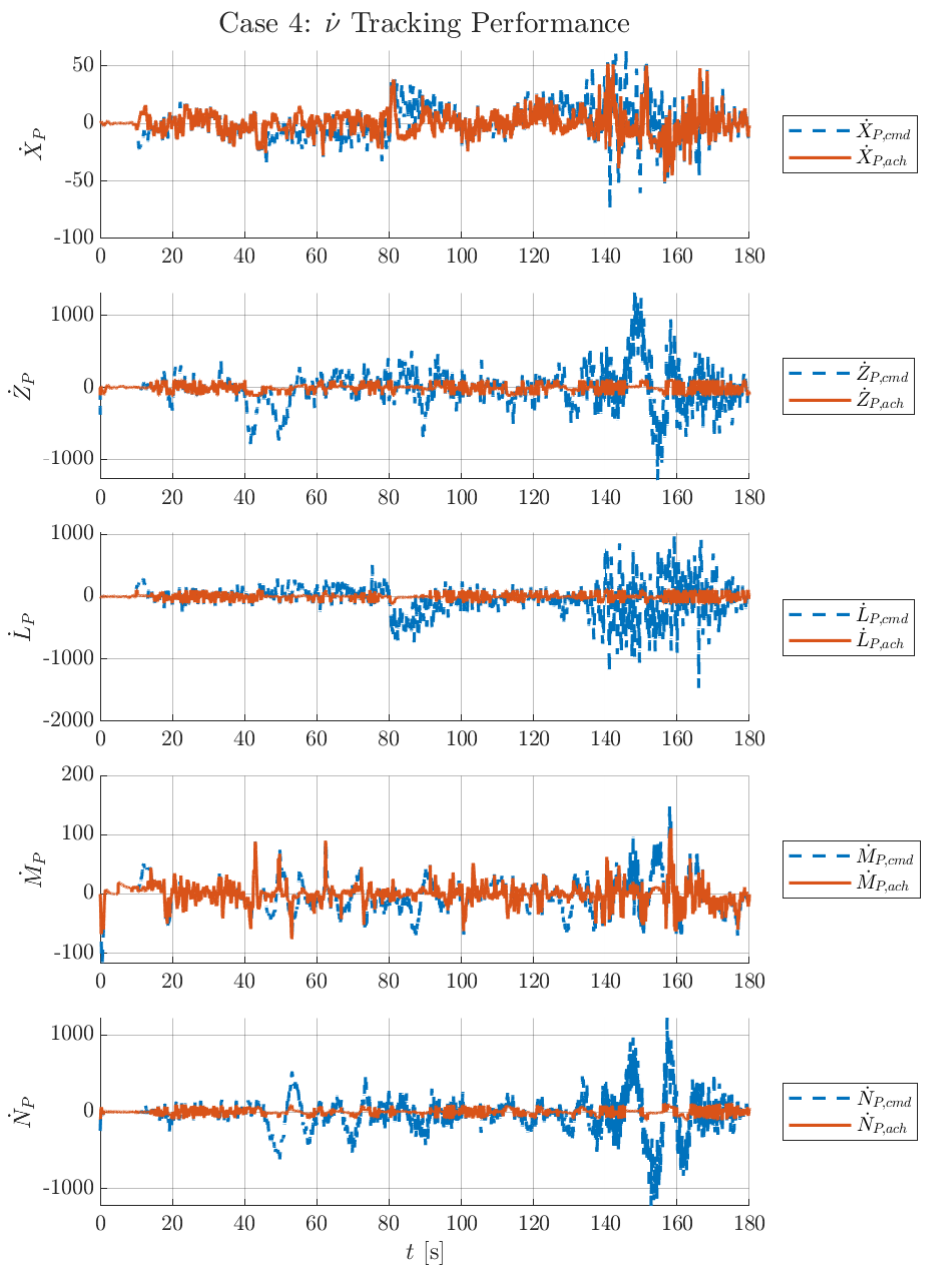}
    \caption{Controller 2, Case 4: $\pseudodot$ Tracking}
    \label{fig:ctrl2_case4_nu_dot}
\end{figure}

\begin{figure}
    \centering
    \includegraphics{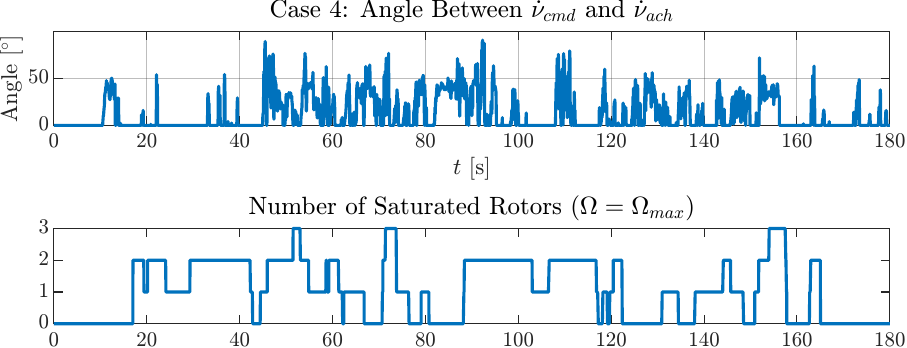}
    \caption{Controller 2, Case 4: Angle between $\pseudodot_{cmd}$ and $\pseudodot_{ach}$}
    \label{fig:ctrl2_case4_nu_dot_angle}
\end{figure}

\begin{figure}
    \centering
    \includegraphics{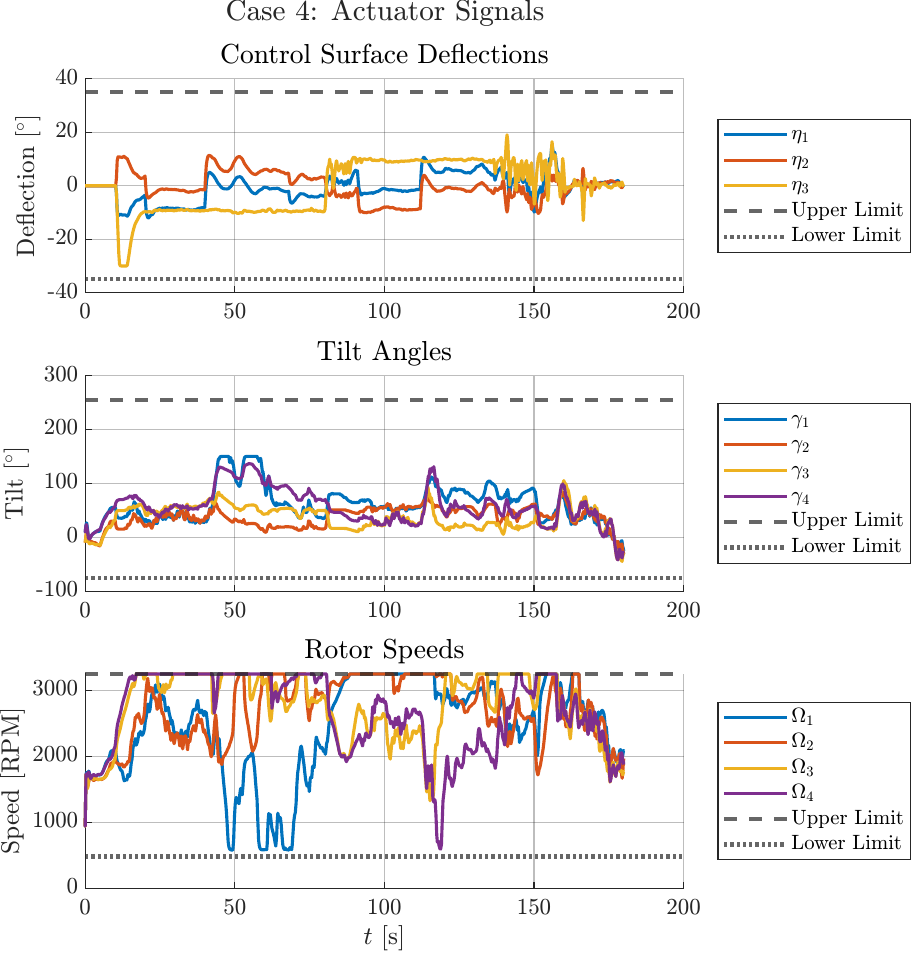}
    \caption{Controller 2, Case 4: Actuator Positions}
    \label{fig:ctrl2_case4_actuators}
\end{figure}
  \chapter{Conclusion}
This thesis presented the design, implementation, and evaluation of a robust flight control strategy for a vectored-thrust airship equipped with four tiltable propellers. The unique flight dynamics of airships, driven primarily by buoyant lift and characterized by high sensitivity to environmental disturbances and low control authority at low speeds, demanded a controller capable of operating reliably across the entire flight envelope, including hover, transition, and forward flight.

To address these requirements, an Extended Incremental Nonlinear Dynamic Inversion control strategy was employed. This controller combines an inner loop that controls linear and angular velocity with an outer loop responsible for controlling attitude and generating the required inner loop commands. The controller employs a control allocation strategy, ensuring efficient use of the tilt rotors, and incorporates a nullspace-based optimization to steer actuators toward desirable trim positions during steady flight.

The controller was rigorously tested and compared in a series of simulation-based scenarios to validate its performance and robustness. These test cases included:

\begin{itemize}
    \item Aggressive maneuvers requiring maximum performance from all control channels
    \item Gust rejection
    \item Operation in Dryden-modeled atmospheric turbulence
    \item Significant model mismatches simulating parameter uncertainties of up to \qty{50}{\percent}
\end{itemize}

Across all scenarios, the proposed controller demonstrated very good tracking performance, effective disturbance rejection, and strong robustness to model uncertainties, even under very demanding scenarios. In contrast, a second controller developed at the institute, based on direct inversion at the jerk level, showed slightly worse performance, particularly when subjected to turbulence and model uncertainties.

\end{mainContent}


\begin{backMatter}


  \makeBibliography{}

  \begin{appendix}
    \chapter{Additional Material}
\section{Derivation of the jerk-level equations of motion}\label{app:jerk}
\begin{figure}
    \centering
    \includegraphics[width=\linewidth]{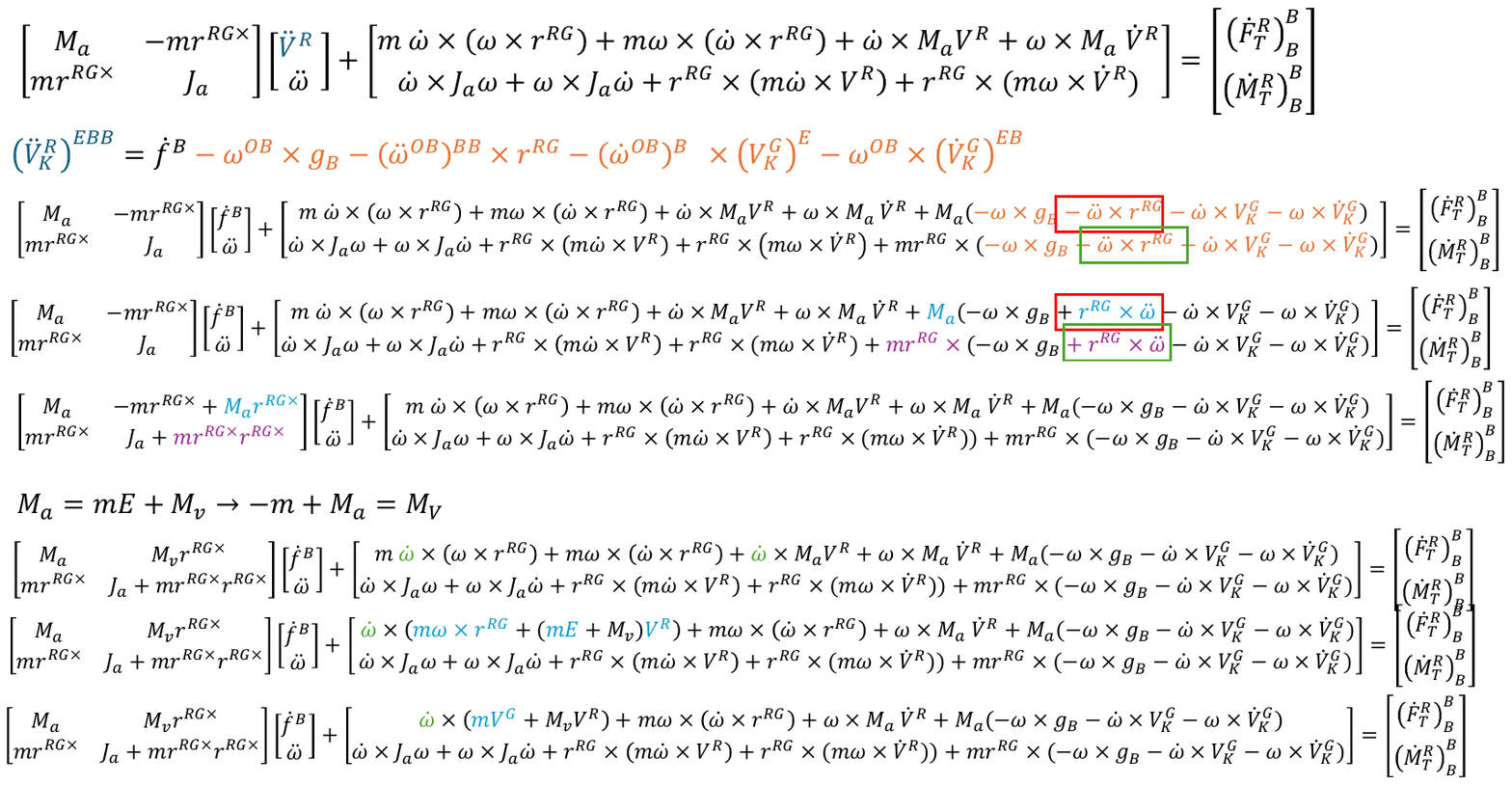}
    \caption{Derivation of the Jerk-Level Equation of Motion Part 1}
\end{figure}

\begin{figure}
    \centering
    \includegraphics[width=\linewidth]{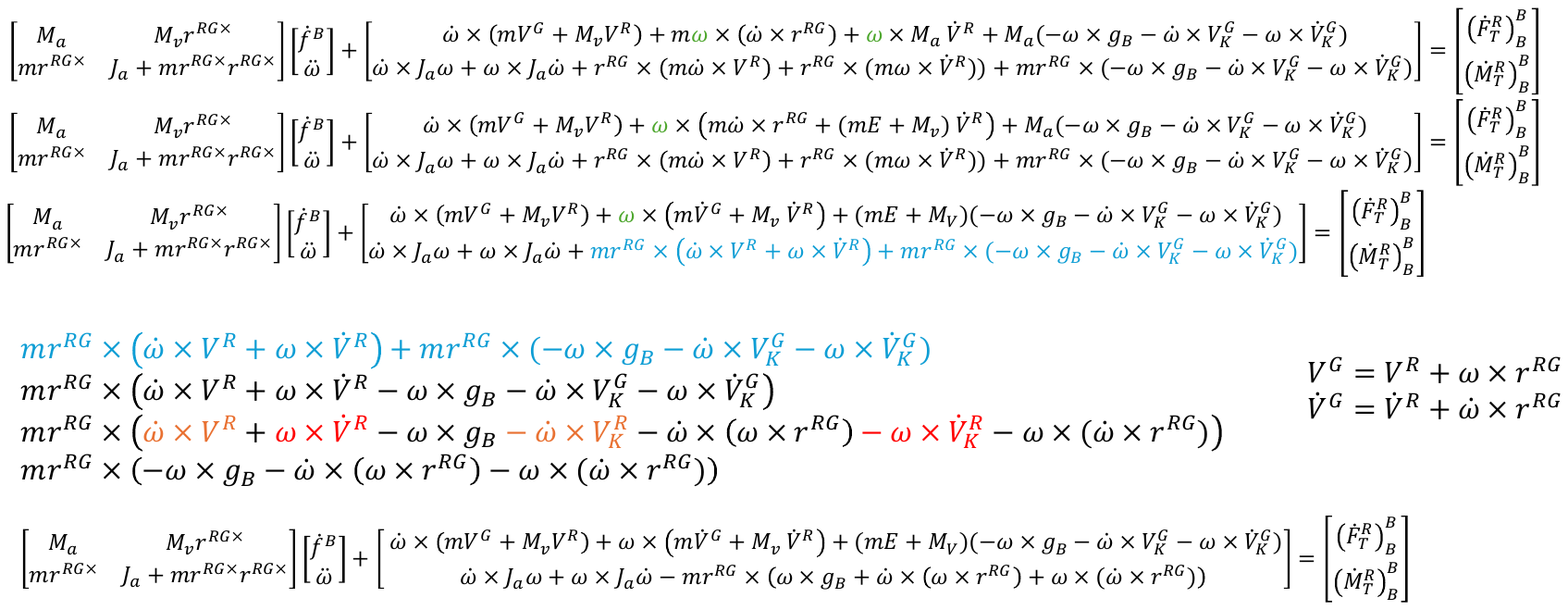}
    \caption{Derivation of the Jerk-Level Equation of Motion Part 2}
\end{figure}

\begin{figure}
    \centering
    \includegraphics[width=\linewidth]{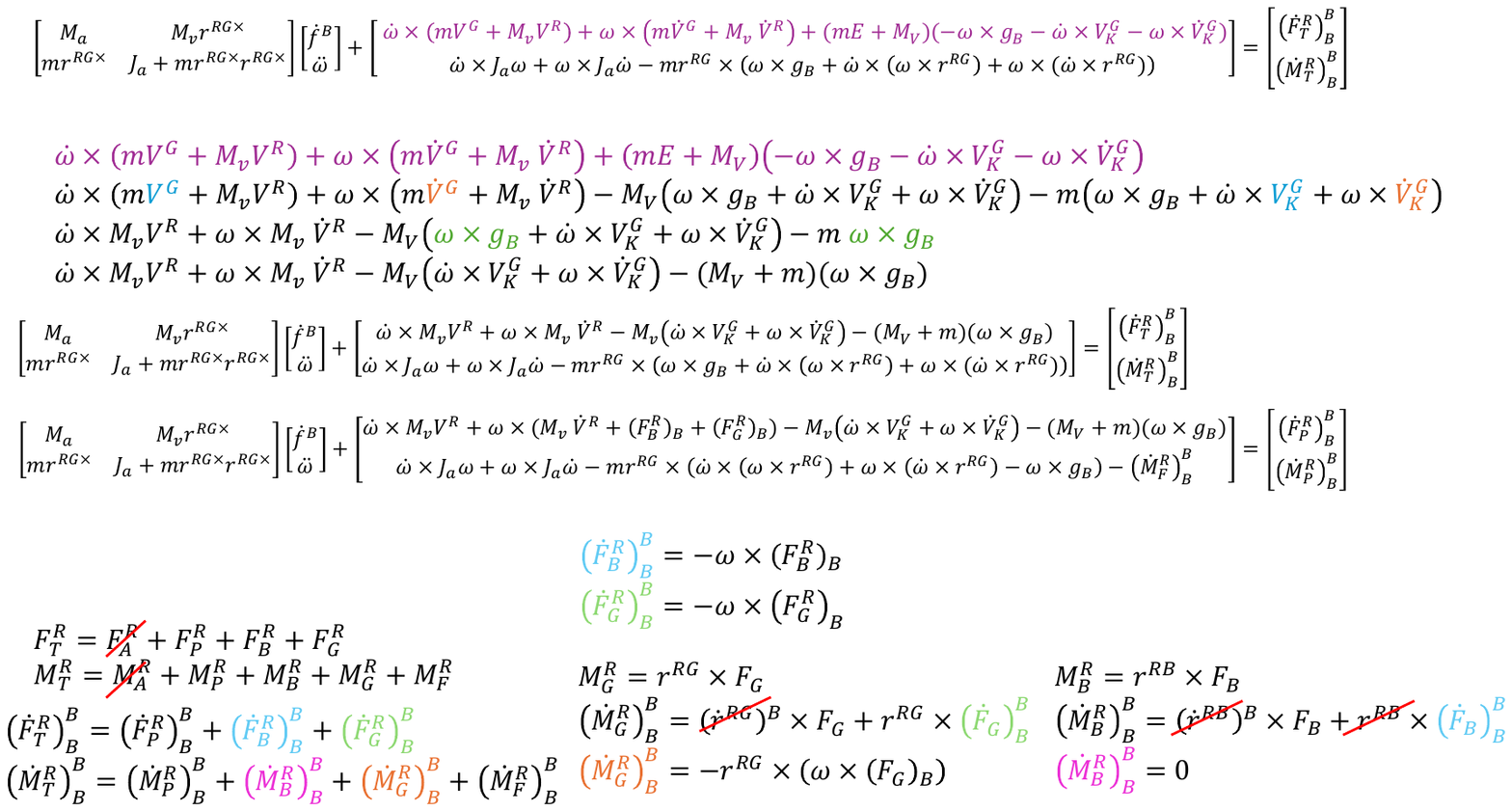}
    \caption{Derivation of the Jerk-Level Equation of Motion Part 3}
\end{figure}

  \end{appendix}

\end{backMatter}

\end{document}